\documentclass{article}[times, 12pt]
\usepackage{graphicx} 
\usepackage{natbib}
\usepackage{tikz}
\usetikzlibrary{shapes,arrows}
\usepackage{pdflscape}
\usepackage{longtable,booktabs,array}
\usepackage{tabu}
\usepackage{amssymb}
\usepackage[margin=2cm]{geometry}
\usepackage{float}
\usepackage{subcaption}
\usepackage{rotating}
\usepackage{hyperref}
\usepackage{amsmath}
\usepackage{xfrac}
\usepackage{adjustbox}
\usepackage{nicefrac}
\usepackage{pdflscape}

\hypersetup{
	colorlinks=true,
	linkcolor=blue,
	filecolor=blue,      
	urlcolor=blue,
	citecolor=blue
}

\title{Sample Design and Cross-sectional Weights for the Brazilian PCSVDF-Mulher Study: Integrating a Refreshment Sample with an Ongoing Longitudinal Wave to Calculate IPV Prevalence}

\author{Jos\'{e} Raimundo Carvalho\thanks{Professor in the Graduate Program in Economics, CAEN/UFC and PI of the PCSVDF-Mulher Project. Av. da Universidade, 2762 - Prédio CAEN/História - 1º e 2º andares - Benfica - CEP: 60.020-181 - Fortaleza-CE. E-mail: josecarv@ufc.br. ORCID: 0000-0001-5774-5925 (Author for correspondence).} \and Diego de Maria André\thanks{Professor in the Graduate Program in Economics (PPECO), Federal University of Rio Grande do Norte (UFRN). Av. Sen. Salgado Filho, 3000, Lagoa Nova, Natal/RN, CEP 59078-970, Brazil. E-mail:diego.andre@ufrn.br. ORCID: 0000-0003-3142-8336.}}

\usepackage{setspace}
\begin{document}

\maketitle

\begin{abstract}
\noindent Addressing unit nonresponse between waves of longitudinal studies using sampling design in weighting has recently incorporated a new strategy based on the availability of supplemental samples, collecting either refreshment or replacement samples on an ongoing larger sample. We implement an approach for calculating individual cross-sectional weights and apply them to the 2016 and 2017 waves of the PCSVDF-Mulher study, a large ($\approxeq 10,000$), complex, multipurpose, population-representative and interdisciplinary longitudinal household dataset in Brazil to study Intimate Partner Violence (IPV). We developed a set of weights that combines a refreshment sample collected in 2017 with the ongoing sample started in 2016. With this set of individual weights, we calculated IPV prevalence for nine capital cities in Brazil. As far as we know, this is the first attempt to calculate cross-sectional weights with supplemental samples applied to a representative sample in a developing country. Our analysis produced a set of weights that shows neglected survey design methodological improvements on IPV measurement. One of our findings pointed out that, even in well-designed longitudinal surveys, the indiscriminate use of unweighted designs to calculate IPV prevalence might inadvertently inflate values, bringing distortions with scientific implications. Also, by not incorporating weights, the analysis neglects variance reduction gains.
\vspace{0.5cm}
\\
\noindent \textit{Keywords}: Survey Design, Intimate Partner Violence, Longitudinal Data, Refreshment Sample \\
\end{abstract}

\newpage

\section{Introduction}
Intimate Partner Violence (IPV) is one of the most persistent social scourges in societies and a significant public health issue, as well as a violation of human rights. The World Health Organization (WHO), see, \citep{Sardinha2022}, asserts that more than one-third of women globally experience violence perpetrated by their partners or ex-partners and that high prevalence has impacted on women's and their offsprings' health, labor productivity, educational achievement, and many other socioeconomic dimensions.

However, only more recently, there is a movement of intellectual convergence on the interdisciplinary scholarship voicing the need for looking for real ``causal explanations'' for IPV to back up the design of strategies to address the problem; see, \citep{Jewkes2002}, \citep{Hsiao2007}, \cite{Averett2016}, \citep{Mulla2018}, \citep{Rose2018},  and \citep{Rothman2018}. These same authors and many others (see, \citep{Johnson2005} and \citep{Vaisey2016}) have already recognized the importance of collecting better data, invariably meaning collecting longitudinal evidence. The situation, regrettably, is of a paucity of large, population representative IPV longitudinal data sets, especially in developing countries, \citep{NAP2002}.

Even if a researcher of IPV can operate under an experimental setup or if the study objective is to calculate time changes in prevalence values, the importance of collecting and modeling longitudinal data to either make ``causal claims'' or to improve the quality of estimates is indisputable, see, \citep{Rose2000-po}, \citep{Singer2003-ks}, \citep{Frees2004-qp}, and \citep{Andress2015-kr}. Notwithstanding the value of longitudinal designs, this type of data brings its own set of challenges, among them the issue on how to best incorporate the sampling design in weighting adjustments for panels stand out, especially if unit nonresponse (attrition or missing data) is a likely occurrence, see, \citep{Valliant2018}, and \citep{LynnLongitudinal2021}.

More recently, the approach to addressing unit nonresponse between waves of longitudinal studies by means of sampling design in weighting has focused on modern missing data analysis procedures that incorporate a strategy based on the availability of supplemental samples, either collecting refreshment or replacement samples on an ongoing larger sample, \citep{hirano_combining_2001, Vehovar2003, taylor_evaluating_2020, Watson2021}.

These samples can be useful in many dimensions. It can restore initial sample sizes in such a way that sample representativeness is brought back again; address non-coverage, increase overall sample size, and increase the sample size of particular sub-groups, \citep{Watson2021}. As stated by \citep{hirano_combining_2001}, these supplemental samples can also be helpful in mitigating the effects of attrition in two ways. First, it can make the estimation of conventional attrition models more robust and precise, and allow for testing them. Second, it allows for the estimation of richer models, potentially resolving differences between selection models common in the statistical and econometric literature.

The paper has two main objectives. The foremost aim is to implement a strategy for calculating individual cross-sectional weights and apply them to the 2016 and 2017 waves of the PCSVDF-Mulher (Pesquisa de Condições Socioeconômicas e Violência Doméstica e Familiar contra a Mulher - Survey of Socioeconomic Conditions and Domestic and Family Violence against Women), a large and interdisciplinary longitudinal data set in Brazil to study violence against women, its causes and consequences. Indeed, the PCSVDF-Mulher Project already collected two more additional waves, in 2019 and in 2021.

Following the proposed solutions of \citep{o2002combining}, \citep{Watson_2014}, and \citep{Watson2021}, we have devised a set of weights that combines a refreshment sample collected in 2017 with the ongoing longitudinal sample started in 2016. Armed with this set of individual weights, we calculate IPV values of prevalence for nine capital cities in Brazil for the years 2016 and 2017. As far as we know, this is the first attempt to calculate cross-sectional weights with the aid of supplemental samples applied to a large population ($\approx 10,000$) sample focused on IPV.

The second objective is to present to a broader audience the PCSVDF-Mulher project, emphasizing how it might serve as an empirical source to back up studies interested in disentangling among the many theories about the etiology of IPV under a ``causal framework'' as well as measuring the effect of IPV on individual life trajectories.

Besides this Introduction, Section \ref{SECTION_PcsvdfStudy} provides more details about the PCSVDF-Mulher project and how IPV is measured. Section \ref{SECTION_StudyPopulation} describes the population under study, the available sample for analysis, and some basic information about the refreshment sample; and Section \ref{SECTION_Ethics} points out the ethical and methodological guidelines incorporated all over the study, especially at the data collection phase. Section \ref{SECTION_SamplingPlan} outlines the PCSVDF-Mulher complex, stratified, multistage probability cluster sampling
design; Section \ref{SECTION_SampleSelecModels} describes the way the various stages of sampling were determined; and Section \ref{SECTION_SampleScheme} calculates the probability of inclusion in the sample for each individual observation at each stage of the sampling process. 

Section \ref{SECTION_WeightsCalculation} defines and calculates the cross-sectional weights for wave 2016; addresses non-response (attrition) around the ``IPV Section'' of the questionnaire (not all women accept to answer that section); and calculate both the weighted and non-weighted IPV values of prevalence for all cities and for all combinations of type of IPV, and time window. A similar exercise is performed for the 2017 wave, with an emphasis on how the supplemental sample was numerically incorporated into the weighting design. Section \ref{SECTION_ComparingDesigns} is the core analytical exercise of the paper. There, we perform two simple exercises of comparing the weighted ($PREV_{j}^{w}$) and unweighted ($PREV_{j}^{unw}$) designs for $j = 2016 \textrm{ and } 2017$, in order to assess how the difference (if any) presents itself. 

Our attempt to calculate individual cross-sectional weights for the PCSVDF-Mulher project produced a set of operational weights whose comparison between weighted and unweighted designs shows clearly neglected trends in the literature of IPV measurement. Actually, one of our key results pointed out to the fact that, even in well-designed longitudinal household surveys as the PCSVDF-Mulher project, the indiscriminate use of unweighted designs to calculate IPV prevalence might artificially and inadvertently inflate their values, which brings distortions and likely political, social, budgetary, and scientific implications. Also, unweighted designs seem to be throwing away variance reduction gains brought by weighting. Section \ref{SECTION_Final} concludes with some remarks on how our analysis might help the quest for measurement improvements of IPV prevalence in Brazil and other similarly developing countries.

\section{The PCSVDF-Mulher Study}\label{SECTION_PcsvdfStudy}
\subsection{An Overall View}
The PCSVDF-Mulher is an interdisciplinary effort to build empirical longitudinal evidence that enables the study of violence against women (including Intimate Partner Violence - IPV), its causes and consequences to direct and indirect victims; the allocation of resources in the household; women and children's health; child development; and the interrelationships among them through an ethical and methodological sound approach, see, \citep{Carvalho2018}.

The Project has a strong empirical orientation, recognising important scientific gaps in IPV conceptualization, measurement and modelling; and aiming at closing them by collecting data to address issues such as: 1) lack of individually disaggregated violence against women surveillance information using uniform definitions and survey methods; 2) paucity of frequent nationally and sub-nationally (state level) representative data, measured consistently over time to monitor trends; and 3) need for micro longitudinal data at the individual and household levels with improved quality, more detail than previous surveys to increase understanding of nature, context, severity, and consequences of domestic violence.

Funded by the Special Secretariat of Policies for Women/Ministry of Justice, Brazil, the PCSVDF-Mulher used a CAPI (\textit{Computer-Assisted Personal Interviewing}) methodology to gather information from more than 10,000 women aged between 15 and 50 years old ($15 \leq age < 50$) who lived in all nine capitals of northeastern Brazilian States in two waves, 2016 and 2017 \citep{Carvalho2018}. Wave 1 was administered between May 03 - August 01, 2016, and Wave 2 was administered between March 16 - October 25, 2017.

Besides information on violence against women, the PCSVDF-Mulher provides data about socioeconomic characterization of household members; general and reproductive health; norms, awareness/knowledge about violence against women and the ``Maria da Penha Law''; bargaining power and empowerment; match evaluation; labor market; participation in state and federal social programs; history of pregnancies; history of partnerships; experiences of violence (current partner, ex-partner (most recent) or any other ex-partner); women's subjective expectations and beliefs relative to the her welfare and partner's abuse; match valuation, subjective expectations about risk of victimization, subjective counterfactuals; and many others \citep{Carvalho2018}. Look at Table \ref{secao_ondas} and the questionnaires (English versions) available at \href{https://ufrnedubr-my.sharepoint.com/:f:/g/personal/diego_andre_ufrn_br/EiksJ-qWespJpGVQ7bMrQpEBd4snxrMHeqrD_xH5Pm5rtg?e=tMG6Lu}{Additional Files\_Cross-Sectional Weights PCSVDF-Mulher} for details. Also, interviewers answered a questionnaire at each wave which included questions on socioeconomic background, interviewers experience in survey data collection, attitudes, behavior and others. 

The breadth and interdisciplinary structure of the PCSVDF-Mulher questionnaire supports our vision of approaching violence aginst women from a modern, rigorous, and interdisciplinary perspective. We employed a methodology based on the best international studies on victimization and gender violence. More specifically, as to aspects of gender violence questions and interview protocols, we want to mention the World Health Organization - WHO study ``Multi-country Study on Women's Health and Domestic Violence against Women'', see, \citep{MorenoJansen2005}; and the ``Demographic and Health Surveys (DHS) Program'' from the United States Agency for International Development (USAID), see, \citep{Yount2022}.

\begin{table}[!ht]
\caption{Questionnaire Sections in the PCSVDF-Mulher 2016 and 2017 Waves}
\label{secao_ondas}
\normalsize{
\begin{tabular}{p{12cm}|c|c}

SECTION & Wave 2016 & Wave 2017\\
\hline
Household Visited & & X\\ \hline
Administration Form - Random Numbers & X & X \\
\hline
Woman's Selection Form (WSF) - Characterization of Household Members & X & X \\
\hline
Identification of the Woman who Participated in the First Wave & & X\\
\hline
Household Selection Form & X & X\\
\hline
Woman's Questionnaire (WQ) - General and Reproductive Health & X & X\\
\hline
Norms, Awareness/Knowledge About Violence Against Women and the ``Maria da Penha Law'' & X & X\\
\hline
Respondent and her Partner & X & X\\
\hline
Bargaining Power/Bargaining and Empowerment & X & X \\
\hline
Experiences of Violence (Current Partner, Ex-Partner (Most Recent) or Any Other Ex-Partner & X & X \\
\hline
Experiences of Violence not Related to Partner & X & X\\
\hline
Match Valuation, Subjective Expectations and Counterfactuals & X & X\\
\hline
Supplementary Section & X & X \\
\hline
Resuls & X & X \\
\hline            
\end{tabular}}
\end{table}

\noindent The PCSVDF-Mulher was designed with such perspective and can effectively add to this existing literature. Its longitudinal format allow us to further address identification issues inherent to attempts to estimate causal effects. In essence, not only economists but sociologists, statisticians, criminologists, psychologists, public health, epidemiologists, gender studies specialists and justice and law scholars can effectively find in PCSVDF-Mulher a good empirical benchmark to advance their analysis on the causes and consequences of domestic violence.

\subsection{Measuring the Prevalence of Intimate Partner Violence}
Before we calculate IPV prevalences, it is important to understand the conceptual, statistical and behavioural nuances related to the methodology followed by us. From a conceptual perspective, we need first to examine with detail the concept of intimate partner violence, how to operationalize it and how to calculate it within PCSVDF-Mulher.

We followed as close as possible \citep{Renzetti2011} and \citep{BreidingEtAll2015} who put forward the following two basic definitions. First, Intimate Partner Violence includes physical violence, sexual violence, stalking and psychological aggression (including coercive tactics) by a current or former intimate partner (i.e., spouse, boyfriend/girlfriend, dating partner, or ongoing sexual partner). Second, an Intimate Partner is a person with whom one has a close personal relationship that may be characterized by the partners' emotional connectedness, regular contact, ongoing physical contact and sexual behavior, identity as a couple, and familiarity and knowledge about each other's lives. The relationship need not involve all of these dimensions. Intimate partner relationships include current or former spouses (married spouses, common-law spouses, civil union spouses, domestic partners), boyfriends/girlfriends, dating partners, or ongoing sexual partners. Intimate partners may or may not be cohabiting. Intimate partners can be opposite or same sex.

As to discussions about measurement scales for IPV appearing in \citep{Thompsn2006MeasuringIP}, \citep{Costa2016} and \citep{TarrioConcejero2022}, it is worth to mention that there are three main dominant scales to measure IPV, say, the Conflict Tactic Scale - CTS, original or revised (\citep{Straus1996}); the Abuse Assessment Screen (\citep{McFarlane1995}); and the WHO scale which is a CTS-type scale, (\citep{MorenoJansen2005}). The PCSVDF-Mulher project adhered to the WHO scale.

From a statistical perspective, IPV has three dimensions (physical, sexual, and emotional) where each one is measured by a set of multi-item dichotomous response questions: 1) Physical violence by an intimate partner, with eight items with ``yes'' or ``no'' answers each (Was slapped or had something thrown at her that could hurt her; Was pushed or shoved; Was hit with fist or something else that could hurt; Was kicked, dragged or beaten up; Was strangled;Was choked or burnt on purpose; Perpetrator threatened to use used a gun, knife or other weapon against her; and Perpetrator actually used a gun, knife or other weapon against her); 2) Sexual violence by an intimate partner, with three items with ``yes'' or ``no'' answers each (Was physically forced to have sexual intercourse when she did not want to; Had sexual intercourse when she did not  want to because she was afraid of what partner  might do; and Was forced to do something sexual that she found degrading or humiliating); and 3) Emotional abuse by an intimate partner, with five items with ``yes'' or ``no'' answers each (Was insulted or made to feel bad about herself; Was belittled or humiliated in front of  other people; Perpetrator had done things to scare or  intimidate her on purpose, e.g. by the way he  looked at her, by yelling or smashing things; and Perpetrator had threatened to hurt someone she cared about), see, \citep{Straus1996}, \citep{MorenoJansen2005}, \citep{BreidingEtAll2015}, and \citep{Chapman2019}.

The most common approach to measure IPV prevalence is to assign a status of victim of IPV if a woman answers ``yes'' to at least one of the corresponding items for a specific dimension of IPV. Last, the literature considered two important operational time frames when asking these questions, as there is the need to explicitly define a retrospective interval in the interviewer's past when questioning him/her: ``Lifetime'' and ``12 Past Months'', see \citep{MorenoJansen2005}. 

Estimating the prevalence of IPV is a challenge as victims' and perpetrators' behaviors interact in complex ways especially in developing countries such as Brazil, giving rise to the notorious under-reporting of both prevalence and its intensity. The literature on violence against women repeatedly emphasizes that the vast majority of victims do not seek help, and those who do not seek support tend to resort to informal networks of friends, neighbors, relatives, religious institutions, or community organizations (see, \citep{Ruiz-Perez2007} and \citep{UN2014}).

\section{Study Population and Available Sample}\label{SECTION_StudyPopulation}

\subsection{Population}

\begin{figure}[h!]
\centering
\includegraphics[scale=0.60]{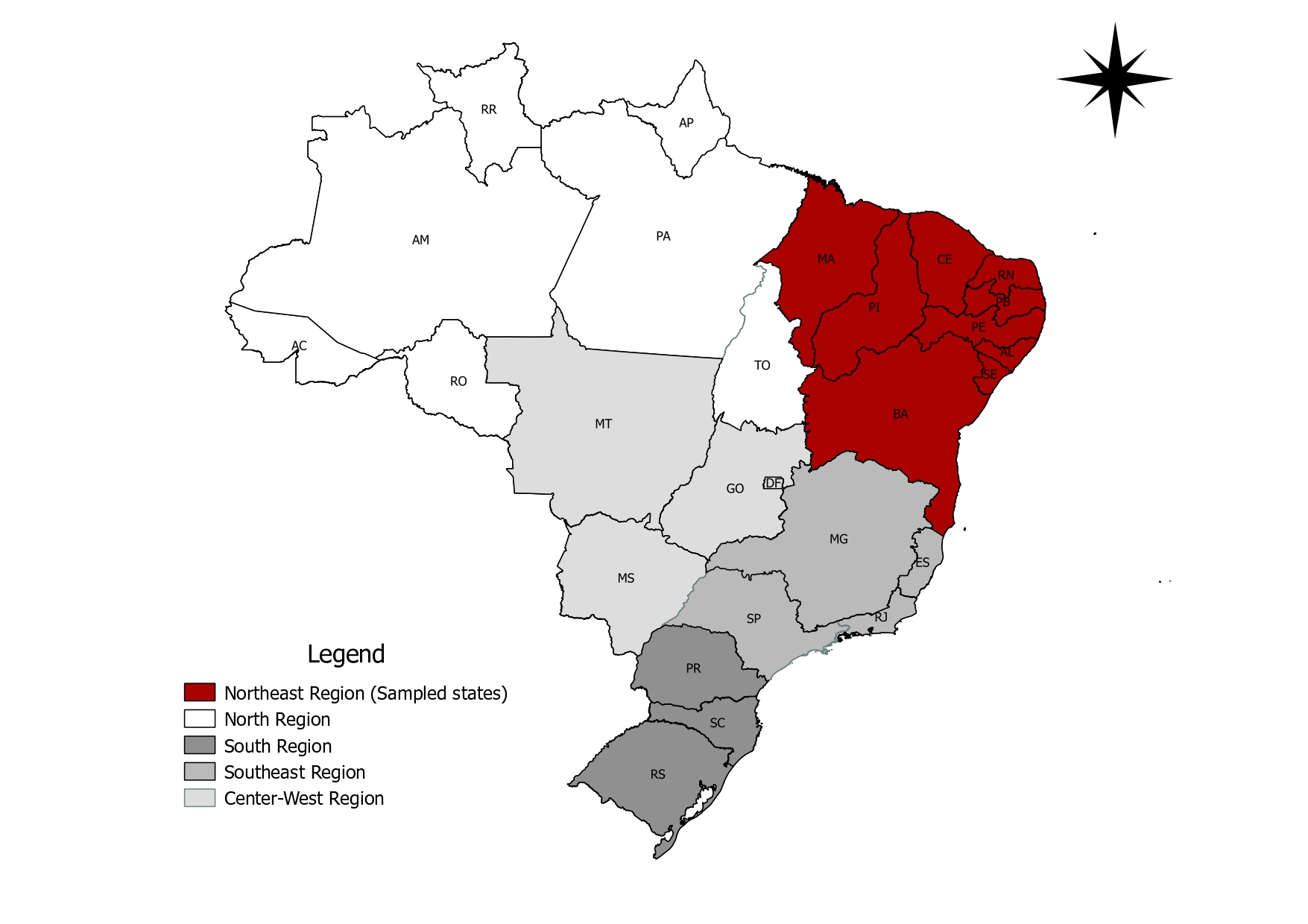}
\caption{Sample Distribution of PCSVDF-Mulher 2016 and 2017 Samples}
\label{fig:mapa}
\end{figure}

The study population was defined as all women between 15 and 50 years of age residing in permanent private households in the capitals (Aracaju/SE, Fortaleza/CE, João Pessoa/PB, Maceió/AL, Natal/RN, Recife/PE, Salvador/BA, São Luís/MA, and Teresina/PI) of all states in the northeast region of Brazil (see, Figure \ref{fig:mapa}).

\noindent The northeastern region of Brazil occupies an area of approximately 1.5 million km², making up 18\% of Brazil's territory. Its population is 57,071,654 inhabitants, and its demographic density is 39.64 inhabitants per km². Northeastern states have low educational achievement compared to those in the South or South-west states. Moreover, Northeastern states also present smaller per capita household incomes than the rest of the country. The average monthly per capita household income in  2022 for Brazilian regions are: Nordeste (Northeast)	$R\$$ 1.053, Norte (North)	 $R\$$ 1.143, Sudeste (Southeast) $R\$$ 1.842,  Sul $R\$$ 1.983, and Centro-Oeste (Center-West) $R\$$ 2.011 (see, \cite{IBGE2022}).

\subsection{Available Sample for Analysis}

In PCSVDF-Mulher Wave 1 (2016), the original sample was composed of 10,094 observations. In order to carry out the weighting, we needed the GPS coordinates of the residences and, unfortunately, 258 (2.56\%) observations did not have their coordinates retrieved, leaving 9,836 observations after this first cleaning up. After calculating the weights, we carried out a post-stratification process using variables age, race, and education. During this step, we lost another 218 (2.22\% of the 9,836) observations.

Thus, our final weighted base has 9,618 observations, with a loss of 476 observations, equivalent to 4.72\% of the original sample collected. In Wave 2 (2017), the original base collected has 10,518 observations. A total of 251 observations were lost due to lack of GPS and a further 422 in the post-stratification process, resulting in a sample of 9,845 observations (loss of 673 observations, equivalent to 6.4\%).

Table \ref{tab:attrition_refreshment} depicts the original sample, attrition level, and refreshment sample by city and by type of refreshment (either at the same or at a different household). The PCSVDF-Mulher refreshment sample protocol resembles the Understanding Society study (\citep{Burton2012}), at least as regards the strategy of selecting a new sample closer to the same primary sampling units as the original sample in order to result in efficient field workloads, \citep{Burton2012}, and \citep{Watson2021}.

\begin{table}[h!] 
\caption{\label{tab:attrition_refreshment}Collected, Attrited, and Refreshed Unit Samples, 2016 and 2017}
\centering
\begin{tabu} to \linewidth {>{\raggedright\arraybackslash}p{3cm}>{\centering}X>{\centering}X>{\centering}X>{\centering}X>{\centering}X>{\centering}X>{\centering}X}
\toprule
  City & W2016 & $ATT_{OUT}$ & $ATT_{IN}$ & ATT & PAIR & $REF_{OUT}$ & W2017 \\
  &  &  &  &  &  &  & \\
  & a & b & c &  d=b+c & e=a-d & f & g=f+c+e \\
\midrule
  Aracaju/SE     & 986   & 506   & 88  & 594 & 392 & 521 & 1,001\\ 
  Fortaleza/CE   & 1,172 & 473   & 31  & 504 & 668 & 534 & 1,233\\
  J Pessoa/PB    & 1,072 & 599   & 73  & 672 & 400 & 522 & 995\\
  Maceio/AL      & 943   & 469   & 36  & 505 & 438 & 599 & 1,073\\ 
  Natal/RN       & 1,052 & 582   & 106 & 688 & 364 & 621 & 1,091\\
  Recife/PE      & 1,245 & 301   & 332 & 633 & 612 & 268 & 1,212\\
  Salvador/BA    & 1,104 & 506   & 53  & 559 & 545 & 589 & 1,187\\   
  S Luís/MA      & 1,082 & 504   & 112 & 616 & 466 & 519 & 1,097\\ 
  Teresina/PI    & 962   & 428   & 103 & 531 & 431 & 422 & 956\\ 
\midrule
  Total (Nordeste)         & 9,618 & 4,368 & 934 & 5,302 & 4,316 & 4,595 & 9,845\\
\bottomrule
\multicolumn{8}{l}{\rule{0pt}{1em}\textit{Source:} Elaborated by the authors.}\\
\multicolumn{8}{l}{\rule{0pt}{1em}\textit{Note 1:} The unit of observation is a woman in a given household.} \\
\multicolumn{8}{l}{\rule{0pt}{1em}\textit{Note 2:} W2016 and W2017 = collected questionnaires with GPS coordinates in 2016 and 2017.} \\
\multicolumn{8}{l}{\rule{0pt}{1em}\textit{Note 3:} $ATT_{OUT}$ = attrition outside the 2016 household; $ATT_{IN}$ = attrition inside the 2016 household.} \\
\multicolumn{8}{l}{\rule{0pt}{1em}\textit{Note 4:} PAIR = woman present at both waves; and $REF_{OUT}$ = refreshment outside the 2016 household.} \\
\multicolumn{8}{l}{\rule{0pt}{1em}\textit{Note 5:} An attrition inside the 2016 household is coupled with a refreshment, say, her substitute.} \\
\multicolumn{8}{l}{\rule{0pt}{1em}\textit{Note 6:} Our calculated attrition rate was $55.13\%$, $\sfrac{5,302}{9,618}$.}
\end{tabu}
\end{table}

\noindent At first glance, a $55.13\%$ attrition rate seems too high or even disappointing, however, a more detailed analysis do not corroborate that. First, the study asks a considerable amount of ``sensitive questions’’ such as IPV, use of contraceptive, abortion, alcohol and drug consumption, infidelity and others. A fair amount of literature related to the accuracy of survey reports about topics such as illicit drug use, abortion, and sexual behavior has been established since the seminal paper of \citep{Tourangeau2007}. These concerns are also related to attrition rates in longitudinal studies due to these type of questions. So, considering the amount of sensitiveness present at our study, we believe we fare reasonably well.

Second, our attrition was encouraging, indeed, considering the length of our questionnaire (with an average duration of 48.37 min (s.d. 29.01) for Wave 1 and 49.47 min (s.d. 29.78) for Wave 2, and the fact that there was no financial incentives for women to participate in the study. Incentives to participate in such type of household studies are still hotly contested in Brazil.

Third, if you compare the PCSVDF-Mulher attrition rate to those in the (few) longitudinal studies with similar characteristics another picture emerges. For instances, \citep{Davis1997} an RCT-type study that assigned families who reported domestic incidents in New York to receive police patrol follow-up visits after the incident. Participation in domestic violence education programs was also randomized (sample size = 436; inter-wave interval = 6 months; and attrition rate $28.0\%$); \citep{Gondolf2001} an RCT-type study where male perpetrators of domestic violence were randomized into two groups: i)
experimental, where they were sentenced to 1-year probation and 26 weeks of group
counseling sessions; and ii) control, where they were sentenced to only one year of
probation. The study followed both men and their female victims (sample size = 404; inter-wave interval = 12 months; and attrition rate $79.0\%$); \citep{mertin2001follow} a study that evaluates recovery from post-traumatic stress disorder (PTSD), anxiety and depression consequences of domestic violence (sample size = 100; inter-wave interval = 12 months; and attrition rate $41.0\%$); and \citep{MCHUGO2005} a cooperative study to evaluate new service
models for women with mental health and substance use disorders and a history of physical and/or sexual abuse (sample size = 2,729; inter-wave interval = 6 months; and attrition rate $26.5\%$).

\section{Ethical Aspects}\label{SECTION_Ethics}

The PCSVDF-Mulher follows the ethical and methodological guidelines from the World Health Organization (WHO)'s report `Putting Women First,' which focuses explicitly on the ethics of research related to domestic violence interventions \citep{WHO2001}, as well as those from the DHS – Demographic and Health Surveys program (\citep{DHS2020}). We adhered to the following recommendations from the  \citep{WHO2001} study:

\begin{itemize}
\item Full informed consent, privacy, and confidentiality;
\item Use of female interviewers only
\item When communicating about the study to men, not framing the study as a domestic violence study but rather a study on decision-making in the household;
\item Procedures to manage conflict situations and ensure privacy
\item Training to interviewers on how to deal with privacy, sensitivity, confidentiality, and logistics;
\item Anonymization and data-protection;
\item Referral to local services and sources of support for women experiencing domestic violence. 
\item Measuring and monitoring harm related to the research.
\end{itemize}

\noindent The Brazilian National Committee for Ethics in Research (CONEP - \url{https://conselho.saude.gov.br/comissoes-cns/conep?view=default}) approved the first two waves of the study under number CAAE 53690816.5.0000.5054. 

All interviews were conducted by trained female interviewers whose educational background was at least that of incomplete higher education. The training took place in partnership with the Maria da Penha Institute which discussed issues related to basic gender definitions, intimate partner violence concepts and specificities, field behavior, use of laptops, and field operations' risk management. Founded in 2009, with headquarters in Fortaleza/CE and representation in Recife/PE, the Maria da Penha Institute (IMP) is a non-profit, non-governmental organization that aims to contribute and strengthen mechanisms to curb and prevent domestic and family violence against women, according to Art. 1 of Law no. 11,340/2006. For more information, see: \url{https://www.institutomariadapenha.org.br}. 

The entire training lasted for 40 hours and was supported by a well-developed set of 5 Training and Field Manuals (Field Manual;
MODULE 1 - Gender and Domestic Violence against Women; MODULE 2 - Gender, Diversity and Ethnic-Racial Relations; MODULE 3 - Maria da Penha Law and Public Policies; and MODULE 4 - Ethics in Research on Domestic Violence and Conflict Management) available at \href{https://ufrnedubr-my.sharepoint.com/:f:/g/personal/diego_andre_ufrn_br/EiksJ-qWespJpGVQ7bMrQpEBd4snxrMHeqrD_xH5Pm5rtg?e=tMG6Lu}{Additional Files\_Cross-Sectional Weights PCSVDF-Mulher}.

\section{Sampling Plan and Stratification}\label{SECTION_SamplingPlan}

The PCSVDF-Mulher sampling plan used a complex, stratified, multistage probability cluster sampling design with unequal selection probabilities to select Wave I participants. The primary sampling unit (PSU) was a neighborhood (\textit{bairro} in Portuguese), and the elementary sampling unit was always a resident woman. In each selected household, all residents were enrolled, and the study's target data were recorded for a randmoly selected woman aged between 15 and 50 years ($15 \leq age < 50$).

At each city sampled, strata were formed by allocating all neighborhoods in four blocks according to the number of permanent private households. To define each block, we rank all neighborhoods in decreasing order of the number of households and stratify them into four quartiles according to the number of households in each. The separation criteria for each quartile is roughly 25\% of the accumulated number of households. If adding the very next neighborhood adds up more than 25\% to the quartile under formation, this neighborhood will be the first of the very next quartile. Of course, the four strata have no identical sizes. We avoid splitting neighborhoods into different quartiles, though.

We finish the first procedure with four strata, which divide the whole city into sets of four quartiles with approximately the same number of households. The data for the total number of permanent private households were obtained from the 2010 Brazilian census. At each one of the 4 strata formed at each of the nine cites, the primary sampling unit (PSU) was a neighborhood, the secondary sampling unit (SSU) was a census track, the tertiary sampling unit (TSU) was a household, and the quaternary sampling unit (QSU) was an eligible woman. These four strata define four stages of sample selection, which are described next.

\section{Sample Selection Methods in the Various Stages}\label{SECTION_SampleSelecModels}

In the first stage, neighborhoods were chosen through systematic sampling in each stratum by a process we will assume can be approximated by random sampling. Then, knowing the number of neighborhoods selected at each stratum is sufficient to calculate the probability of sampling that neighborhood.

In the second stage, some census tracks were selected at each previously selected neighborhood. By design, our research group has stipulated that no less than 4 and no more than 12 questionnaires should be collected at each census track. Hence, to fulfill the number of questionnaires at each selected neighborhood, a variable number of census tracks was selected with probability proportional to their number of households. 

In the third stage, the interviewer received a random geographic starting point within the census tract from which she would start approaching households in an attempt to conduct interviews with eligible women. Between one interview and the next, the interviewer should maintain a minimum distance of 4 houses between interviews. For example, if she conducts an interview in the first house of the block, she should skip the second, third, fourth, and fifth houses and then attempt the subsequent interview in the sixth house. This skip considers both sides of the street, i.e., the interviewer cannot conduct an interview across the street within 4 houses of the first interview conducted. She could, however, skip a greater number of squares before conducting the subsequent interview. Although this skipping pattern can induce bias, we are convinced it is insufficient to compromise the sampling process.

The fourth stage corresponds to selecting a woman aged between 15 and 50. Inside the household, interviewers collected some initial socio-demographic information for all residents before implementing the full questionnaire. If there is no eligible woman, the interviewer ends the interview process and returns to the protocol to choose the next household to be approached. If there is only one eligible woman, she is invited to complete the survey. If there are multiple eligible women, we conduct a random drawing using the tablet's own pre-programmed random device to select the woman who will be invited to participate in the survey. The interviewer does not participate in the randomized drawing process.

\section{Probabilistic Sampling Scheme}\label{SECTION_SampleScheme}

The probability of inclusion in the sample of neighborhood $i$ in stratum $k$, represented by $P(N_{ki})$, is proportional to the number of households in the neighborhood, as indicated by equation (\ref{prob_bairro}):


\begin{equation}\label{prob_bairro}
P(N_{ki}) = n_{k} * \frac{NHouseholds_{ki}}{NHouseholds_{k}}
\end{equation}

\noindent where \(n_{k}\) is the number of neighborhoods sampled in stratum \(k\), \(NHouseholds_{ki}\) is the number of households in neighborhood \(i\) of stratum \(k\), and \(NHouseholds_{k}\) is the number of households in stratum \(k\).

The conditional probability of inclusion in the sample of census tract $j$ in neighborhood $i$ in stratum $k$, conditioned by the selection of neighborhood $ki$, represented by $P(S_{kij}|N_{ki})$, is given by equation (\ref{prob_setor}):


\begin{equation}\label{prob_setor}
P(S_{kij}|N_{ki}) = s_{ik} * \frac{NHouseholds_{kij}}{NHouseholds_{ki}}
\end{equation}

\noindent where \(s_{ik}\) is the number of census tracts sampled in neighborhood \(i\) in stratum \(k\), \(NHouseholds_{kij}\) is the number of households in census tract \(j\), in neighborhood \(i\) and stratum \(k\).

In census tracts $S_{kij}$, the conditional probability of interviewing household $D_{kijh}$ is:


\begin{equation}\label{prob_res}
P(D_{kijh}|S_{kij}) = \frac{NVHouseholds_{kij}}{NHouseholds_{kij}}
\end{equation}

\noindent where \(NVHouseholds_{kij}\) is the number of households visited in census tract \(j\), neighborhood \(i\), and stratum \(k\), regardless of whether the interview attempt was successful or not.

Since even after contacting the residence, the interview may not be carried out, either because of refusal by the residents or because there is no eligible woman, it is necessary to carry out a non-response adjustment. Accordingly to \citet{PAD_MG_2010} and \citet{Souza2015}, the non-response adjustment followed the usual practice in household surveys and was performed separately in each sector. This approach is equivalent to assuming a model in which non-response probability is constant within a sector and varies for households in different sectors. The non-response adjustment is given by:


\begin{equation}\label{prob_nr}
P(DNR_{kijh}) = \frac{NVQuestionnaires_{kij}}{NVHouseholds_{kij}}
\end{equation}

\noindent Where \(NVQuestionnaires_{kij}\) is the number of valid questionnaires in census tract $S_{kij}$. Here, it is assumed that a questionnaire is valid if the eligible woman interviewed reaches the women's health section. Finally, the conditional probability of interviewing woman $W_{kijhw}$ in household $D_{kijh}$ is:

\begin{equation}\label{prob_mulher}
P(W_{kijhw}|D_{kijh}) = \frac{1}{E_{kijh}}
\end{equation}

\noindent where \(E_{kijh}\) is the number of eligible women, i.e., women aged between 15 and 49, in household \(h\), in census tract \(j\), neighborhood \(i\), and stratum \(k\).

Thus, the probability of inclusion in the sample of woman $W_{kijhw}$ is expressed by:

\begin{equation}\label{prob_final}
P(W_{kijhw}) =  P(N_{ki}) \times P(S_{kij}|N_{ki}) \times P(D_{kijh}|S_{kij}) \times P(DNR_{kijh}) \times P(W_{kijhw}|D_{kijh})
\end{equation}

\section{Weights and IPV Prevalence}\label{SECTION_WeightsCalculation}

\subsection{The 2016 Wave}

\subsubsection{Sample Weighting}

The PCSVDF-Mulher's complex sampling design and its unequal selection probabilities require the definition of sample weighting for data analysis. Following an approach similar to the Brazilian National Health Survey (PNS 2013) (\cite{Souza2015}), the final weight of the basic design is a product of the inverse selection probabilities at each stage of the sampling plan (plus the nonresponse correction processes) and calibration adjustments to the known population totals. The basic weight for the eligible woman is then calculated according to the following expression:

\begin{equation}\label{weights}
bw_{kijhw} = 1/P(W_{kijhw}) =  1/[P(N_{ki}) \times P(S_{kij}|N_{ki}) \times P(D_{kijh}|S_{kij}) \times P(DNR_{kijh}) \times P(W_{kijhw}|D_{kijh})]
\end{equation}

\noindent As usual, after calculating the basic weights, we performed a trimming in order to exclude outliers. We established the 5\% lowest values and the 5\% highest values of the weight distributions as the limit for the ``trimming''. All calculations were performed using the R software \citep{RCoreTeam}, and we used the \emph{trimWeights} function from the \emph{survey} package (for more details, see \cite{Lumley2020}). In that function, the total trimmed amount is divided among the observations that were not trimmed so that the total weight remains the same.

To adjust the sample to known populations totals, we apply the Raking method using variables age (Young {[}15-29{]} and Adult{[}30-59{]}), race (White and Non-white), and education (elementary, which also includes people with no education; high school; and undergraduate), from the Brazilian 2016 Pesquisa Nacional por Amostra de Domicílios (National Household Sample Survey), see, \cite{IBGE2016}.

According to \citet{Grande2015}, raking is an iterative process applied to the proportional adjustment of the weights. In each step, one variable is used to gradually adjust the data (or survey margin totals) to fit a specific characteristic of the population (or control totals). The process is repeated until the differences between all the categories' proportions from the population data and raked weights from the survey data margin are convergent. To do that, we counted on the function \textbf{rake}, from the \textbf{survey} package from the R software.

After the raking, we scaled the weights to have a mean of 1. After this process, we identified extreme outlying weights and trimmed the weights again. At this time, we established only the 5\% highest values of the weight distributions as the limit for the ``trimming''. Finally, we re-scaled the weights to have a mean of 1, giving us the final weight of the basic design. Table (\ref{tab:a14}) shows the distribution of these weights.

\begin{table}[h!]
\caption{\label{tab:a14}Sample Summary of Final Weights of the Basic Design by City - 2016}
\centering
\begin{tabu} to \linewidth {>{\raggedright\arraybackslash}p{2.5cm}>{\centering}X>{\centering}X>{\centering}X>{\centering}X>{\centering}X>{\centering}X>{\centering}X}
\toprule
  City & N & Mean & Sd & Min & Max & IQR & CV \\ 
\midrule
  Aracaju/SE   & 986 & 1.00 & 1.71 & 0.05 & 16.92 & 0.64 & 1.71 \\ 
  Fortaleza/CE & 1,172 & 1.00 & 1.35 & 0.08 & 13.45 & 0.84 & 1.35 \\
  J Pessoa/PB  & 1,072 & 1.00 & 1.00 & 0.09 & 7.54 & 0.80 & 1.00 \\ 
  Maceio/AL    & 943 & 1.00 & 1.01 & 0.07 & 11.50 & 0.92 & 1.01 \\ 
  Natal/RN     & 1,052 & 1.00 & 1.15 & 0.07 & 8.82 & 0.85 & 1.15 \\ 
  Recife/PE    & 1,245 & 1.00 & 1.37 & 0.06 & 13.71 & 0.88 & 1.37 \\
  Salvador/BA  & 1,104 & 1.00 & 1.78 & 0.03 & 18.01 & 0.76 & 1.78 \\ 
  S Luís/MA    & 1,082 & 1.00 & 1.19 & 0.07 & 8.56 & 0.62 & 1.19 \\ 
  Teresina/PI  & 962 & 1.00 & 1.02 & 0.07 & 7.77 & 0.79 & 1.02 \\ 
\bottomrule
\multicolumn{8}{l}{\rule{0pt}{1em}\textit{Source: } Elaborated by the authors.}\\
\end{tabu}
\end{table}
  
\subsubsection{Correction for the Questionnaire ``IPV Section'' Nonresponse}

In order to calculate the prevalence rates of intimate partner violence, it was necessary to create weights for non-response for the section on domestic violence against women (\emph{Experiences of Violence (Current Partner, Ex-Partner (Most Recent) or Any Other Ex-Partner)}), see, Table \ref{secao_ondas}. For this, in each city, the following logit model (weighted by the final weight of the basic design) was estimated:

\begin{equation}\label{logit}
Prob(answer\_violence_{i} = 1) = Logit \big( \beta_{0} + \beta_{1} cohab_{i} + \beta_{2} know\_victim_{i} + \beta_{3} children_{i}\big)
\end{equation}

\noindent Where \(answer\_violence_{i}\) is a \emph{dummy} variable that takes the value 1 if the woman answered the violence section and 0 if she did not; \(cohab_{i}\) is a \emph{dummy} variable that takes the value 1 if the woman lives in the same household with a partner; \(know\_victim_{i}\) is a \emph{dummy} variable that takes the value 1 if any woman in the interviewee's social circle has suffered domestic violence in the last 12 months; and \(children_{i}\) is a \emph{dummy} variable that takes the value 1 if the interviewee has at least one child. The results can be seen in the table \ref{eq:logit2016} in the appendix. The non-response weight is the inverse of the probability estimated by the logit model given by equation (\ref{logit}).

Thus, the final weight for estimating the prevalence rates of domestic violence is given by multiplying the final weight of the basic design $bw_{kijhw}$ by the non-response propensity to answer the domestic violence section estimated by the Logit in equation \ref{logit}. We also scaled these weights to have a mean of 1. 

\begin{table}[h!]
\caption{\label{tab:a15} Sample Summary of Final Weights for Calculating Prevalence by city - 2016}
\centering
\begin{tabu} to \linewidth {>{\raggedright\arraybackslash}p{2.5cm}>{\centering}X>{\centering}X>{\centering}X>{\centering}X>{\centering}X>{\centering}X>{\centering}X}
\toprule
  City & N & Mean & Sd & Min & Max & IQR & CV \\ 
\midrule
  Aracaju/SE   & 745 & 1.00 & 1.27 & 0.08 & 8.10 & 0.54 & 1.27 \\ 
  Fortaleza/CE & 897 & 1.00 & 1.07 & 0.12 & 6.74 & 0.73 & 1.07 \\ 
  J Pessoa/PB  & 836 & 1.00 & 0.90 & 0.15 & 4.62 & 0.86 & 0.90 \\ 
  Maceio/AL    & 739 & 1.00 & 0.88 & 0.17 & 4.44 & 0.87 & 0.88 \\ 
  Natal/RN     & 630 & 1.00 & 0.85 & 0.17 & 4.23 & 0.86 & 0.85 \\ 
  Recife/PE    & 783 & 1.00 & 1.03 & 0.05 & 5.49 & 0.79 & 1.03 \\ 
  Salvador/BA  & 922 & 1.00 & 1.33 & 0.05 & 10.10 & 0.63 & 1.33 \\ 
  S Luís/MA    & 794 & 1.00 & 0.96 & 0.19 & 5.34 & 0.68 & 0.96 \\ 
  Teresina/PI  & 632 & 1.00 & 0.82 & 0.19 & 5.28 & 0.82 & 0.82 \\
\bottomrule
\multicolumn{8}{l}{\rule{0pt}{1em}\textit{Source: } Elaborated by the authors.}\\
\end{tabu}
\end{table}
  

\subsubsection{Prevalence Rates}

Now, we compare prevalence rates of ``lifetime'' emotional violence, emotional violence ``in the last 12 months'', ``lifetime'' physical violence, physical violence ``in the last 12 months'', ``lifetime'' sexual violence, and sexual violence ``in the last 12 months''. We show these results by city, for both the original dataset (``Original dataset'' column) and for the cleaned dataset with complete information on GPS coordinates. For the later sample, we calculated prevalences without (``Unweighted dataset'' column) and with (``Weighted dataset'' column) final weights summarized by Table \ref{tab:a14}. We apply Taylor Series Linearization to calculate design-adjusted standard errors, as well as clusters and stratum information.

\begin{table}[H]

\caption{\label{tab:p1}Prevalence of  Emotional Violence Lifetime - 2016}
\centering
\begin{tabu} to \linewidth {>{\raggedright\arraybackslash}p{2cm}>{\centering}X>{\centering}X>{\centering\arraybackslash}p{2cm}|>{\centering}X>{\centering}X>{\centering\arraybackslash}p{2cm}|>{\centering\arraybackslash}p{1cm}>{\centering\arraybackslash}p{2cm}}
\toprule
\multicolumn{1}{c}{ } & \multicolumn{3}{c}{Original dataset} & \multicolumn{3}{c}{Unweighted dataset} & \multicolumn{2}{c}{Weighted dataset} \\
\cmidrule(l{3pt}r{3pt}){2-4} \cmidrule(l{3pt}r{3pt}){5-7} \cmidrule(l{3pt}r{3pt}){8-9}
  & n & Prev & CI & n & Prev & CI & Prev & CI\\
\midrule
Aracaju/SE & 545 & 27.52 & 23.93 - 31.43 & 536 & 27.80 & 20.22 - 36.91 & 27.16 & 20.06 - 35.66\\
Fortaleza/CE & 745 & 27.92 & 24.81 - 31.26 & 710 & 28.17 & 24.46 - 32.20 & 25.77 & 21.62 - 30.40\\
J Pessoa/PB & 721 & 34.81 & 31.42 - 38.37 & 686 & 35.28 & 30.01 - 40.93 & 32.67 & 26.69 - 39.28\\
Maceió/AL & 521 & 35.32 & 31.32 - 39.53 & 471 & 35.03 & 29.54 - 40.95 & 33.46 & 27.47 - 40.05\\
Natal/RN & 354 & 39.27 & 34.30 - 44.46 & 342 & 38.30 & 33.03 - 43.87 & 35.59 & 28.59 - 43.27\\
Recife/PE & 464 & 32.97 & 28.84 - 37.39 & 447 & 33.56 & 28.49 - 39.03 & 28.38 & 23.51 - 33.81\\
Salvador/BA & 837 & 24.37 & 21.58 - 27.40 & 769 & 24.58 & 16.36 - 35.18 & 21.82 & 14.66 - 31.19\\
S Luís/MA & 586 & 22.18 & 19.00 - 25.73 & 565 & 22.30 & 18.34 - 26.83 & 21.98 & 17.14 - 27.74\\
Teresina/PI & 422 & 24.41 & 20.54 - 28.74 & 410 & 24.39 & 20.15 - 29.20 & 22.91 & 18.93 - 27.46\\
Nordeste & 5195 & 29.30 & 28.08 - 30.55 & 4936 & 29.42 & 27.24 - 31.69 & 27.31 & 25.19 - 29.54\\
\bottomrule
\multicolumn{9}{l}{\rule{0pt}{1em}\textit{Source: } Elaborated by the authors.}\\
\end{tabu}
\end{table}

\begin{table}[H]

\caption{\label{tab:p2}Prevalence of  Emotional Violence in the last 12 months - 2016}
\centering
\begin{tabu} to \linewidth {>{\raggedright\arraybackslash}p{2cm}>{\centering}X>{\centering}X>{\centering\arraybackslash}p{2cm}|>{\centering}X>{\centering}X>{\centering\arraybackslash}p{2cm}|>{\centering\arraybackslash}p{1cm}>{\centering\arraybackslash}p{2cm}}
\toprule
\multicolumn{1}{c}{ } & \multicolumn{3}{c}{Original dataset} & \multicolumn{3}{c}{Unweighted dataset} & \multicolumn{2}{c}{Weighted dataset} \\
\cmidrule(l{3pt}r{3pt}){2-4} \cmidrule(l{3pt}r{3pt}){5-7} \cmidrule(l{3pt}r{3pt}){8-9}
  & n & Prev & CI & n & Prev & CI & Prev & CI\\
\midrule
Aracaju/SE & 543 & 14.00 & 11.32 - 17.18 & 534 & 14.23 & 8.73 - 22.36 & 12.43 & 7.26 - 20.47\\
Fortaleza/CE & 731 & 13.27 & 10.99 - 15.93 & 697 & 13.49 & 10.97 - 16.48 & 12.78 & 10.09 - 16.05\\
J Pessoa/PB & 706 & 14.59 & 12.17 - 17.39 & 671 & 14.61 & 12.18 - 17.42 & 13.70 & 10.74 - 17.30\\
Maceió/AL & 504 & 16.87 & 13.84 - 20.40 & 455 & 16.26 & 12.36 - 21.10 & 17.12 & 12.39 - 23.18\\
Natal/RN & 348 & 19.25 & 15.44 - 23.74 & 336 & 18.15 & 15.60 - 21.02 & 16.74 & 11.93 - 22.98\\
Recife/PE & 451 & 14.86 & 11.86 - 18.45 & 434 & 15.44 & 11.85 - 19.87 & 12.84 & 8.98 - 18.02\\
Salvador/BA & 832 & 9.74 & 7.90 - 11.95 & 764 & 9.95 & 4.81 - 19.44 & 9.02 & 4.18 - 18.38\\
S Luís/MA & 575 & 8.70 & 6.65 - 11.30 & 556 & 8.81 & 6.25 - 12.28 & 9.61 & 6.35 - 14.28\\
Teresina/PI & 422 & 11.37 & 8.67 - 14.78 & 410 & 11.46 & 8.34 - 15.56 & 11.16 & 7.51 - 16.26\\
Nordeste & 5112 & 13.18 & 12.28 - 14.14 & 4857 & 13.22 & 11.80 - 14.78 & 12.50 & 11.04 - 14.13\\
\bottomrule
\multicolumn{9}{l}{\rule{0pt}{1em}\textit{Source: } Elaborated by the authors.}\\
\end{tabu}
\end{table}

\begin{table}[H]

\caption{\label{tab:p3}Prevalence of  Physical Violence Lifetime - 2016}
\centering
\begin{tabu} to \linewidth {>{\raggedright\arraybackslash}p{2cm}>{\centering}X>{\centering}X>{\centering\arraybackslash}p{2cm}|>{\centering}X>{\centering}X>{\centering\arraybackslash}p{2cm}|>{\centering\arraybackslash}p{1cm}>{\centering\arraybackslash}p{2cm}}
\toprule
\multicolumn{1}{c}{ } & \multicolumn{3}{c}{Original dataset} & \multicolumn{3}{c}{Unweighted dataset} & \multicolumn{2}{c}{Weighted dataset} \\
\cmidrule(l{3pt}r{3pt}){2-4} \cmidrule(l{3pt}r{3pt}){5-7} \cmidrule(l{3pt}r{3pt}){8-9}
  & n & Prev & CI & n & Prev & CI & Prev & CI\\
\midrule
Aracaju/SE & 534 & 16.29 & 13.39 - 19.68 & 525 & 16.38 & 12.32 - 21.46 & 13.00 & 9.73 - 17.16\\
Fortaleza/CE & 734 & 19.75 & 17.03 - 22.80 & 700 & 20.14 & 17.27 - 23.36 & 17.19 & 13.90 - 21.07\\
J Pessoa/PB & 702 & 19.52 & 16.75 - 22.62 & 667 & 19.79 & 15.69 - 24.64 & 16.69 & 12.60 - 21.78\\
Maceió/AL & 504 & 22.22 & 18.80 - 26.07 & 456 & 22.81 & 17.39 - 29.31 & 17.36 & 12.06 - 24.34\\
Natal/RN & 351 & 22.51 & 18.43 - 27.18 & 338 & 22.19 & 17.78 - 27.33 & 17.31 & 12.42 - 23.60\\
Recife/PE & 453 & 20.53 & 17.05 - 24.50 & 436 & 20.64 & 16.13 - 26.02 & 18.40 & 13.30 - 24.89\\
Salvador/BA & 834 & 20.02 & 17.44 - 22.88 & 765 & 19.74 & 15.78 - 24.40 & 13.28 & 10.35 - 16.87\\
S Luís/MA & 561 & 14.62 & 11.92 - 17.79 & 541 & 14.60 & 11.10 - 18.97 & 16.40 & 11.57 - 22.73\\
Teresina/PI & 412 & 16.02 & 12.78 - 19.89 & 400 & 15.75 & 11.83 - 20.67 & 13.67 & 9.68 - 18.96\\
Nordeste & 5085 & 19.04 & 17.98 - 20.14 & 4828 & 19.08 & 17.67 - 20.57 & 15.80 & 14.40 - 17.32\\
\bottomrule
\multicolumn{9}{l}{\rule{0pt}{1em}\textit{Source: } Elaborated by the authors.}\\
\end{tabu}
\end{table}

\begin{table}[H]

\caption{\label{tab:p4}Prevalence of  Physical Violence in the last 12 months - 2016}
\centering
\begin{tabu} to \linewidth {>{\raggedright\arraybackslash}p{2cm}>{\centering}X>{\centering}X>{\centering\arraybackslash}p{2cm}|>{\centering}X>{\centering}X>{\centering\arraybackslash}p{2cm}|>{\centering\arraybackslash}p{1cm}>{\centering\arraybackslash}p{2cm}}
\toprule
\multicolumn{1}{c}{ } & \multicolumn{3}{c}{Original dataset} & \multicolumn{3}{c}{Unweighted dataset} & \multicolumn{2}{c}{Weighted dataset} \\
\cmidrule(l{3pt}r{3pt}){2-4} \cmidrule(l{3pt}r{3pt}){5-7} \cmidrule(l{3pt}r{3pt}){8-9}
  & n & Prev & CI & n & Prev & CI & Prev & CI\\
\midrule
Aracaju/SE & 533 & 5.82 & 4.12 - 8.16 & 524 & 5.92 & 3.93 - 8.80 & 4.97 & 2.95 - 8.26\\
Fortaleza/CE & 715 & 5.87 & 4.37 - 7.86 & 682 & 6.01 & 4.45 - 8.07 & 5.13 & 3.44 - 7.60\\
J Pessoa/PB & 696 & 6.61 & 4.98 - 8.72 & 661 & 6.81 & 5.08 - 9.07 & 5.65 & 3.65 - 8.66\\
Maceió/AL & 489 & 9.20 & 6.94 - 12.11 & 442 & 9.28 & 6.09 - 13.89 & 6.79 & 3.98 - 11.37\\
Natal/RN & 335 & 7.16 & 4.84 - 10.47 & 322 & 6.83 & 4.85 - 9.55 & 4.96 & 2.41 - 9.90\\
Recife/PE & 442 & 7.01 & 4.97 - 9.81 & 426 & 7.28 & 5.05 - 10.38 & 5.55 & 3.19 - 9.48\\
Salvador/BA & 826 & 4.84 & 3.57 - 6.54 & 758 & 4.88 & 2.52 - 9.26 & 3.18 & 1.33 - 7.45\\
S Luís/MA & 550 & 4.36 & 2.94 - 6.43 & 531 & 4.33 & 2.91 - 6.39 & 5.34 & 2.99 - 9.34\\
Teresina/PI & 411 & 4.87 & 3.16 - 7.43 & 399 & 4.76 & 2.92 - 7.66 & 5.08 & 2.94 - 8.65\\
Nordeste & 4997 & 6.06 & 5.43 - 6.76 & 4745 & 6.11 & 5.33 - 7.00 & 5.08 & 4.27 - 6.03\\
\bottomrule
\multicolumn{9}{l}{\rule{0pt}{1em}\textit{Source: } Elaborated by the authors.}\\
\end{tabu}
\end{table}

\begin{table}[H]

\caption{\label{tab:p5}Prevalence of  Sexual Violence Lifetime - 2016}
\centering
\begin{tabu} to \linewidth {>{\raggedright\arraybackslash}p{2cm}>{\centering}X>{\centering}X>{\centering\arraybackslash}p{2cm}|>{\centering}X>{\centering}X>{\centering\arraybackslash}p{2cm}|>{\centering\arraybackslash}p{1cm}>{\centering\arraybackslash}p{2cm}}
\toprule
\multicolumn{1}{c}{ } & \multicolumn{3}{c}{Original dataset} & \multicolumn{3}{c}{Unweighted dataset} & \multicolumn{2}{c}{Weighted dataset} \\
\cmidrule(l{3pt}r{3pt}){2-4} \cmidrule(l{3pt}r{3pt}){5-7} \cmidrule(l{3pt}r{3pt}){8-9}
  & n & Prev & CI & n & Prev & CI & Prev & CI\\
\midrule
Aracaju/SE & 543 & 8.29 & 6.24 - 10.93 & 534 & 8.24 & 3.91 - 16.53 & 5.98 & 3.02 - 11.48\\
Fortaleza/CE & 755 & 7.02 & 5.40 - 9.08 & 718 & 6.96 & 4.83 - 9.95 & 5.52 & 3.62 - 8.33\\
J Pessoa/PB & 700 & 9.57 & 7.60 - 11.99 & 665 & 9.62 & 6.63 - 13.77 & 7.16 & 4.58 - 11.03\\
Maceió/AL & 500 & 10.40 & 8.01 - 13.40 & 453 & 10.82 & 7.39 - 15.57 & 8.35 & 4.70 - 14.40\\
Natal/RN & 343 & 9.91 & 7.16 - 13.56 & 332 & 9.94 & 6.83 - 14.25 & 7.75 & 5.14 - 11.52\\
Recife/PE & 450 & 6.00 & 4.14 - 8.61 & 434 & 6.22 & 4.17 - 9.19 & 5.56 & 2.98 - 10.17\\
Salvador/BA & 835 & 7.90 & 6.26 - 9.94 & 766 & 8.09 & 4.47 - 14.22 & 7.12 & 4.07 - 12.18\\
S Luís/MA & 588 & 4.08 & 2.75 - 6.02 & 567 & 4.06 & 2.61 - 6.25 & 5.09 & 2.96 - 8.60\\
Teresina/PI & 419 & 7.16 & 5.05 - 10.06 & 407 & 7.37 & 4.58 - 11.66 & 7.64 & 4.10 - 13.79\\
Nordeste & 5133 & 7.75 & 7.05 - 8.52 & 4876 & 7.83 & 6.69 - 9.16 & 6.59 & 5.58 - 7.78\\
\bottomrule
\multicolumn{9}{l}{\rule{0pt}{1em}\textit{Source: } Elaborated by the authors.}\\
\end{tabu}
\end{table}

\begin{table}[H]

\caption{\label{tab:p6}Prevalence of  Sexual Violence in the last 12 months - 2016}
\centering
\begin{tabu} to \linewidth {>{\raggedright\arraybackslash}p{2cm}>{\centering}X>{\centering}X>{\centering\arraybackslash}p{2cm}|>{\centering}X>{\centering}X>{\centering\arraybackslash}p{2cm}|>{\centering\arraybackslash}p{1cm}>{\centering\arraybackslash}p{2cm}}
\toprule
\multicolumn{1}{c}{ } & \multicolumn{3}{c}{Original dataset} & \multicolumn{3}{c}{Unweighted dataset} & \multicolumn{2}{c}{Weighted dataset} \\
\cmidrule(l{3pt}r{3pt}){2-4} \cmidrule(l{3pt}r{3pt}){5-7} \cmidrule(l{3pt}r{3pt}){8-9}
  & n & Prev & CI & n & Prev & CI & Prev & CI\\
\midrule
Aracaju/SE & 541 & 4.44 & 2.99 - 6.54 & 532 & 4.51 & 1.21 - 15.39 & 2.43 & 0.63 - 8.86\\
Fortaleza/CE & 753 & 1.99 & 1.20 - 3.28 & 716 & 2.09 & 1.29 - 3.39 & 1.40 & 0.70 - 2.79\\
J Pessoa/PB & 699 & 3.15 & 2.08 - 4.74 & 664 & 3.16 & 2.01 - 4.93 & 2.25 & 1.22 - 4.13\\
Maceió/AL & 498 & 4.22 & 2.76 - 6.38 & 451 & 3.99 & 2.45 - 6.44 & 2.86 & 1.45 - 5.56\\
Natal/RN & 340 & 3.24 & 1.80 - 5.75 & 329 & 3.04 & 1.50 - 6.07 & 2.89 & 1.26 - 6.49\\
Recife/PE & 450 & 1.78 & 0.89 - 3.52 & 434 & 1.84 & 0.81 - 4.16 & 0.89 & 0.35 - 2.27\\
Salvador/BA & 831 & 2.17 & 1.37 - 3.41 & 762 & 2.10 & 0.71 - 6.08 & 1.46 & 0.32 - 6.52\\
S Luís/MA & 586 & 0.85 & 0.36 - 2.04 & 565 & 0.71 & 0.31 - 1.59 & 0.96 & 0.32 - 2.81\\
Teresina/PI & 417 & 2.40 & 1.29 - 4.40 & 405 & 2.47 & 0.95 - 6.25 & 3.23 & 1.08 - 9.26\\
Nordeste & 5115 & 2.62 & 2.22 - 3.10 & 4858 & 2.59 & 1.92 - 3.49 & 1.94 & 1.40 - 2.68\\
\bottomrule
\multicolumn{9}{l}{\rule{0pt}{1em}\textit{Source: } Elaborated by the authors.}\\
\end{tabu}
\end{table}


\subsection{The 2017 Wave}

\subsubsection{The PCSVDF-Mulher 2017 Refreshment Sample}

There are two broad set of strategies to handle missing data for both cross-sectional and longitudinal designs. First, prevention, usually focused on best practices applied to survey design; to questionnaire content and protocol of application development; to production of detailed documentation of the study in the form of the manual of operations;  training of data collectors; to piloting the study; to creating monitoring indicators of missingness; and to the use of incentives to participation and retention (in longitudinal studies), among others, \citep{graham_missing_2009} and \citep{kang_prevention_2013}.

Second, treatment, applied after the missingness has occurred and detected which encompasses various techniques of imputation, deletion; and more recently modelling where the missingness mechanism is hypothesized by means of a given probabilistic structure and statistical analysis is applied, \citep{graham_missing_2012}.

More recently, the approach to unit nonresponse between waves of longitudinal studies (attrition) has moved towards an approach based on the availability of supplemental samples, either collecting refreshment or replacement samples on an ongoing larger sample, \citep{hirano_combining_2001, Vehovar2003, taylor_evaluating_2020, Watson2021}. Indeed, the use of supplemental samples is usually planned before and included as part of a careful survey design protocol which puts that approach under the label of a prevention technique to mitigate attrition. The PCSVDF-Mulher project is one of the few longitudinal studies worldwide related to violence against women that planned and collected a supplemental sample.

The PCSVDF-Mulher supplemental sample had a straightforward protocol, pre-defined in 2016, when its sample design for a two-wave longitudinal study (2016 and 2017) was specified:

\begin{enumerate}
    \item A target sample of 10,000 women at 2016 and 2017 allocated among the nine capitals cities accordingly to a pre-specified level of significance.
    \item The size of the supplemental sample in 2017 is, thus, implicitly defined by the difference between 10,000 and the number of those who attrited between waves.
    \item If a woman (who originally participated in 2016) refused to participate in 2016, interviewers had to ask if another eligible woman at the same household would like to participate. Only after exhausting the list of all eligible woman (by a sequence of refusals) at a given household, a interviewer would be allowed to go after another woman from a different household.
    \item A supplemental sample collected in 2017 either inside or outside the original household (2016 wave) used the same selection criteria as the initial 2016 cross-sectional wave. For the former, the selection is based on random choice of another eligible woman, for the latter, systematic sample is used.
\end{enumerate}

\noindent These four steps bring important statistical nuances and consequences for modelling and calculating the 2017 cross-sectional weights if ones wants to seriously consider using that supplemental sample. As a matter of fact, the study collected a refreshment sample and not a replacement sample; as the former uses the same selection criteria as the initial study and the latter uses auxiliary variables that may help explain patterns of missingness and select new samples based on those characteristics.

\subsubsection{Sample Weighting}

As PCSVDF-Mulher study is longitudinal, we attempted to interview in 2017 the same group of women who participated in 2016. Unfortunately, we were unable to interview all of them again. Out of 9,618 women interviewed in 2016, we managed to re-interview 4,316, resulting in an attrition rate of 55.13\%, see, Table \ref{tab:attrition_refreshment}. However, the PCSVDF-Mulher has a replacement protocol to address this issue. Following this protocol, we obtained a replacement sample of 5,529 women. Of these, 934 came from replacements within the household and 4,595 from other households.

As we have two sub-samples for 2017, i.e., the replacement sample and the non-attrited sample, we need to find a way to combine these two samples and calculate the weights for the entire (combined) 2017 cross-section sample. There are basically two different approaches to integrating two samples through developing a set of weights (\citep{o2002combining}, \citep{Watson_2014}, and \citep{Watson2021}): 1) ``combining estimates,'' where the combined estimate of a statistics is just the weighted average of the statistics over both samples; or 2) ``pooling samples,'' where each observation weight is the probability of being selected in each of the two samples.

To integrate our `refreshment and non-attrited samples, we opted for the ``pooling samples,'' as it seems to be the method more frequently applied in longitudinal household studies worldwide (see \citep{Watson_2014}, and \citep{Watson2021}). Let us set some notations and definitions:

\begin{itemize}
    \item There are two non-overlapping samples that the analyst wants to combine. in our case, let us represent each sample by RE (REFRESHMENT) and NAT (NON-ATTRITED). We can have overlapping samples, but it complicates the development of the weights. Indeed, at the household level, the PCSVDF-Mulher study presents overlapping households: remember that the first criterion to substitute a woman is to try substituting inside the household. We have to redefine the samples to avoid that issue or correct the case of overlapping. More on that issue later in our paper.
    \item We index each member of each sample by $i = 1, 2, \dots, N_{RE}$ and $j = 1, 2, \dots, N_{NAT}$, respectively.
    \item $p_{RE,i}$ is the probability that the refreshment observation $i$ was selected into the RE sample.
    \item $p_{NAT,i}$ is the probability that the refreshment observation $i$ was selected into the NAT sample.
    \item $p_{RE,j}$ is the probability that the non-attrited observation $j$ was selected into the RE sample.
    \item $p_{NAT,j}$ is the probability that the non-attrited observation $j$ was selected into the NAT sample.
    \item The probability of a given sample element from RE to be in the combined $(RE \cup NAT)$ sample is given by the $p_{RE, i}  + p_{NAT, i}$, and its associated weight is $\frac{1}{p_{RE, i} + p_{NAT, i}}$.
    \item The probability of a given sample element from NAT to be in the combined  $(RE \cup NAT)$ sample is given by the $p_{RE,j}  + p_{NAT,j}$, and its associated weight is $\frac{1}{p_{RE,j} + p_{NAT,j}}$.
\end{itemize}

\noindent The quantities $p_{RE, i}$ and $p_{NAT, j}$ are calculated directly from the refreshment and non-attrited samples, respectively, while calculating their weights. The other two quantities are tricky, as they represent counterfactual probabilities in a specific sense. For instance, $p_{NAT, i}$ is the probability that the refreshment observation $i$ was selected into the non-attrited sample (indeed, it was not selected, but we need an expression for the ``would-be'' probability of selection, say, the probability of having observation $i$ selected in the NAT sample). The case $p_{RE, j}$ is analogous. So, $p_{NAT, i}$ and $p_{RE, j}$ can not be calculated but can be estimated. We represent those estimated probabilities $\widehat{p_{NAT, i}}$ and $\widehat{p_{RE, j}}$, respectively.

As we mentioned before, the protocol for substitution adopted by the PCSVDF-Mulher study requires, as a priority, that any substitution should be made first at the household as long as any eligible woman accepts to participate. So, the protocol basically sets the probability that a household is present in both the RE and NAT samples to a value strictly positive, i.e., $p_{(RE and NAT), i}  > 0$. Again, we need to estimate that, so we represent that estimative by $p_{\widehat{(RE and NAT)}} \in (0,1)$, a fixed number. As we include the possibility that both samples can overlap, we need to adjust the final weights to:

\begin{itemize}
    \item $\frac{1}{p_{RE,i}  + \widehat{p_{NAT, i}} - p_{\widehat{(RE and NAT)}}}$ for RE sample
    \item $\frac{1}{\widehat{p_{RE, j}} + p_{NAT, j} - p_{\widehat{(RE and NAT)}}}$ for NAT sample
\end{itemize}

\noindent How to estimate these three quantities? For $\widehat{p_{NAT, i}}$ and $\widehat{p_{RE, j}}$, we follow the proposal of \citet[p. 201]{Watson_2014}. We describe here the process for obtaining $\widehat{p_{RE, j}}$:

\begin{itemize}
    \item Apply a logistic transformation on $p_{RE,i}$, i.e., $ln \left( \frac{p_{RE,i}}{1 - p_{RE,i}} \right)$.
    \item Run a linear regression, where the dependent variable is $ln \left( \frac{p_{RE, i}}{1 - p_{RE, i}} \right)$ and the vector of independent variables are some characteristics of the census tract, the household, and from the head of the household.
    \item Now, to calculate $\widehat{p_{RE, j}}$, for each $j$ from the NAT sample, just use the prediction from the regression estimated above applied to each $j$ independent variables value. Do not forget to ``back-transform'' the logit transformation so we can have $\widehat{p_{RE, j}} \in (0,1)$.
    \item For $\widehat{p_{NAT, i}}$, the process is entirely analogous. 
\end{itemize}

\noindent After this integration step, we followed the same steps as we did before for the 2016 weights: trimming, calibration, and scaling.

\begin{table}[h!]
\caption{\label{tab:a17}Sample Summary of Final Weights of the Basic Design by City - 2017}
\centering
\begin{tabu} to \linewidth {>{\raggedright\arraybackslash}p{2.5cm}>{\centering}X>{\centering}X>{\centering}X>{\centering}X>{\centering}X>{\centering}X>{\centering}X}
\toprule
  City & N & Mean & Sd & Min & Max & IQR & CV  \\ 
\midrule
  Aracaju/SE     & 1,001 & 1.00 & 0.99 & 0.20 & 8.04 & 0.52 & 0.99 \\ 
  Fortaleza/CE   & 1,233 & 1.00 & 1.10 & 0.27 & 11.12 & 0.63 & 1.10 \\ 
  J Pessoa/PB    & 995 & 1.00 & 1.05 & 0.08 & 6.77 & 0.78 & 1.05 \\
  Maceio/AL      & 1,073 & 1.00 & 0.76 & 0.33 & 7.07 & 0.57 & 0.76 \\ 
  Natal/RN       & 1,091 & 1.00 & 0.67 & 0.08 & 4.44 & 0.39 & 0.67 \\  
  Recife/PE      & 1,212 & 1.00 & 1.13 & 0.10 & 12.71 & 0.59 & 1.13 \\ 
  Salvador/BA    & 1,187 & 1.00 & 1.22 & 0.22 & 15.49 & 0.50 & 1.22 \\ 
  S Luís/MA      & 1,097 & 1.00 & 0.69 & 0.16 & 5.06 & 0.71 & 0.69 \\ 
  Teresina/PI    & 956 & 1.00 & 0.80 & 0.11 & 6.19 & 0.51 & 0.80 \\ 
\bottomrule
\multicolumn{8}{l}{\rule{0pt}{1em}\textit{Source: } Elaborated by the authors.}\\
\end{tabu}
\end{table}
  
\subsubsection{Correction for the Questionnaire ``IPV Section'' Non-response}

In order to calculate the prevalence rates of domestic violence in 2017, it was necessary to create weights for non-response for the section on domestic violence against women, just like we did for 2016. We used the same logit model given by Equation \ref{logit}, which results can be seen in the table \ref{eq:logit2017} in the appendix.

\begin{table}[h!]
\caption{\label{tab:a18}Sample Summary of Final Weights for Calculating Prevalence by City - 2017}
\centering
\begin{tabu} to \linewidth {>{\raggedright\arraybackslash}p{2.5cm}>{\centering}X>{\centering}X>{\centering}X>{\centering}X>{\centering}X>{\centering}X>{\centering}X}
\toprule
  City & N & Mean & Sd & Min & Max & IQR & CV  \\
\midrule
  Aracaju/SE     & 801 & 1.00 & 0.97 & 0.23 & 4.88 & 0.57 & 0.98 \\ 
  Fortaleza/CE   & 1,107 & 1.00 & 0.96 & 0.21 & 5.35 & 0.64 & 0.96 \\ 
  J Pessoa/PB    & 855 & 1.00 & 1.00 & 0.18 & 4.70 & 0.85 & 0.99 \\ 
  Maceio/AL      & 824 & 1.00 & 0.71 & 0.35 & 4.09 & 0.57 & 0.71 \\ 
  Natal/RN       & 869 & 1.00 & 0.74 & 0.26 & 4.22 & 0.56 & 0.74 \\ 
  Recife/PE      & 1,034 & 1.00 & 1.12 & 0.16 & 6.86 & 0.63 & 1.12 \\ 
  Salvador/BA    & 1,079 & 1.00 & 1.13 & 0.22 & 8.27 & 0.47 & 1.13 \\ 
  S Luís/MA      & 837 & 1.00 & 0.65 & 0.30 & 3.12 & 0.84 & 0.65 \\ 
  Teresina/PI    & 763 & 1.00 & 0.87 & 0.24 & 4.84 & 0.67 & 0.87 \\
\bottomrule
\multicolumn{8}{l}{\rule{0pt}{1em}\textit{Source: } Elaborated by the authors.}\\
\end{tabu}
\end{table}
  
\subsubsection{Prevalence Rates}

\begin{table}[H]

\caption{\label{tab:p7}Prevalence of  Emotional Violence Lifetime - 2017}
\centering
\begin{tabu} to \linewidth {>{\raggedright\arraybackslash}p{2cm}>{\centering}X>{\centering}X>{\centering\arraybackslash}p{2cm}|>{\centering}X>{\centering}X>{\centering\arraybackslash}p{2cm}|>{\centering\arraybackslash}p{1cm}>{\centering\arraybackslash}p{2cm}}
\toprule
\multicolumn{1}{c}{ } & \multicolumn{3}{c}{Original dataset} & \multicolumn{3}{c}{Unweighted dataset} & \multicolumn{2}{c}{Weighted dataset} \\
\cmidrule(l{3pt}r{3pt}){2-4} \cmidrule(l{3pt}r{3pt}){5-7} \cmidrule(l{3pt}r{3pt}){8-9}
  & n & Prev & CI & n & Prev & CI & Prev & CI\\
\midrule
Aracaju/SE & 605 & 29.92 & 26.39 - 33.69 & 579 & 29.71 & 24.85 - 35.06 & 27.54 & 23.34 - 32.18\\
Fortaleza/CE & 981 & 31.70 & 28.86 - 34.69 & 941 & 31.35 & 27.44 - 35.55 & 28.91 & 24.63 - 33.60\\
J Pessoa/PB & 728 & 32.69 & 29.38 - 36.19 & 653 & 33.69 & 26.84 - 41.31 & 33.82 & 26.18 - 42.42\\
Maceió/AL & 566 & 30.74 & 27.07 - 34.67 & 527 & 31.50 & 26.75 - 36.67 & 28.35 & 23.32 - 33.97\\
Natal/RN & 503 & 31.01 & 27.12 - 35.20 & 481 & 30.77 & 25.40 - 36.71 & 28.34 & 22.13 - 35.51\\
Recife/PE & 621 & 34.94 & 31.29 - 38.79 & 576 & 34.72 & 29.55 - 40.28 & 31.95 & 26.77 - 37.62\\
Salvador/BA & 864 & 33.45 & 30.38 - 36.67 & 810 & 33.83 & 28.53 - 39.56 & 34.92 & 26.43 - 44.48\\
S Luís/MA & 599 & 24.04 & 20.78 - 27.63 & 563 & 23.98 & 20.13 - 28.30 & 21.40 & 17.50 - 25.90\\
Teresina/PI & 607 & 24.05 & 20.81 - 27.62 & 590 & 24.07 & 20.00 - 28.66 & 22.10 & 17.55 - 27.43\\
Nordeste & 6075 & 30.55 & 29.41 - 31.72 & 5720 & 30.63 & 28.94 - 32.37 & 28.91 & 26.79 - 31.12\\
\bottomrule
\multicolumn{9}{l}{\rule{0pt}{1em}\textit{Source: } Elaborated by the authors.}\\
\end{tabu}
\end{table}

\begin{table}[H]

\caption{\label{tab:p8}Prevalence of  Emotional Violence in the last 12 months - 2017}
\centering
\begin{tabu} to \linewidth {>{\raggedright\arraybackslash}p{2cm}>{\centering}X>{\centering}X>{\centering\arraybackslash}p{2cm}|>{\centering}X>{\centering}X>{\centering\arraybackslash}p{2cm}|>{\centering\arraybackslash}p{1cm}>{\centering\arraybackslash}p{2cm}}
\toprule
\multicolumn{1}{c}{ } & \multicolumn{3}{c}{Original dataset} & \multicolumn{3}{c}{Unweighted dataset} & \multicolumn{2}{c}{Weighted dataset} \\
\cmidrule(l{3pt}r{3pt}){2-4} \cmidrule(l{3pt}r{3pt}){5-7} \cmidrule(l{3pt}r{3pt}){8-9}
  & n & Prev & CI & n & Prev & CI & Prev & CI\\
\midrule
Aracaju/SE & 587 & 15.16 & 12.48 - 18.30 & 562 & 15.48 & 12.57 - 18.91 & 15.69 & 11.88 - 20.44\\
Fortaleza/CE & 973 & 15.21 & 13.09 - 17.61 & 933 & 15.11 & 12.85 - 17.70 & 13.32 & 10.89 - 16.20\\
J Pessoa/PB & 708 & 14.97 & 12.53 - 17.80 & 634 & 15.14 & 11.00 - 20.48 & 15.37 & 10.61 - 21.75\\
Maceió/AL & 555 & 14.05 & 11.40 - 17.21 & 516 & 13.95 & 11.07 - 17.44 & 11.46 & 8.95 - 14.56\\
Natal/RN & 496 & 14.11 & 11.31 - 17.47 & 474 & 13.71 & 10.47 - 17.76 & 12.55 & 9.36 - 16.62\\
Recife/PE & 614 & 15.31 & 12.67 - 18.38 & 569 & 14.59 & 11.35 - 18.56 & 14.13 & 10.02 - 19.55\\
Salvador/BA & 860 & 13.60 & 11.47 - 16.07 & 807 & 14.00 & 10.76 - 18.02 & 14.31 & 10.45 - 19.29\\
S Luís/MA & 588 & 11.56 & 9.22 - 14.42 & 552 & 11.23 & 8.69 - 14.40 & 10.69 & 7.91 - 14.30\\
Teresina/PI & 598 & 11.20 & 8.91 - 14.00 & 581 & 11.36 & 8.52 - 14.99 & 10.66 & 7.75 - 14.51\\
Nordeste & 5980 & 14.00 & 13.14 - 14.90 & 5628 & 13.95 & 12.85 - 15.13 & 13.23 & 11.95 - 14.62\\
\bottomrule
\multicolumn{9}{l}{\rule{0pt}{1em}\textit{Source: } Elaborated by the authors.}\\
\end{tabu}
\end{table}

\begin{table}[H]

\caption{\label{tab:p9}Prevalence of  Physical Violence Lifetime - 2017}
\centering
\begin{tabu} to \linewidth {>{\raggedright\arraybackslash}p{2cm}>{\centering}X>{\centering}X>{\centering\arraybackslash}p{2cm}|>{\centering}X>{\centering}X>{\centering\arraybackslash}p{2cm}|>{\centering\arraybackslash}p{1cm}>{\centering\arraybackslash}p{2cm}}
\toprule
\multicolumn{1}{c}{ } & \multicolumn{3}{c}{Original dataset} & \multicolumn{3}{c}{Unweighted dataset} & \multicolumn{2}{c}{Weighted dataset} \\
\cmidrule(l{3pt}r{3pt}){2-4} \cmidrule(l{3pt}r{3pt}){5-7} \cmidrule(l{3pt}r{3pt}){8-9}
  & n & Prev & CI & n & Prev & CI & Prev & CI\\
\midrule
Aracaju/SE & 573 & 16.58 & 13.75 - 19.86 & 549 & 16.03 & 13.16 - 19.38 & 12.11 & 9.16 - 15.83\\
Fortaleza/CE & 972 & 19.14 & 16.78 - 21.73 & 931 & 19.01 & 16.28 - 22.08 & 14.63 & 11.88 - 17.88\\
J Pessoa/PB & 699 & 18.03 & 15.35 - 21.06 & 627 & 18.98 & 14.70 - 24.15 & 15.99 & 11.59 - 21.65\\
Maceió/AL & 559 & 18.60 & 15.59 - 22.05 & 521 & 19.39 & 16.29 - 22.91 & 14.63 & 11.35 - 18.65\\
Natal/RN & 495 & 16.16 & 13.17 - 19.68 & 473 & 16.07 & 12.33 - 20.67 & 11.30 & 8.58 - 14.74\\
Recife/PE & 615 & 21.79 & 18.70 - 25.23 & 570 & 22.28 & 17.92 - 27.35 & 17.31 & 13.01 - 22.67\\
Salvador/BA & 851 & 19.04 & 16.53 - 21.82 & 800 & 19.38 & 15.17 - 24.41 & 15.95 & 11.52 - 21.65\\
S Luís/MA & 597 & 14.41 & 11.81 - 17.46 & 561 & 14.62 & 11.12 - 18.98 & 11.95 & 9.03 - 15.65\\
Teresina/PI & 591 & 11.34 & 9.02 - 14.16 & 573 & 11.34 & 9.23 - 13.87 & 7.96 & 6.58 - 9.59\\
Nordeste & 5953 & 17.47 & 16.53 - 18.46 & 5605 & 17.66 & 16.40 - 19.00 & 13.75 & 12.50 - 15.11\\
\bottomrule
\multicolumn{9}{l}{\rule{0pt}{1em}\textit{Source: } Elaborated by the authors.}\\
\end{tabu}
\end{table}

\begin{table}[H]

\caption{\label{tab:p10}Prevalence of  Physical Violence in the last 12 months - 2017}
\centering
\begin{tabu} to \linewidth {>{\raggedright\arraybackslash}p{2cm}>{\centering}X>{\centering}X>{\centering\arraybackslash}p{2cm}|>{\centering}X>{\centering}X>{\centering\arraybackslash}p{2cm}|>{\centering\arraybackslash}p{1cm}>{\centering\arraybackslash}p{2cm}}
\toprule
\multicolumn{1}{c}{ } & \multicolumn{3}{c}{Original dataset} & \multicolumn{3}{c}{Unweighted dataset} & \multicolumn{2}{c}{Weighted dataset} \\
\cmidrule(l{3pt}r{3pt}){2-4} \cmidrule(l{3pt}r{3pt}){5-7} \cmidrule(l{3pt}r{3pt}){8-9}
  & n & Prev & CI & n & Prev & CI & Prev & CI\\
\midrule
Aracaju/SE & 560 & 5.89 & 4.22 - 8.18 & 539 & 5.94 & 4.33 - 8.08 & 4.95 & 3.42 - 7.10\\
Fortaleza/CE & 967 & 4.96 & 3.76 - 6.53 & 926 & 4.97 & 3.82 - 6.44 & 3.91 & 2.81 - 5.42\\
J Pessoa/PB & 681 & 5.43 & 3.96 - 7.41 & 609 & 5.25 & 3.49 - 7.84 & 4.34 & 2.59 - 7.19\\
Maceió/AL & 550 & 6.73 & 4.91 - 9.15 & 512 & 6.64 & 4.56 - 9.58 & 5.25 & 3.61 - 7.58\\
Natal/RN & 490 & 4.08 & 2.65 - 6.25 & 468 & 4.06 & 2.33 - 6.97 & 3.78 & 1.89 - 7.42\\
Recife/PE & 606 & 5.78 & 4.17 - 7.94 & 562 & 5.52 & 3.79 - 7.96 & 5.16 & 2.96 - 8.85\\
Salvador/BA & 843 & 5.22 & 3.90 - 6.94 & 794 & 5.29 & 3.52 - 7.87 & 4.98 & 2.23 - 10.73\\
S Luís/MA & 587 & 5.28 & 3.74 - 7.42 & 551 & 5.26 & 3.76 - 7.33 & 4.30 & 3.19 - 5.79\\
Teresina/PI & 578 & 3.63 & 2.38 - 5.51 & 560 & 3.75 & 2.24 - 6.22 & 2.53 & 1.47 - 4.34\\
Nordeste & 5863 & 5.22 & 4.68 - 5.82 & 5521 & 5.18 & 4.55 - 5.89 & 4.35 & 3.65 - 5.18\\
\bottomrule
\multicolumn{9}{l}{\rule{0pt}{1em}\textit{Source: } Elaborated by the authors.}\\
\end{tabu}
\end{table}

\begin{table}[H]

\caption{\label{tab:p11}Prevalence of  Sexual Violence Lifetime - 2017}
\centering
\begin{tabu} to \linewidth {>{\raggedright\arraybackslash}p{2cm}>{\centering}X>{\centering}X>{\centering\arraybackslash}p{2cm}|>{\centering}X>{\centering}X>{\centering\arraybackslash}p{2cm}|>{\centering\arraybackslash}p{1cm}>{\centering\arraybackslash}p{2cm}}
\toprule
\multicolumn{1}{c}{ } & \multicolumn{3}{c}{Original dataset} & \multicolumn{3}{c}{Unweighted dataset} & \multicolumn{2}{c}{Weighted dataset} \\
\cmidrule(l{3pt}r{3pt}){2-4} \cmidrule(l{3pt}r{3pt}){5-7} \cmidrule(l{3pt}r{3pt}){8-9}
  & n & Prev & CI & n & Prev & CI & Prev & CI\\
\midrule
Aracaju/SE & 607 & 8.90 & 6.87 - 11.44 & 580 & 8.45 & 6.70 - 10.59 & 6.69 & 4.97 - 8.94\\
Fortaleza/CE & 978 & 8.90 & 7.26 - 10.85 & 937 & 9.18 & 7.04 - 11.88 & 8.60 & 6.31 - 11.61\\
J Pessoa/PB & 696 & 5.89 & 4.36 - 7.91 & 622 & 5.63 & 3.33 - 9.35 & 5.30 & 2.91 - 9.47\\
Maceió/AL & 563 & 6.75 & 4.95 - 9.14 & 525 & 6.86 & 4.64 - 10.02 & 6.02 & 3.46 - 10.26\\
Natal/RN & 491 & 8.55 & 6.38 - 11.38 & 470 & 8.72 & 6.59 - 11.47 & 7.35 & 5.33 - 10.07\\
Recife/PE & 611 & 6.87 & 5.12 - 9.18 & 566 & 7.07 & 4.90 - 10.08 & 5.29 & 3.26 - 8.46\\
Salvador/BA & 862 & 8.35 & 6.68 - 10.40 & 808 & 8.79 & 6.31 - 12.12 & 9.94 & 6.44 - 15.04\\
S Luís/MA & 616 & 6.17 & 4.52 - 8.37 & 580 & 5.69 & 3.79 - 8.45 & 5.00 & 3.23 - 7.64\\
Teresina/PI & 606 & 4.13 & 2.80 - 6.04 & 588 & 3.91 & 2.38 - 6.37 & 2.92 & 1.72 - 4.93\\
Nordeste & 6030 & 7.28 & 6.65 - 7.96 & 5676 & 7.29 & 6.52 - 8.15 & 6.60 & 5.69 - 7.64\\
\bottomrule
\multicolumn{9}{l}{\rule{0pt}{1em}\textit{Source: } Elaborated by the authors.}\\
\end{tabu}
\end{table}

\begin{table}[H]

\caption{\label{tab:p12}Prevalence of  Sexual Violence in the last 12 months - 2017}
\centering
\begin{tabu} to \linewidth {>{\raggedright\arraybackslash}p{2cm}>{\centering}X>{\centering}X>{\centering\arraybackslash}p{2cm}|>{\centering}X>{\centering}X>{\centering\arraybackslash}p{2cm}|>{\centering\arraybackslash}p{1cm}>{\centering\arraybackslash}p{2cm}}
\toprule
\multicolumn{1}{c}{ } & \multicolumn{3}{c}{Original dataset} & \multicolumn{3}{c}{Unweighted dataset} & \multicolumn{2}{c}{Weighted dataset} \\
\cmidrule(l{3pt}r{3pt}){2-4} \cmidrule(l{3pt}r{3pt}){5-7} \cmidrule(l{3pt}r{3pt}){8-9}
  & n & Prev & CI & n & Prev & CI & Prev & CI\\
\midrule
Aracaju/SE & 598 & 4.01 & 2.70 - 5.92 & 571 & 4.03 & 2.75 - 5.87 & 2.72 & 1.90 - 3.89\\
Fortaleza/CE & 974 & 2.67 & 1.82 - 3.89 & 933 & 2.68 & 1.62 - 4.40 & 2.02 & 1.14 - 3.56\\
J Pessoa/PB & 688 & 2.47 & 1.54 - 3.94 & 615 & 2.60 & 1.29 - 5.17 & 2.58 & 1.28 - 5.16\\
Maceió/AL & 562 & 2.85 & 1.75 - 4.60 & 524 & 2.67 & 1.62 - 4.37 & 2.65 & 1.19 - 5.78\\
Natal/RN & 491 & 2.04 & 1.10 - 3.75 & 470 & 2.13 & 1.11 - 4.05 & 2.17 & 1.17 - 3.98\\
Recife/PE & 611 & 1.47 & 0.77 - 2.81 & 566 & 1.41 & 0.71 - 2.78 & 0.86 & 0.38 - 1.90\\
Salvador/BA & 861 & 2.32 & 1.50 - 3.57 & 807 & 2.48 & 1.64 - 3.72 & 2.72 & 1.46 - 5.01\\
S Luís/MA & 615 & 2.28 & 1.35 - 3.81 & 579 & 2.07 & 1.17 - 3.63 & 1.42 & 0.77 - 2.59\\
Teresina/PI & 602 & 1.00 & 0.45 - 2.20 & 584 & 1.03 & 0.34 - 3.05 & 0.94 & 0.30 - 2.88\\
Nordeste & 6002 & 2.37 & 2.01 - 2.78 & 5649 & 2.37 & 1.97 - 2.85 & 2.03 & 1.63 - 2.53\\
\bottomrule
\multicolumn{9}{l}{\rule{0pt}{1em}\textit{Source: } Elaborated by the authors.}\\
\end{tabu}
\end{table}


\section{Comparing the Weighted and Unweighted Designs}\label{SECTION_ComparingDesigns}

In survey sampling methodology, weighting is one of the critical steps if one wants to produce design-unbiased estimates of parameters of interest, see, \citep{LavalleeBeaumont2015} and \citep{Sarndal2005-ca}. However, practitioners, specially those outside the community of statistical analysis of survey data, usually do not consider weights. If the purpose is to estimate population descriptive statistics (totals, means, ratios, proportions, prevalences or similars) than the use of weights is (almost) mandatory to make the analysis sample representative of the target population. If the objective is to estimate causal effects, the weighting issue is much less frequent, more nuanced as well as controversial, see, \citep{Sarndal2005-ca}, \citep{Solon2015} and \citep{Valliant2018-mc}. We perform two simple exercises of comparing the weighted ($PREV_{j}^{w}$) and unweighted ($PREV_{j}^{unw}$) designs for $j = 2016 \textrm{ and } 2017$: 1) plot the percentage differences between the weighted and unweighted designs; and 2) plot the ratio between the weighted and unweighted variance designs.

\subsection{The Weighted and Unweighted ($Diff$)}

\begin{figure}[h!]
    \centering
    \begin{subfigure}[t]{0.45\linewidth}
        \centering
        \includegraphics[width=\linewidth, height = 5.5cm]{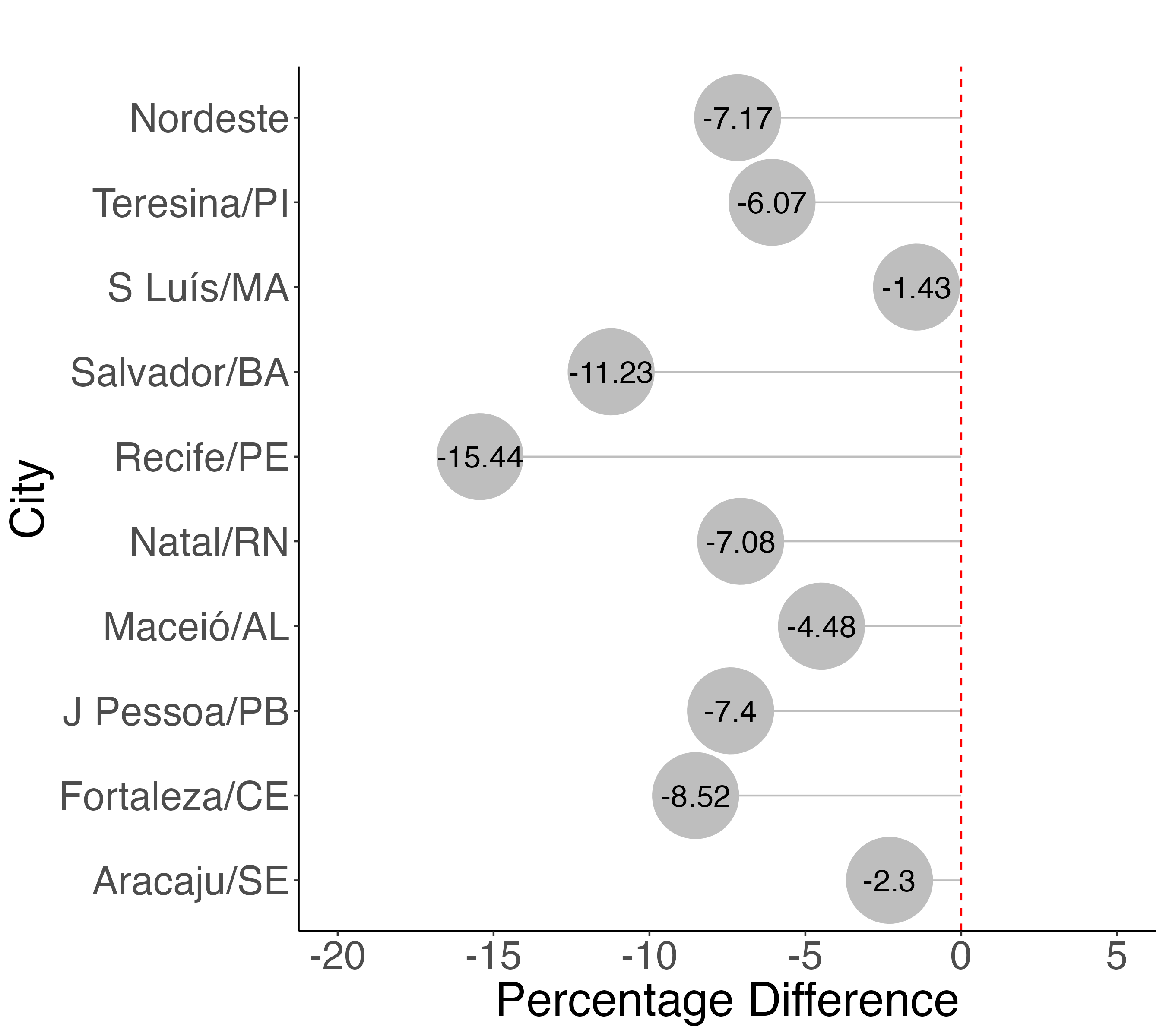}
        \caption{Emotional Violence Lifetime}
    \end{subfigure}%
    ~ \vspace{0.5cm}
    \begin{subfigure}[t]{0.45\linewidth}
        \centering
        \includegraphics[width=\linewidth, height = 5.5cm]{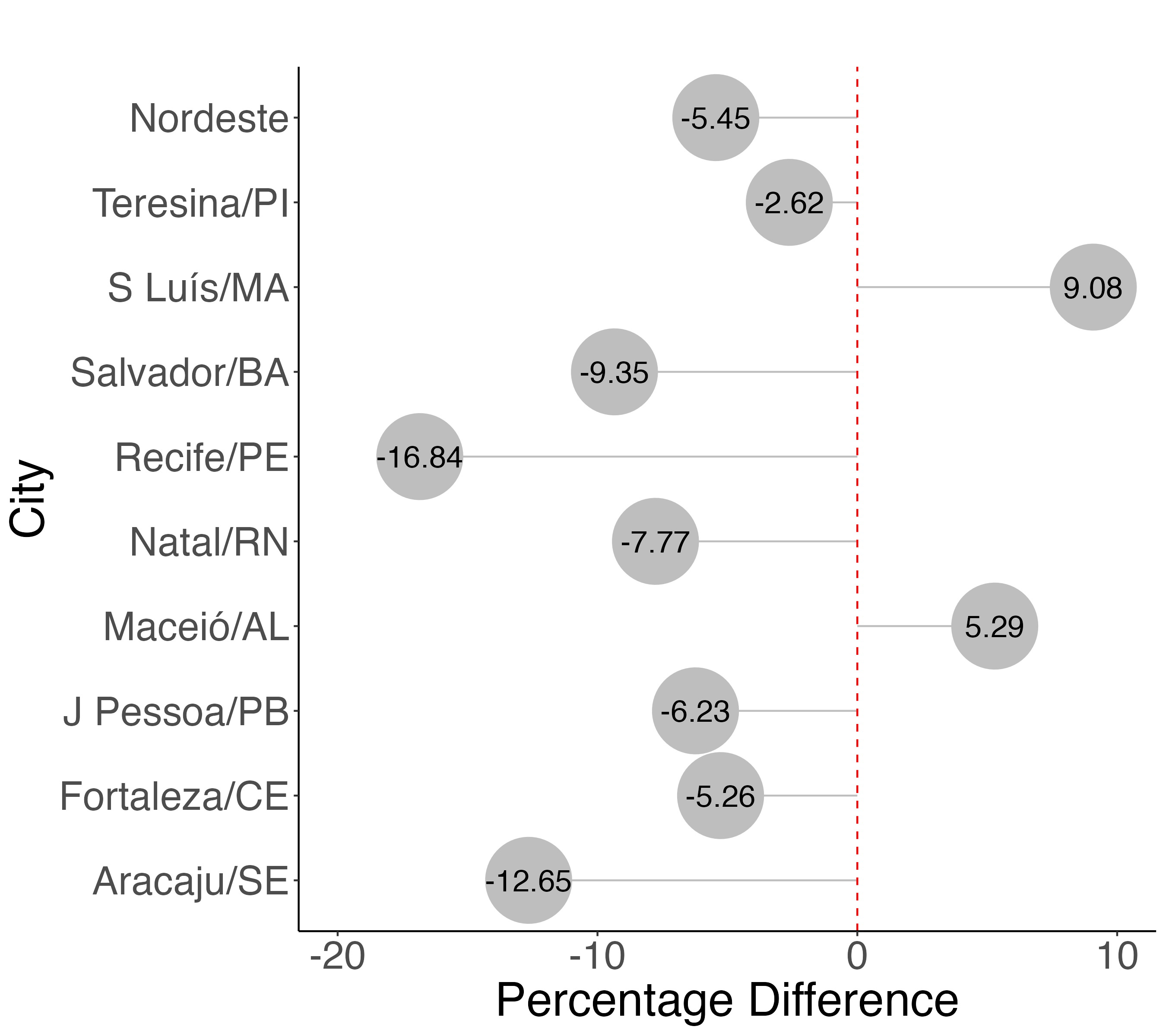}
        \caption{Emotional Violence last 12 months}
    \end{subfigure}
    ~ \vspace{0.5cm}
    \begin{subfigure}[t]{0.45\linewidth}
        \centering
        \includegraphics[width=\linewidth, height = 5.5cm]{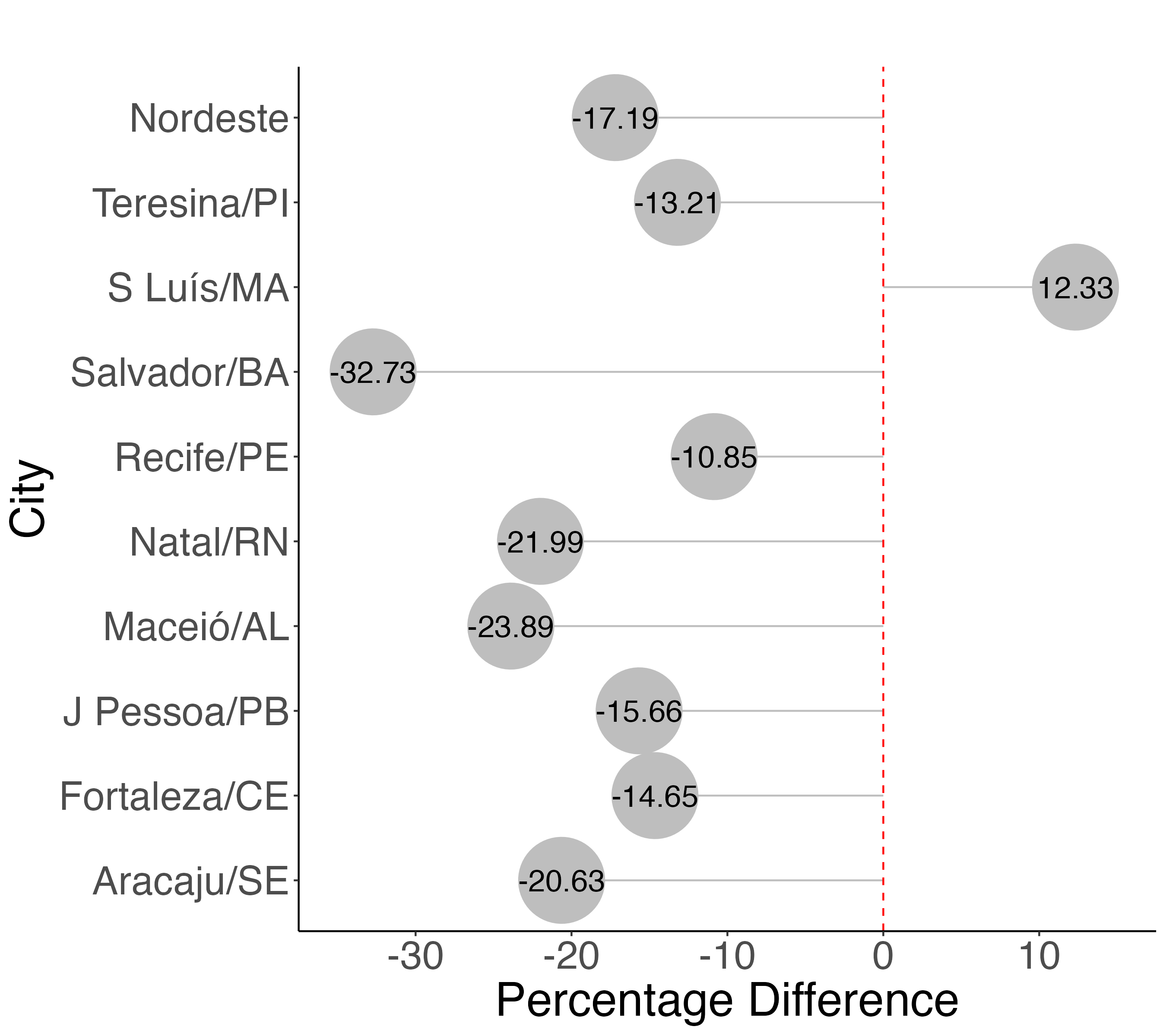}
        \caption{Physical Violence Lifetime}
    \end{subfigure}
    ~ 
    \begin{subfigure}[t]{0.45\linewidth}
        \centering
        \includegraphics[width=\linewidth, height = 5.5cm]{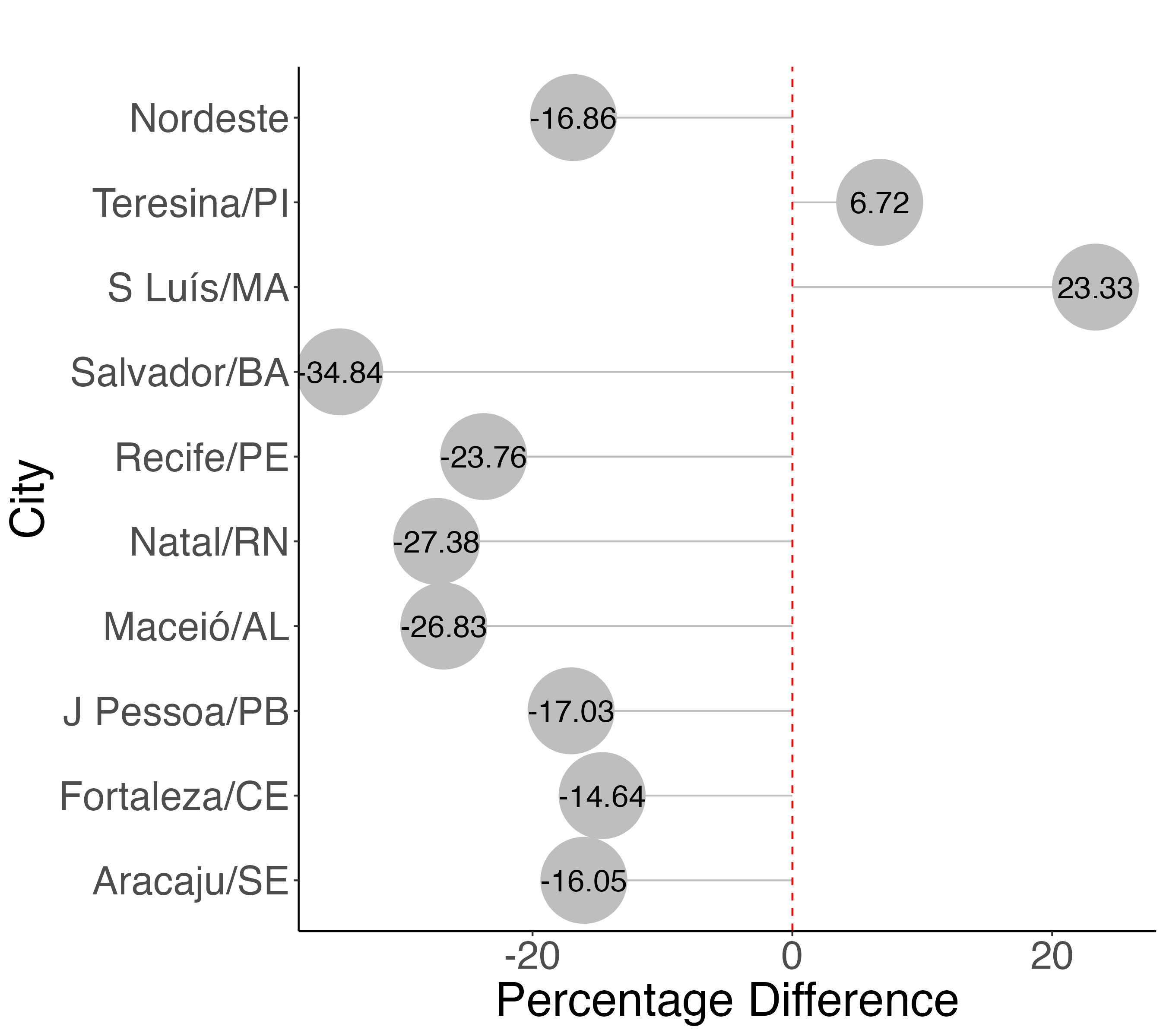}
        \caption{Physical Violence last 12 months}
    \end{subfigure}
    ~
    \begin{subfigure}[t]{0.45\linewidth}
        \centering
        \includegraphics[width=\linewidth, height = 5.5cm]{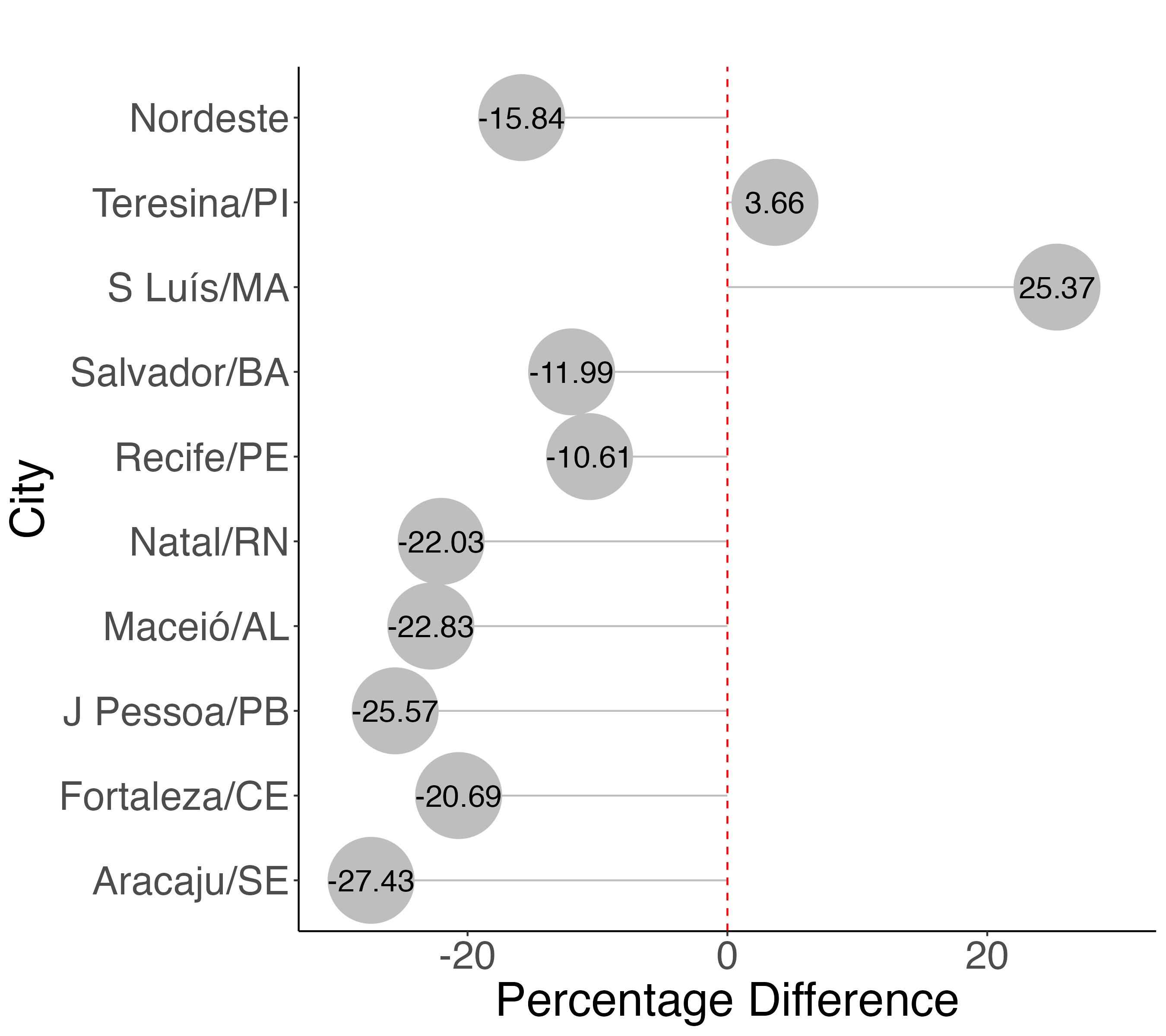}
        \caption{Sexual Violence Lifetime}
    \end{subfigure}
    ~ 
    \begin{subfigure}[t]{0.45\linewidth}
        \centering
        \includegraphics[width=\linewidth, height = 5.5cm]{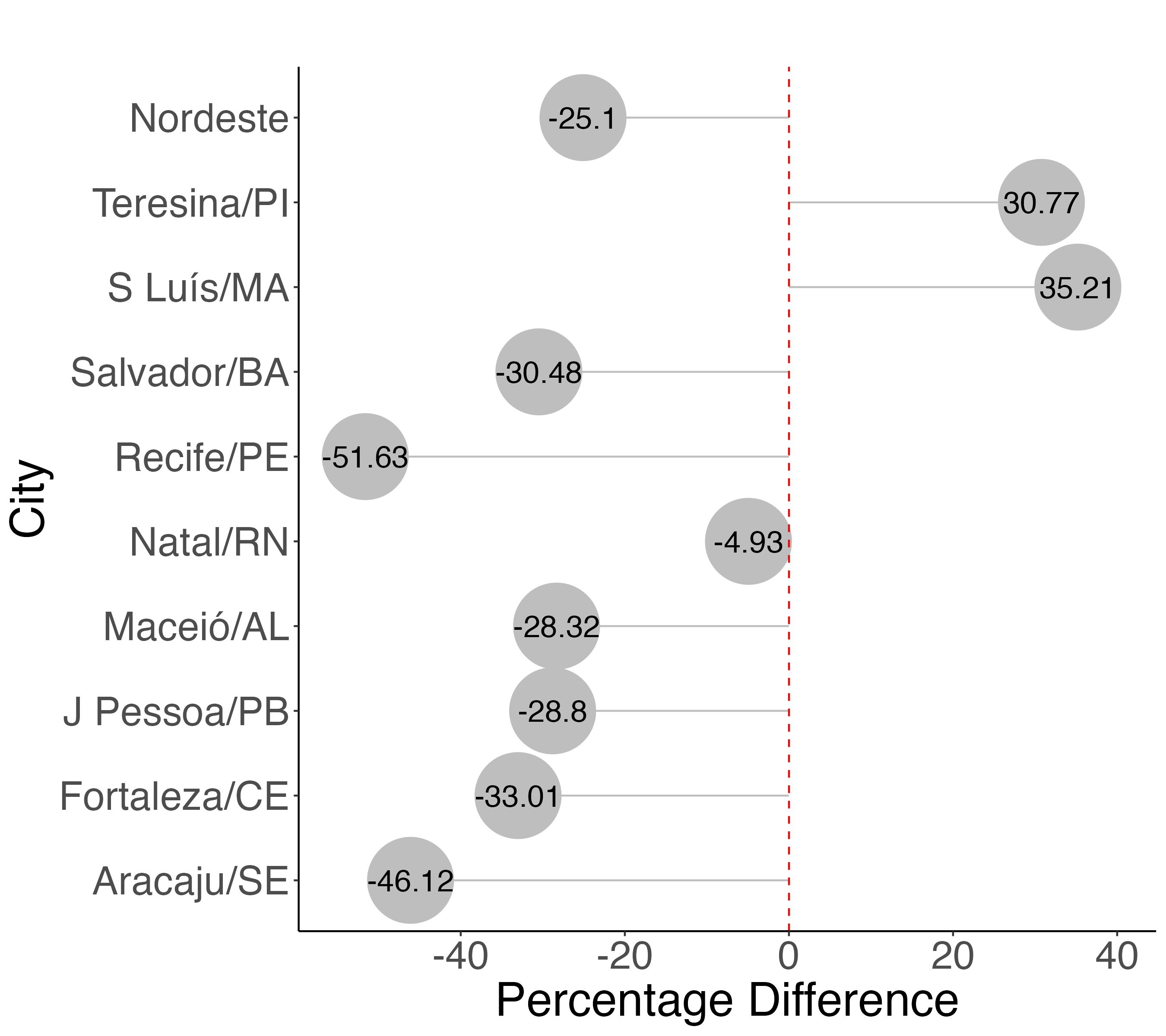}
        \caption{Sexual Violence last 12 months}
    \end{subfigure}
    ~
    \caption{Percentage Difference between Weighted and Unweighted Designs, 2016}\label{fig:c2_2016}
\end{figure}

\begin{figure}[!h]
    \centering
    \begin{subfigure}[t]{0.45\linewidth}
        \centering
        \includegraphics[width=\linewidth, height = 5.5cm]{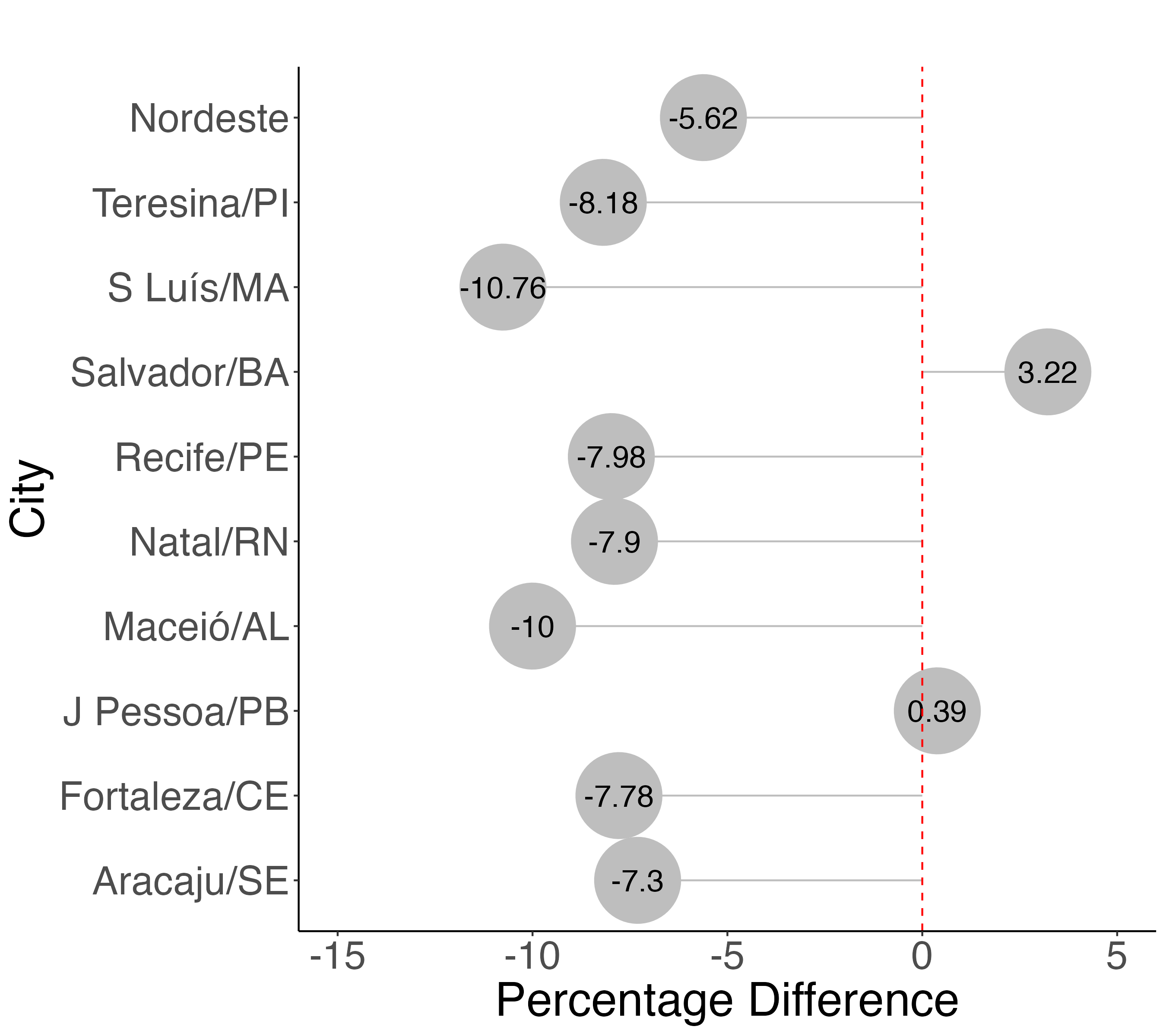}
        \caption{Emotional Violence Lifetime}
    \end{subfigure}%
    ~ \vspace{0.5cm}
    \begin{subfigure}[t]{0.45\linewidth}
        \centering
        \includegraphics[width=\linewidth, height = 5.5cm]{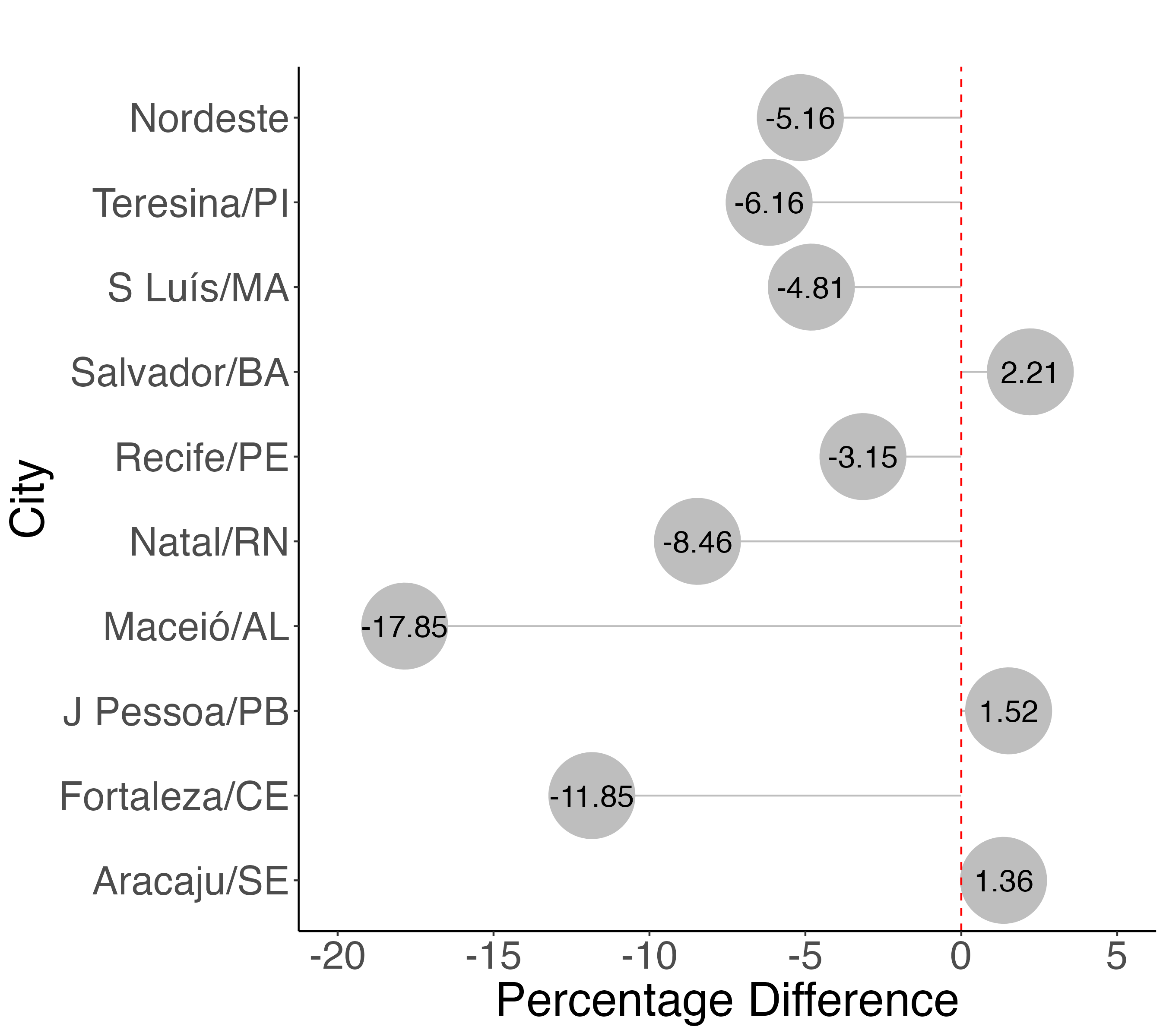}
        \caption{Emotional Violence last 12 months}
    \end{subfigure}
    ~ \vspace{0.5cm}
    \begin{subfigure}[t]{0.45\linewidth}
        \centering
        \includegraphics[width=\linewidth, height = 5.5cm]{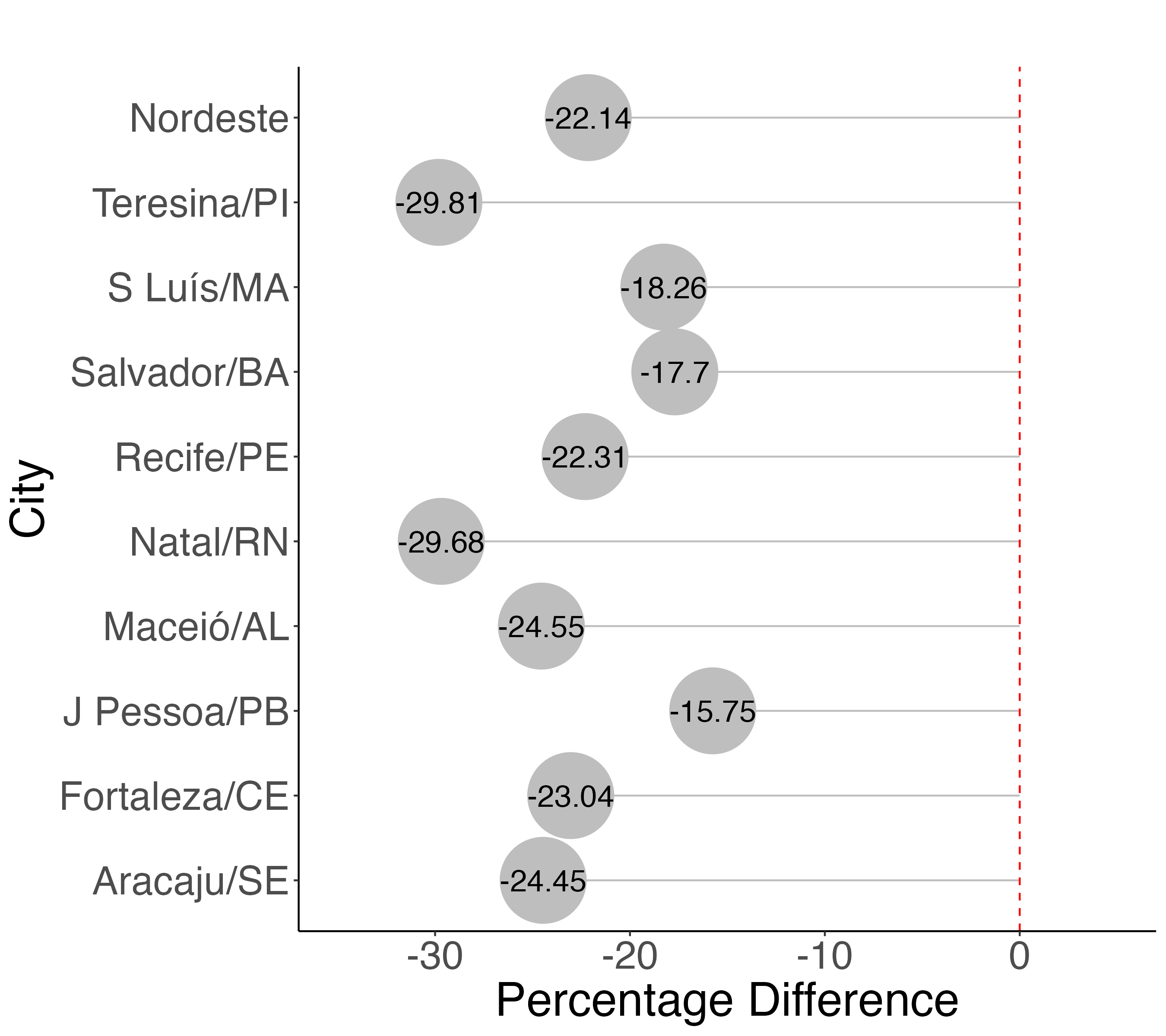}
        \caption{Physical Violence Lifetime}
    \end{subfigure}
    ~ 
    \begin{subfigure}[t]{0.45\linewidth}
        \centering
        \includegraphics[width=\linewidth, height = 5.5cm]{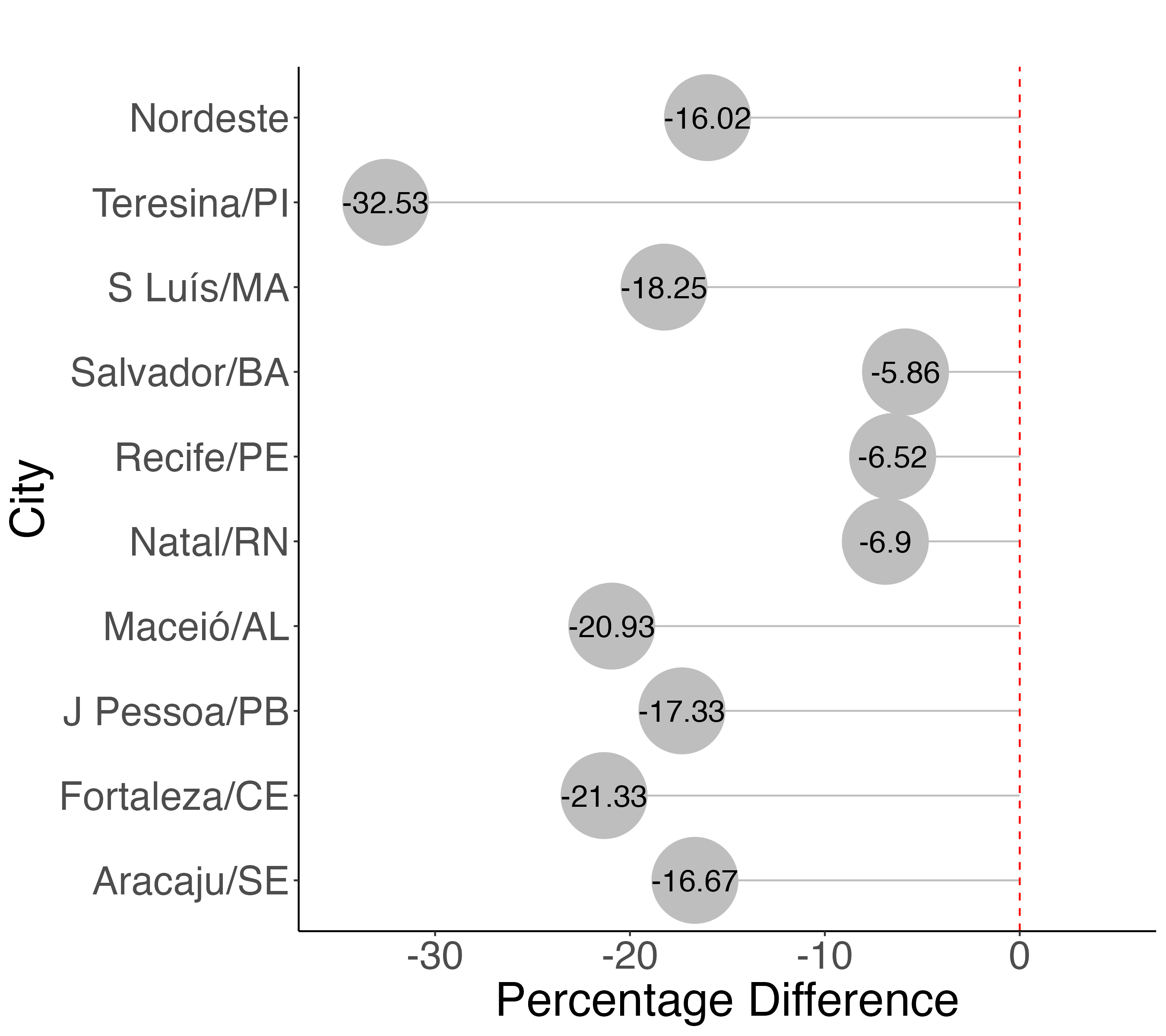}
        \caption{Physical Violence last 12 months}
    \end{subfigure}
    ~
    \begin{subfigure}[t]{0.45\linewidth}
        \centering
        \includegraphics[width=\linewidth, height = 5.5cm]{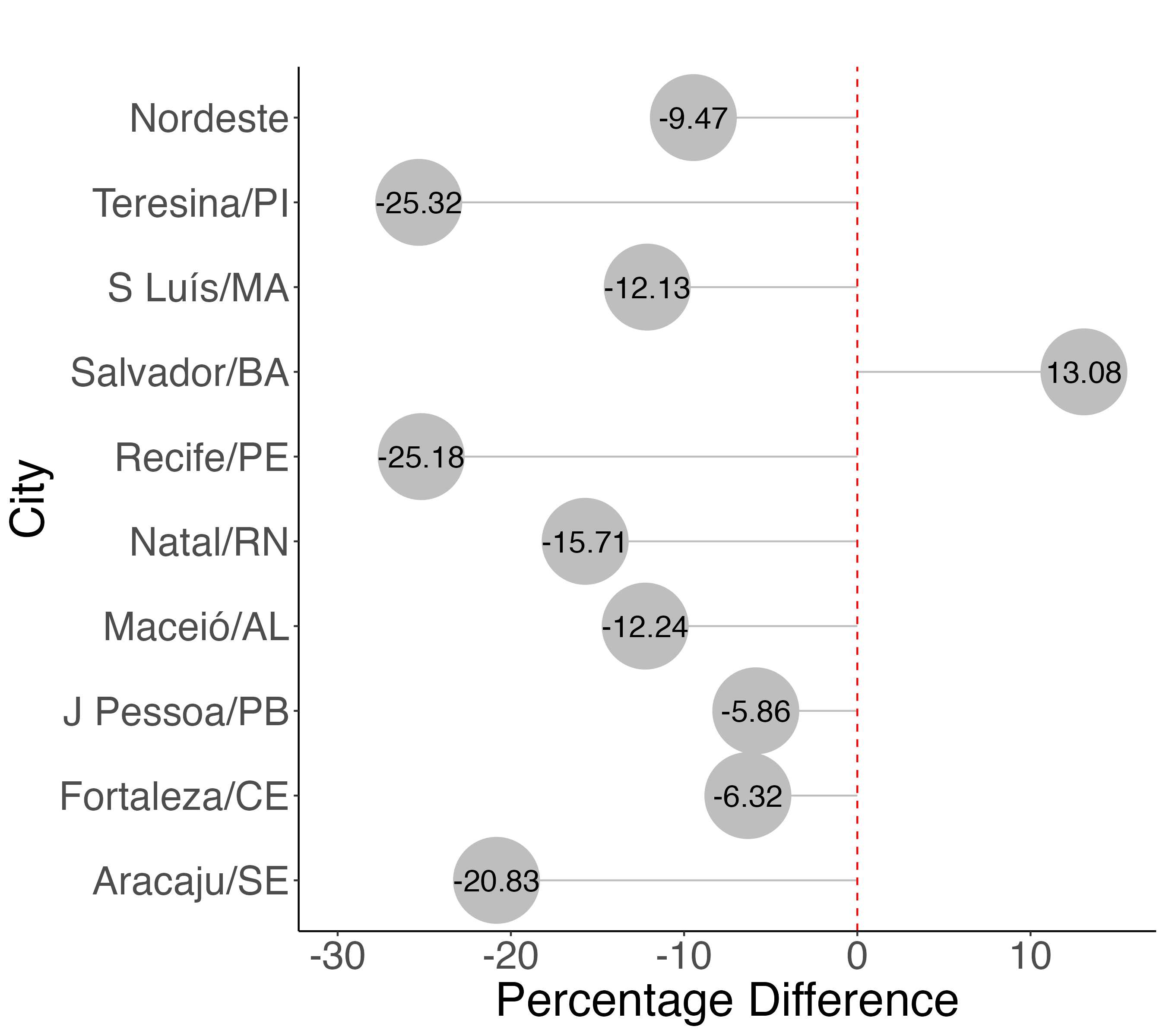}
        \caption{Sexual Violence Lifetime}
    \end{subfigure}
    ~ 
    \begin{subfigure}[t]{0.45\linewidth}
        \centering
        \includegraphics[width=\linewidth, height = 5.5cm]{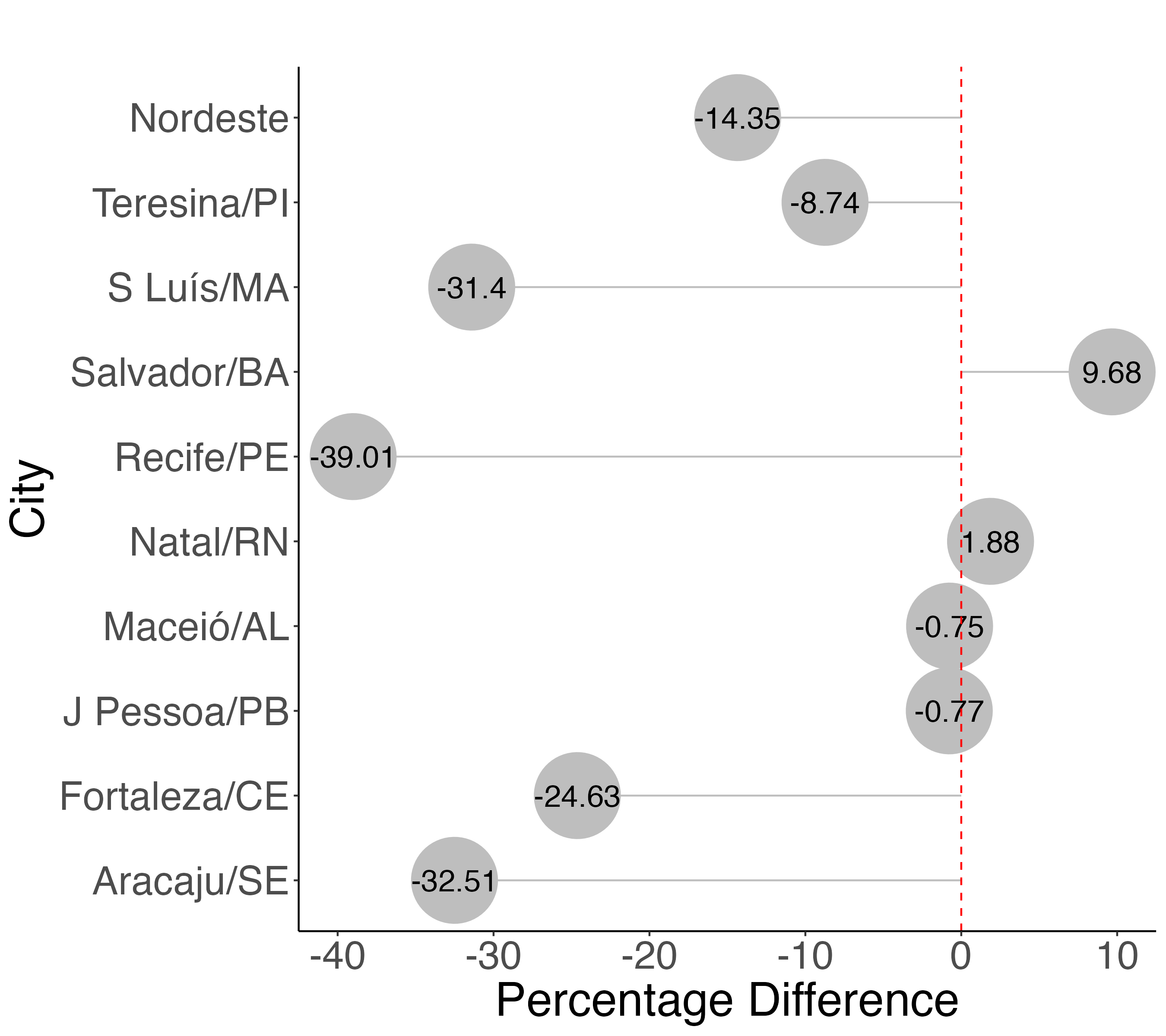}
        \caption{Sexual Violence last 12 months}
    \end{subfigure}
    ~
    \caption{Percentage Difference between Weighted and Unweighted Designs, 2017}\label{fig:c2_2017}
\end{figure}

Figures \ref{fig:c2_2016} and \ref{fig:c2_2017} plot the percentage differences between the weighted and unweighted designs, respectively, where $Diff$ is defined by: $Diff = \nicefrac{\left(PREV_{j}^{w} - PREV_{j}^{unw}\right)}{PREV_{j}^{unw}}$.


\newpage

\noindent Figure \ref{fig:c2_2016} depicts a first clear pattern where the calculated weights decrease the prevalence for all types of IPV (emotional, physical, and sexual) at all ``time windows'', and for all cities, except S\~{a}o Luis/MA \textit{vis a vis} the corresponding values calculated by the unweighted design; and to a much less extent Maceió/AL and Teresina/PI. Sometimes, this decrease is considerable ($51.63 \%$ at the city of Recife/PE in Sub-figure (f)).

A second observed regularity has to due with the scale of the downward correction brought by the weights in 2016. Again, except for S\~{a}o Luis/MA, comparing each type of IPV at different ``time windows'' (looking at each row separately) brings a pattern where the downward effect on prevalence values brought by the weights for the ``12 months'' window are roughly higher (in module) than the corresponding values for the ``Lifetime'' window. As a matter of fact, if we compare the highest percentage decrease at each line (for instance, for sexual violence $-51.63 < -27.43$) and so on, the pattern is definitely seen.

A third, and last, pattern has to do with the emergence of an ordering among types of violence. Sexual violence values of $Diff$ are roughly higher than those of physical violence, which possess higher values than the corresponding $Diff$ values derived for emotional violence at both time windows.  This is so even for S\~{a}o Luis/MA, a city with a positive correction estimated by the weighted design.

Figure \ref{fig:c2_2017} depicts a strong similarity with the first  pattern spotted in Figure \ref{fig:c2_2016}. Some differences appeared, though. For example, a few more cities have their weighted designs inflating their prevalence values, for instances, Salvador/BA and Natal/RN. However, this is a very mild change. Moving to check out Figure \ref{fig:c2_2016} second observed regularity, the same pattern emerges. Overall, the coincidence on both patterns for 2016 and 2017 seems a consistent but surprisingly result to us worth investigating further in the near future.

As to the ordering pattern observed in 2016, we see it again in 2017, at least partially. For the ``12 months'' window, Figure \ref{fig:c2_2017} depicts the ordering $\textrm{sexual } \succ \textrm{ physical } \succ \textrm{ emotional}$. However, that does not happen for the ``Lifetime'' window. Last, but not least, we analyzed inter-year differences among each of the six $Diff$ graphs. With the sole exception of emotional violence ``last 12 months'' $Diff$ graph, all other graphs show a striking regularity: the downward effect from the weighted design in 2016 is larger than observed in 2017.

Our analysis of the performance of our attempt to integrate a refreshment sample with an ongoing longitudinal sample to calculate IPV prevalence in our PCSVDF-Mulher study has delivered four interesting patterns of great importance for the violence against women agenda, especially to those interested in measurement issues: 1) the weights seem to suggest an overall necessity to proceed to a downward adjustment on prevalence values; 2) this downward adjustment should be higher for shorter ``time windows'' of IPV measurement; 3) there seems to be an ordering in the magnitude of design adjustment tied to the type of violence; and 4) behavioral dynamic issues tied to consecutive measurement years might be an important issue to address. As the cross-sectional weights affect longitudinal weights, our future next step of calculating these balanced panel weights should consider seriously the patterns we discovered here (\citep{Watson_2014}, \citep{taylor_evaluating_2020} and \citep{Watson2021}).

\subsection{The Weighted to Unweighted Variance Ratio ($VarRatio$)}

\begin{figure}[h!]
    \centering
    \begin{subfigure}[t]{0.45\linewidth}
        \centering
        \includegraphics[width=\linewidth, height = 5.5cm]{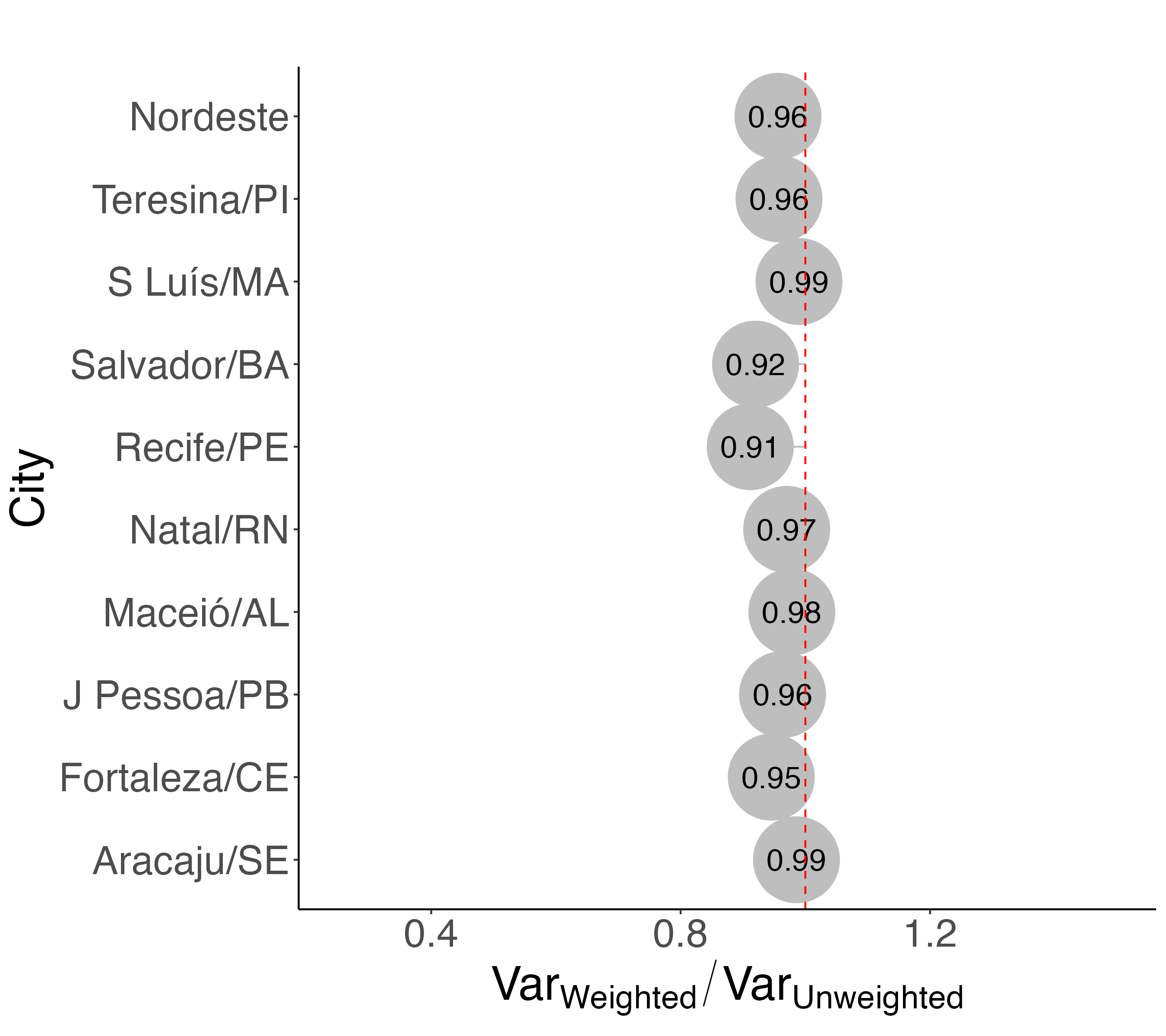}
        \caption{Emotional Violence Lifetime}
    \end{subfigure}%
    ~ \vspace{0.5cm}
    \begin{subfigure}[t]{0.45\linewidth}
        \centering
        \includegraphics[width=\linewidth, height = 5.5cm]{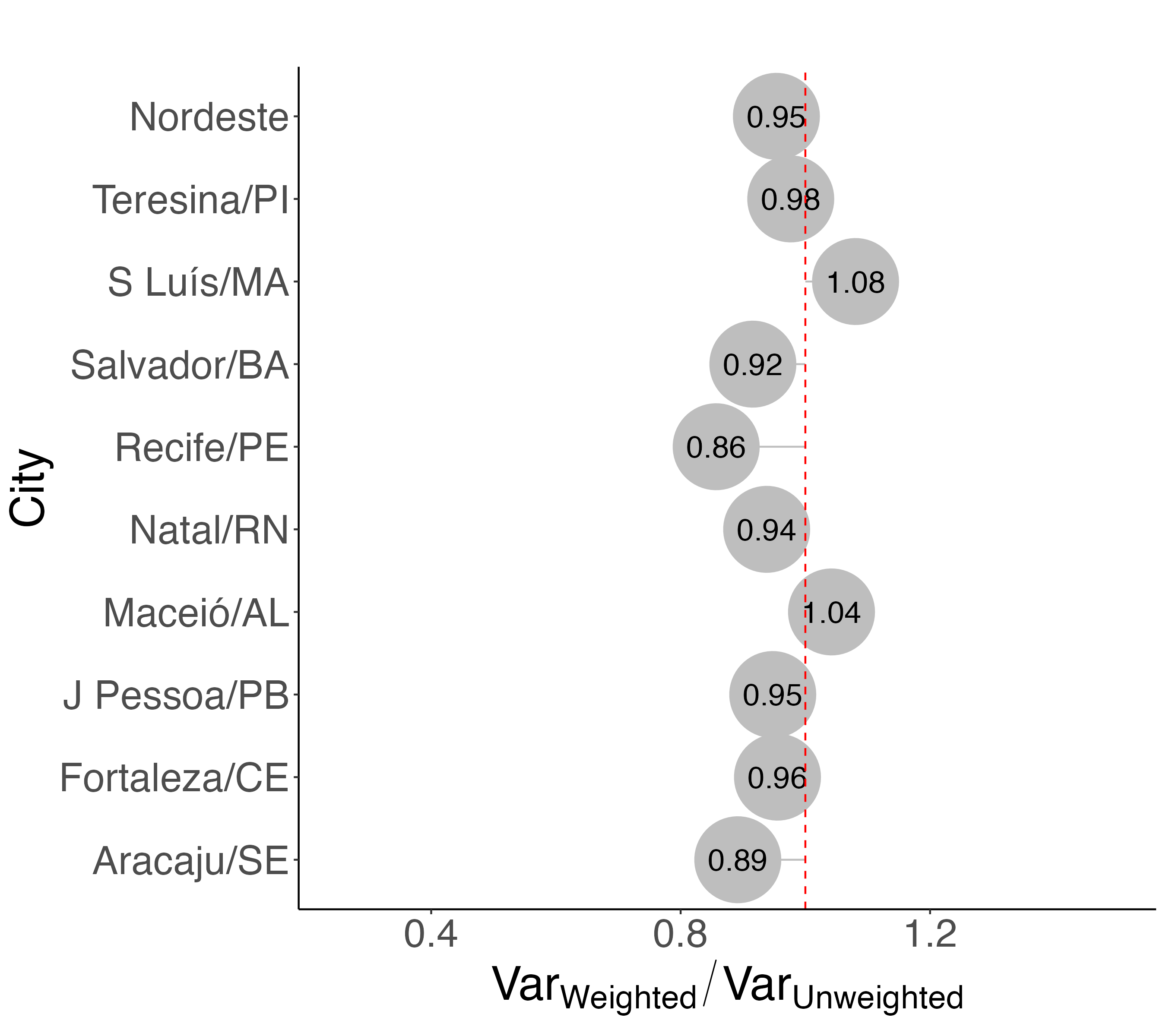}
        \caption{Emotional Violence last 12 months}
    \end{subfigure}
    ~ \vspace{0.5cm}
    \begin{subfigure}[t]{0.45\linewidth}
        \centering
        \includegraphics[width=\linewidth, height = 5.5cm]{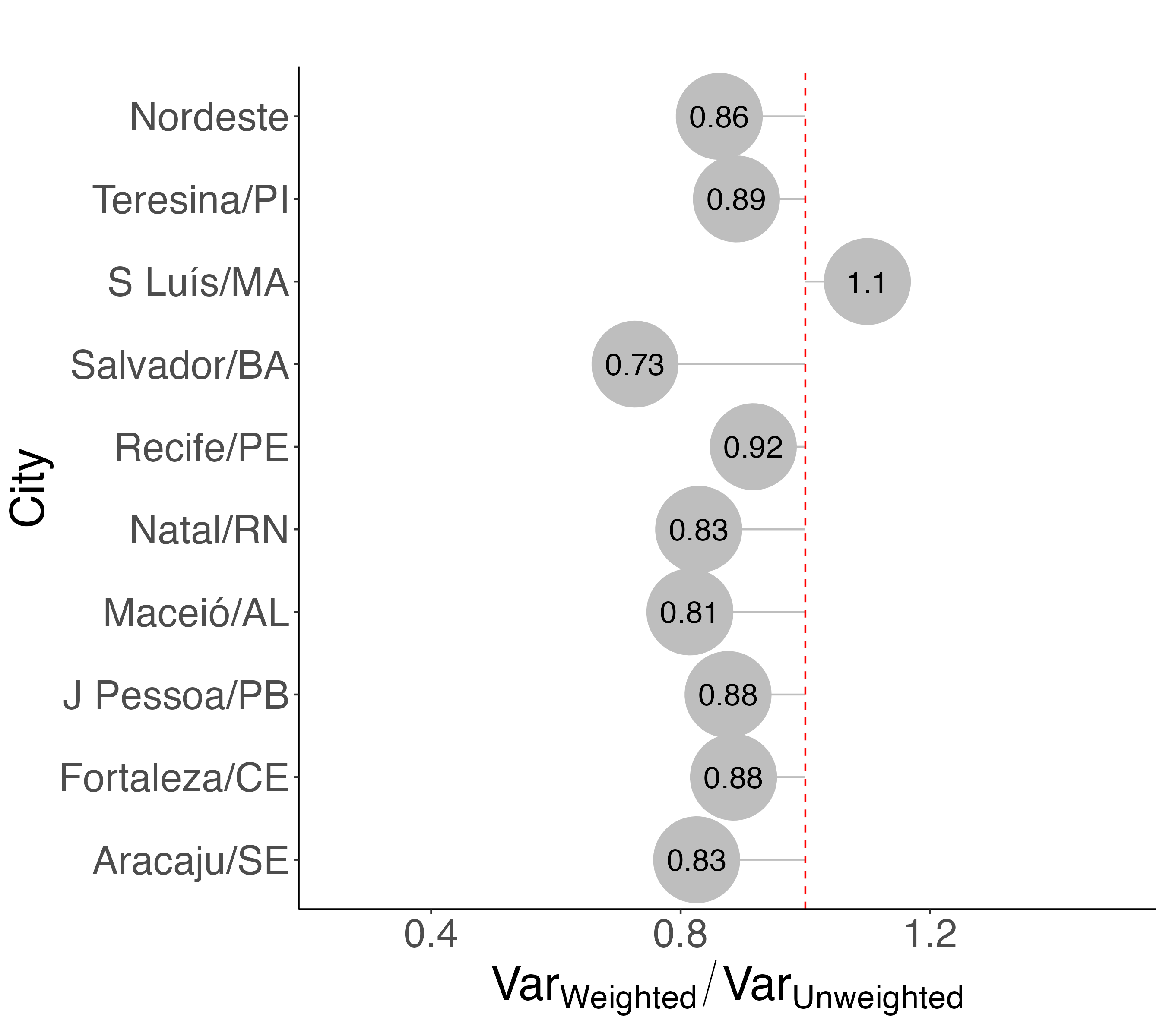}
        \caption{Physical Violence Lifetime}
    \end{subfigure}
    ~ 
    \begin{subfigure}[t]{0.45\linewidth}
        \centering
        \includegraphics[width=\linewidth, height = 5.5cm]{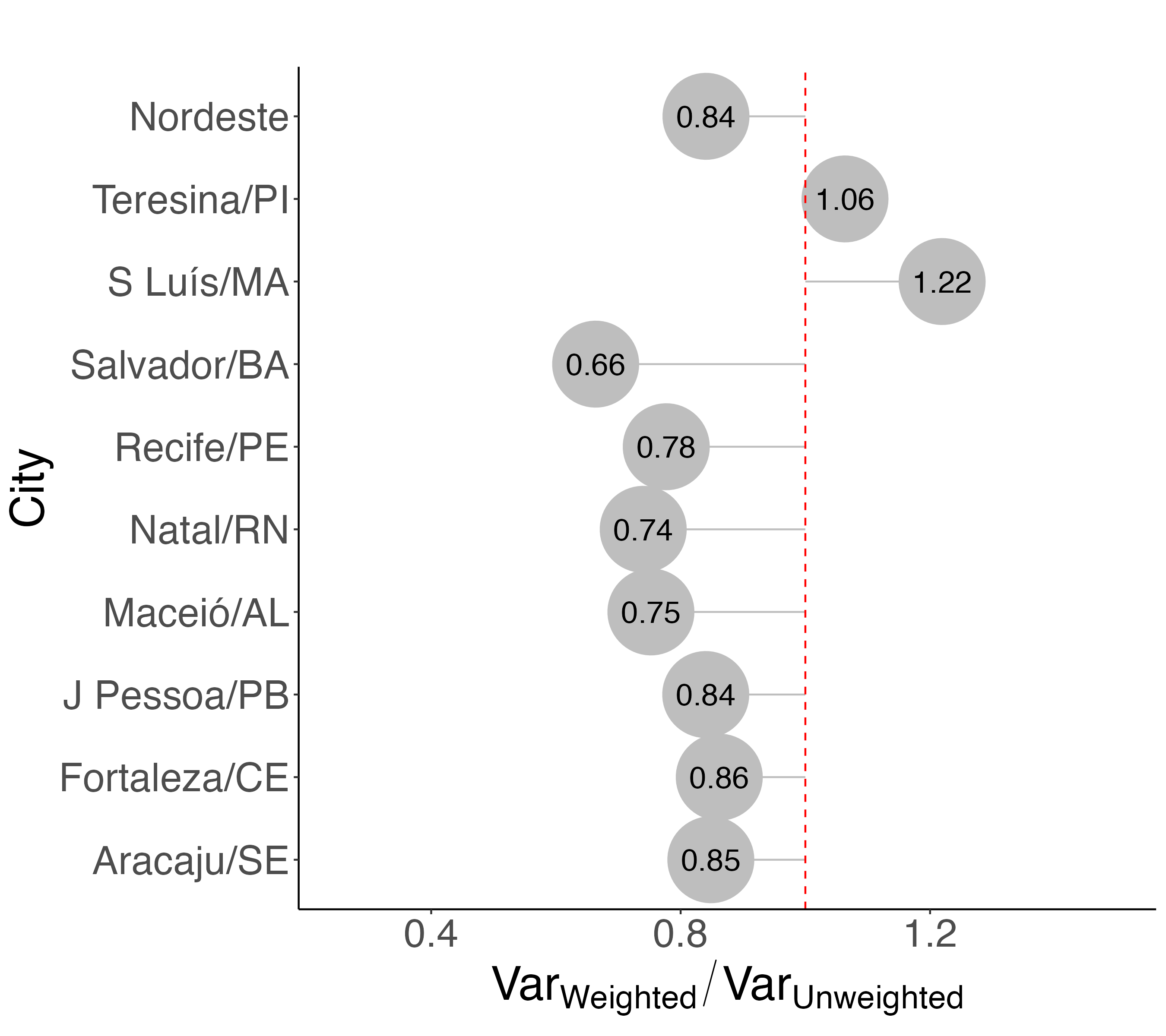}
        \caption{Physical Violence last 12 months}
    \end{subfigure}
    ~
    \begin{subfigure}[t]{0.45\linewidth}
        \centering
        \includegraphics[width=\linewidth, height = 5.5cm]{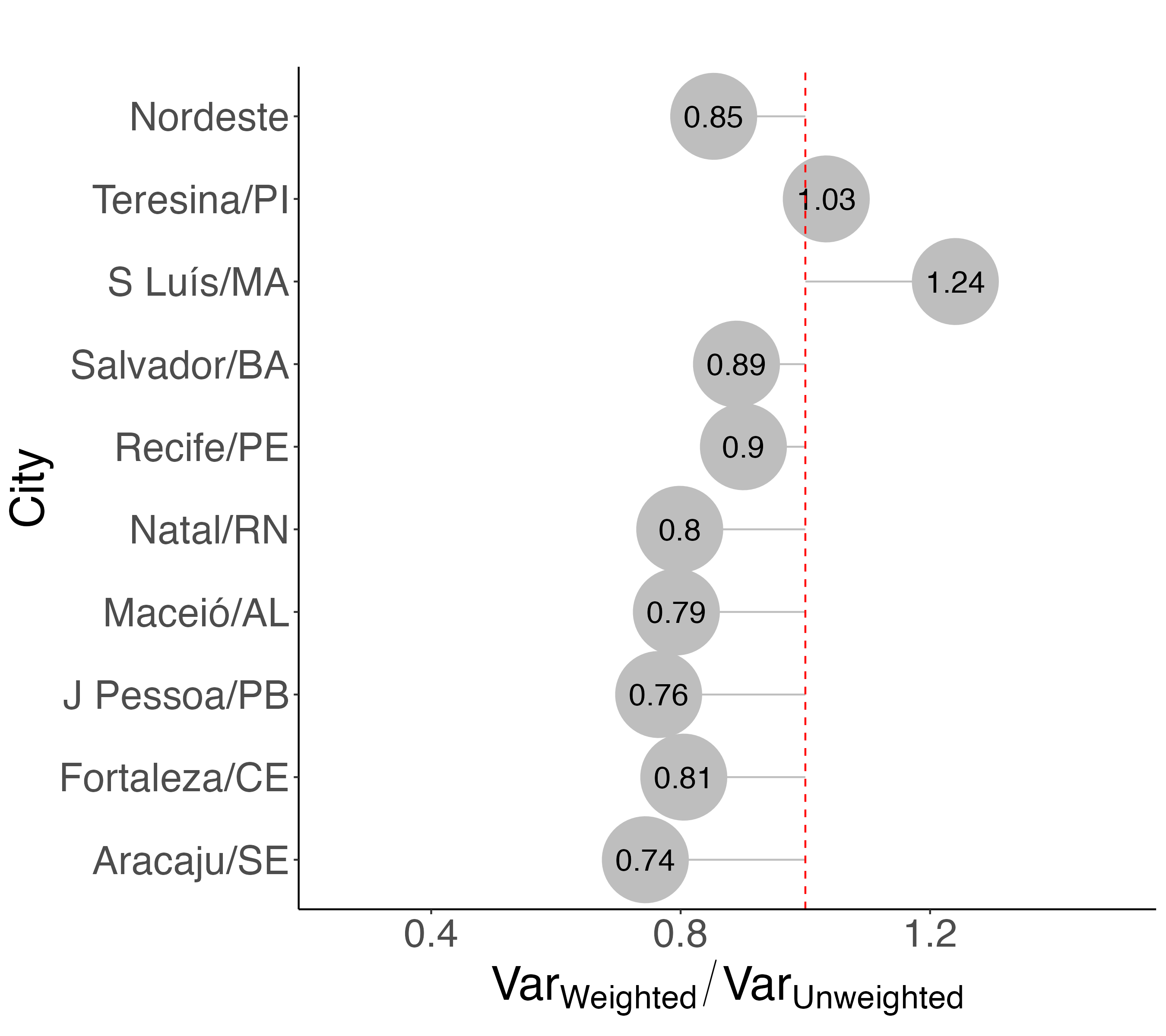}
        \caption{Sexual Violence Lifetime}
    \end{subfigure}
    ~ 
    \begin{subfigure}[t]{0.45\linewidth}
        \centering
        \includegraphics[width=\linewidth, height = 5.5cm]{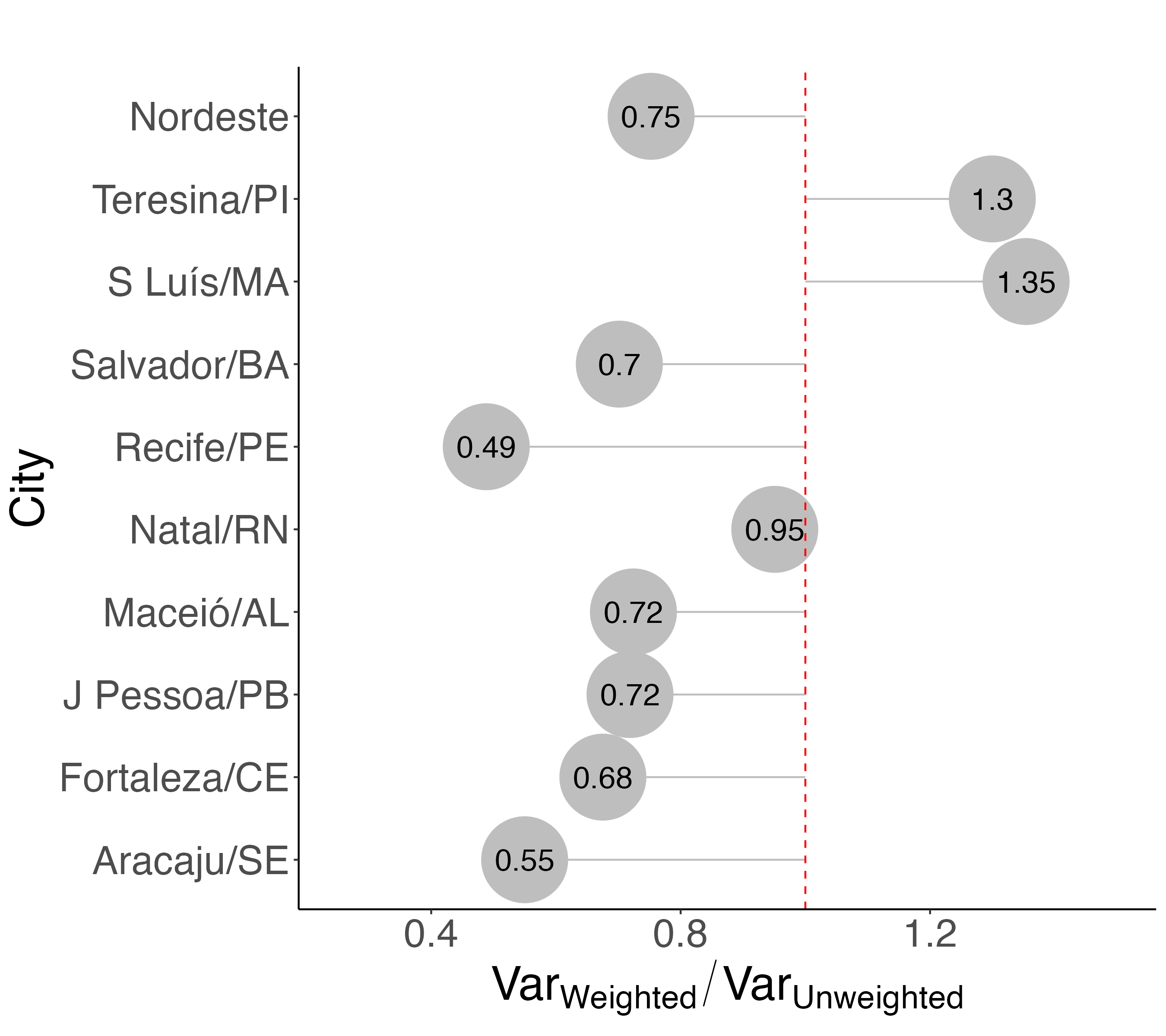}
        \caption{Sexual Violence last 12 months}
    \end{subfigure}
    ~
    \caption{Percentage Difference between Weighted and Unweighted Designs, 2016}\label{fig:VR_2016}
\end{figure}

\begin{figure}[h!]
    \centering
    \begin{subfigure}[t]{0.45\linewidth}
        \centering
        \includegraphics[width=\linewidth, height = 5.5cm]{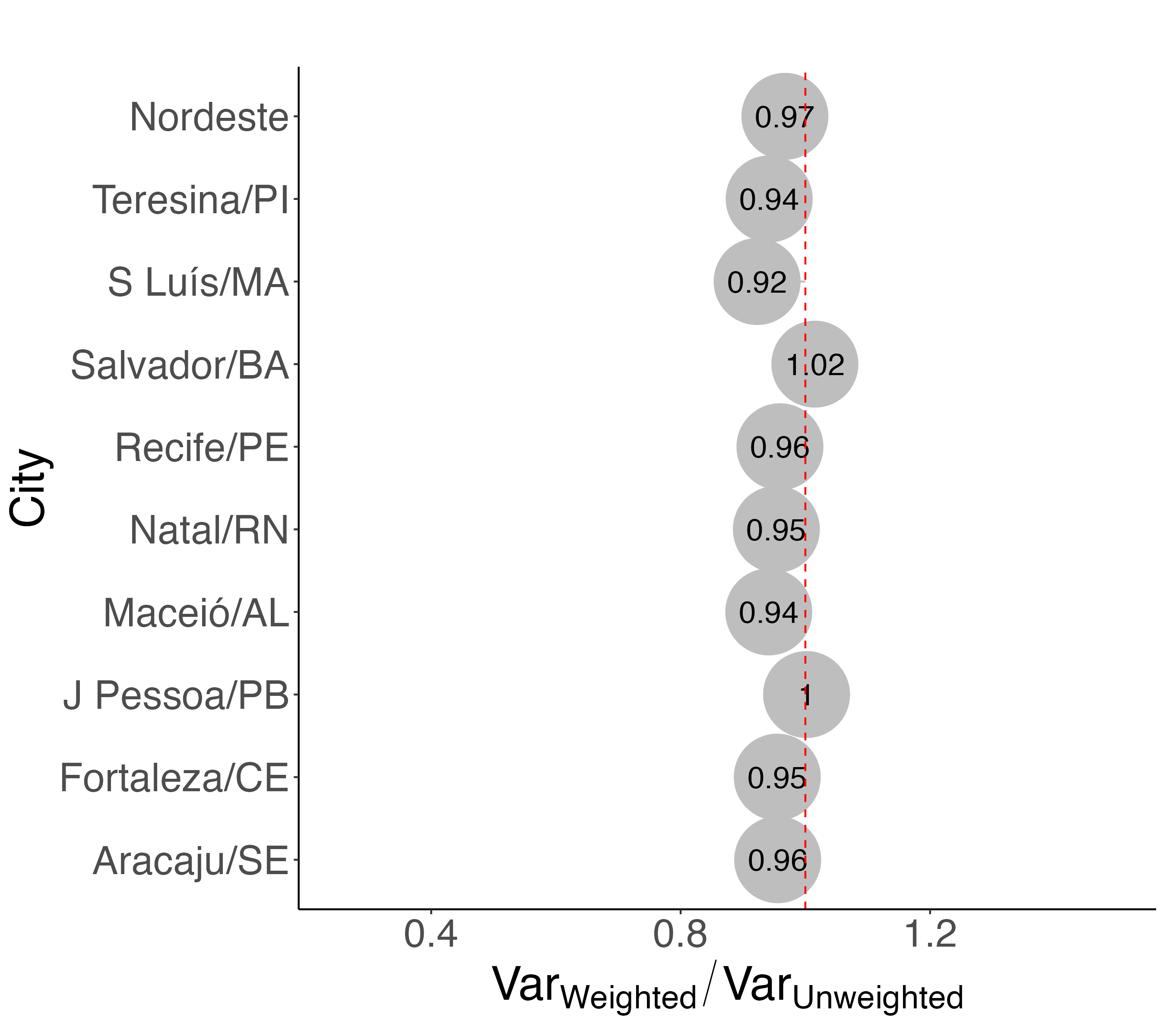}
        \caption{Emotional Violence Lifetime}
    \end{subfigure}%
    ~ \vspace{0.5cm}
    \begin{subfigure}[t]{0.45\linewidth}
        \centering
        \includegraphics[width=\linewidth, height = 5.5cm]{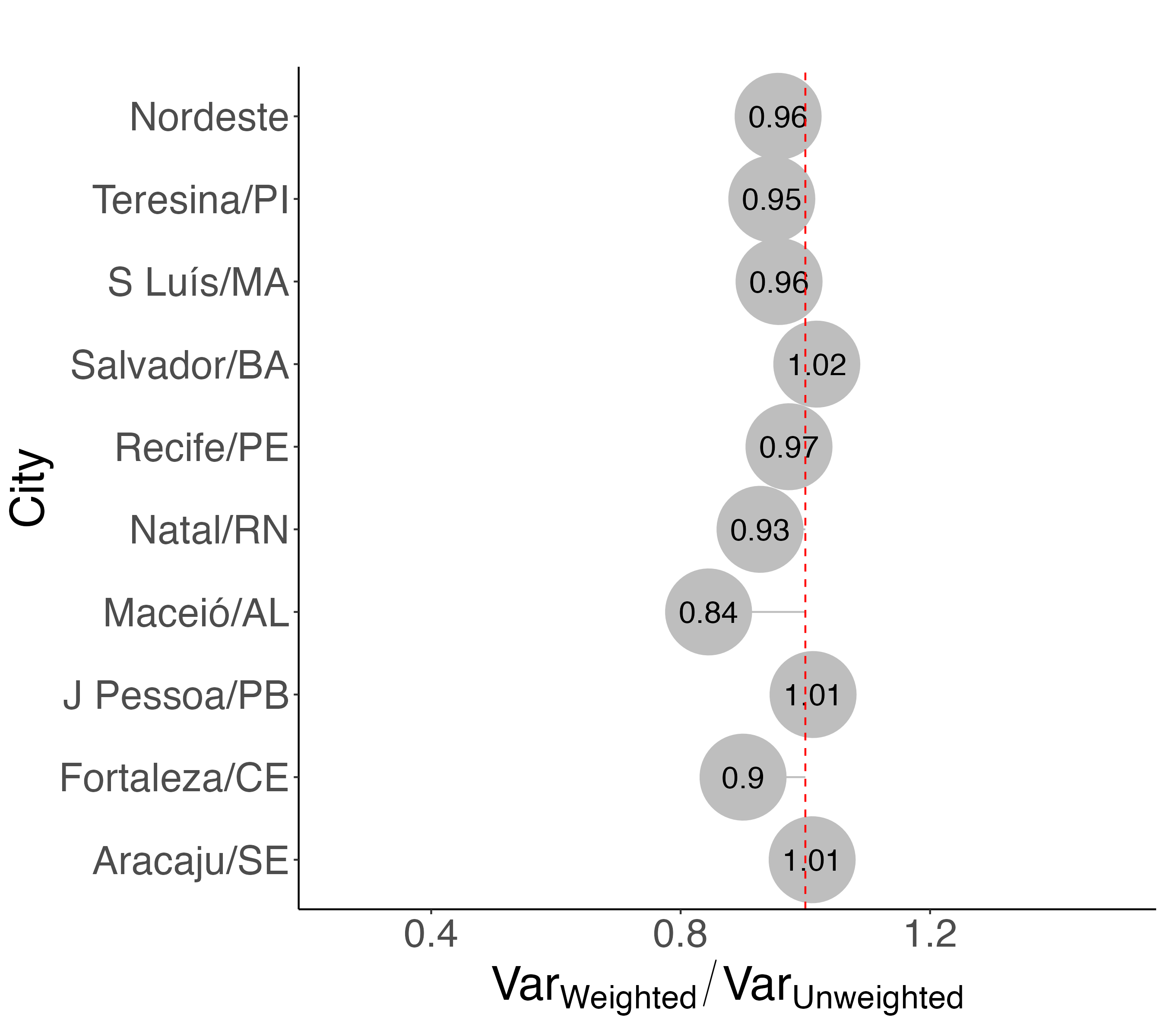}
        \caption{Emotional Violence last 12 months}
    \end{subfigure}
    ~ \vspace{0.5cm}
    \begin{subfigure}[t]{0.45\linewidth}
        \centering
        \includegraphics[width=\linewidth, height = 5.5cm]{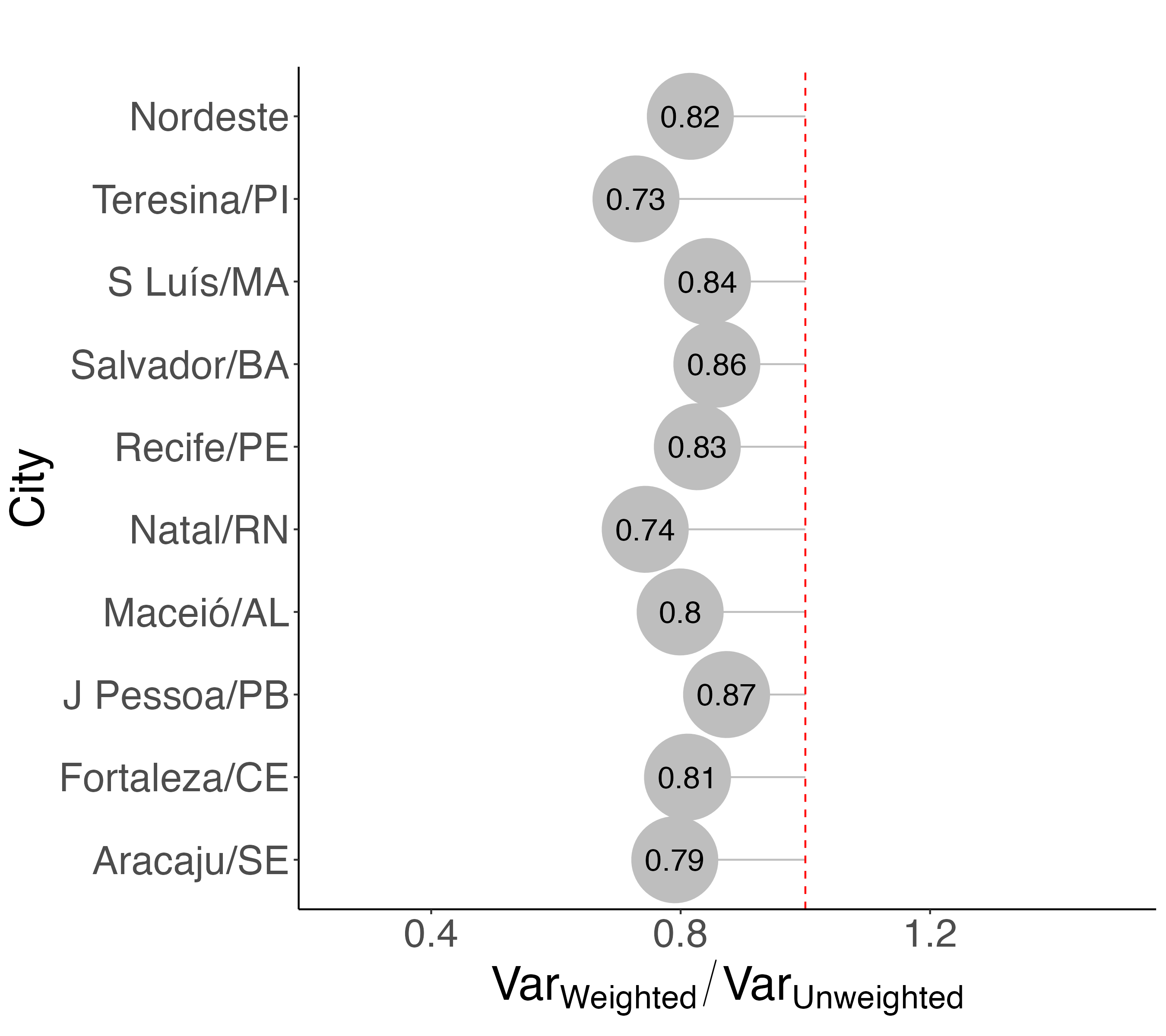}
        \caption{Physical Violence Lifetime}
    \end{subfigure}
    ~ 
    \begin{subfigure}[t]{0.45\linewidth}
        \centering
        \includegraphics[width=\linewidth, height = 5.5cm]{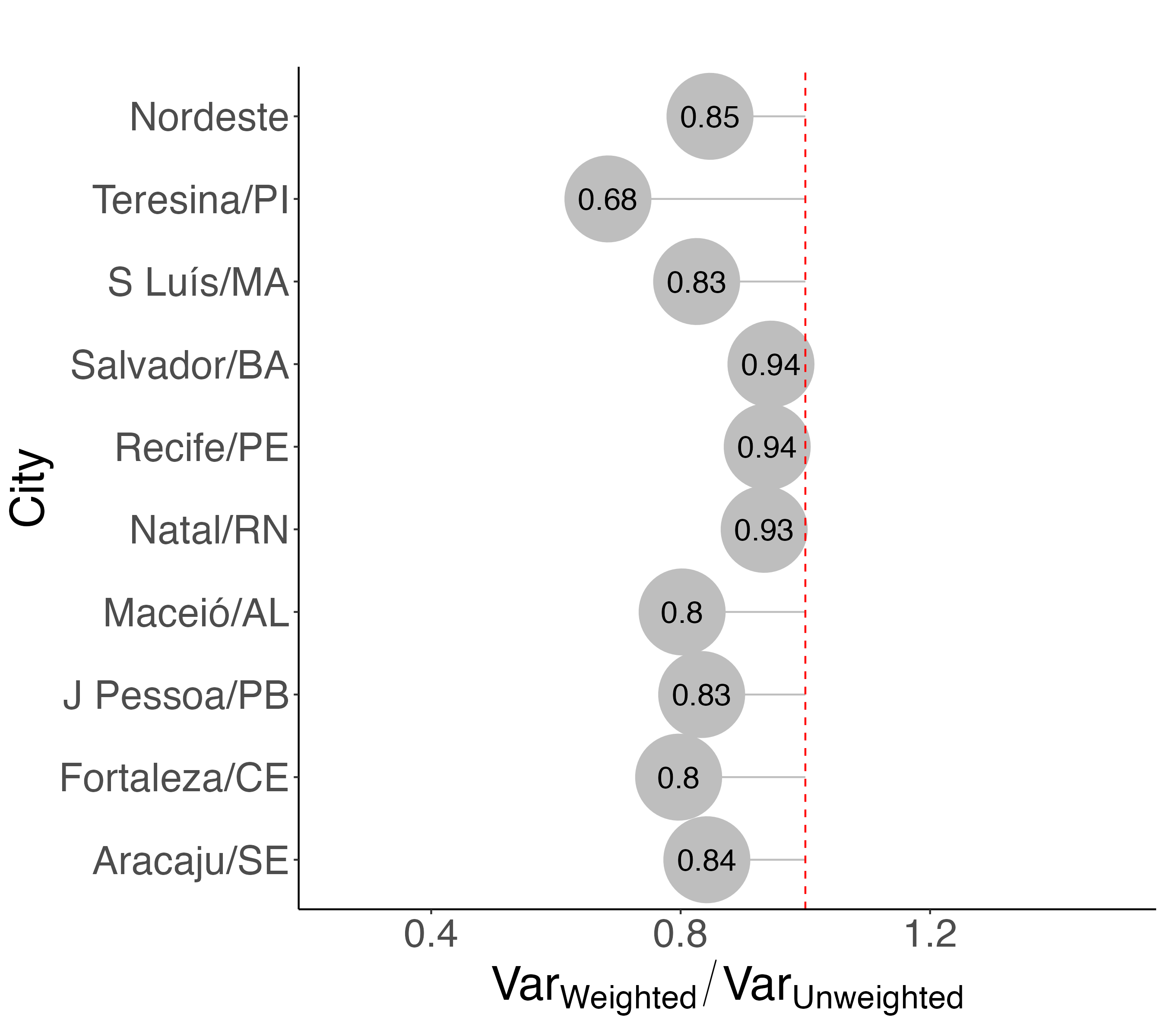}
        \caption{Physical Violence last 12 months}
    \end{subfigure}
    ~
    \begin{subfigure}[t]{0.45\linewidth}
        \centering
        \includegraphics[width=\linewidth, height = 5.5cm]{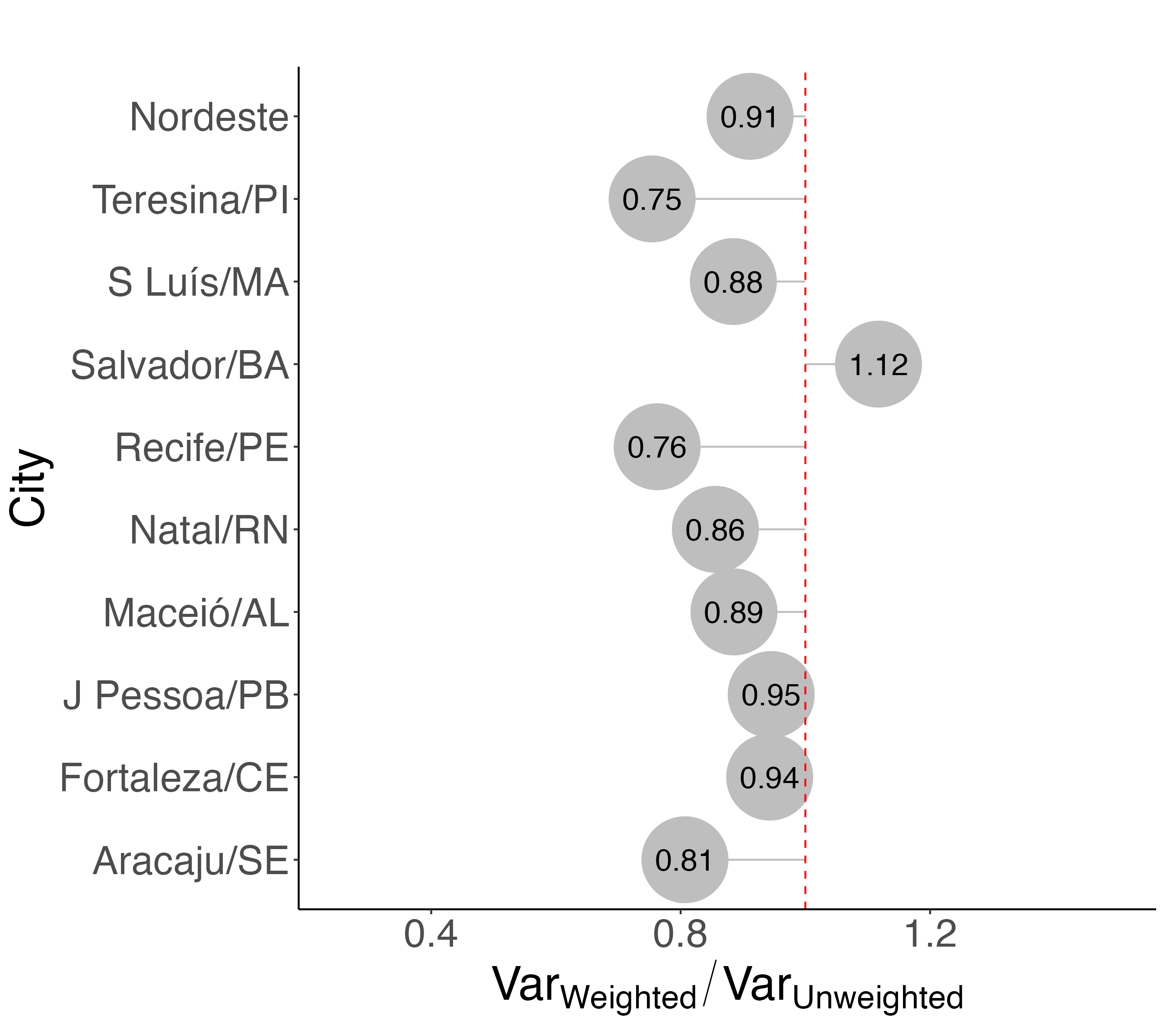}
        \caption{Sexual Violence Lifetime}
    \end{subfigure}
    ~ 
    \begin{subfigure}[t]{0.45\linewidth}
        \centering
        \includegraphics[width=\linewidth, height = 5.5cm]{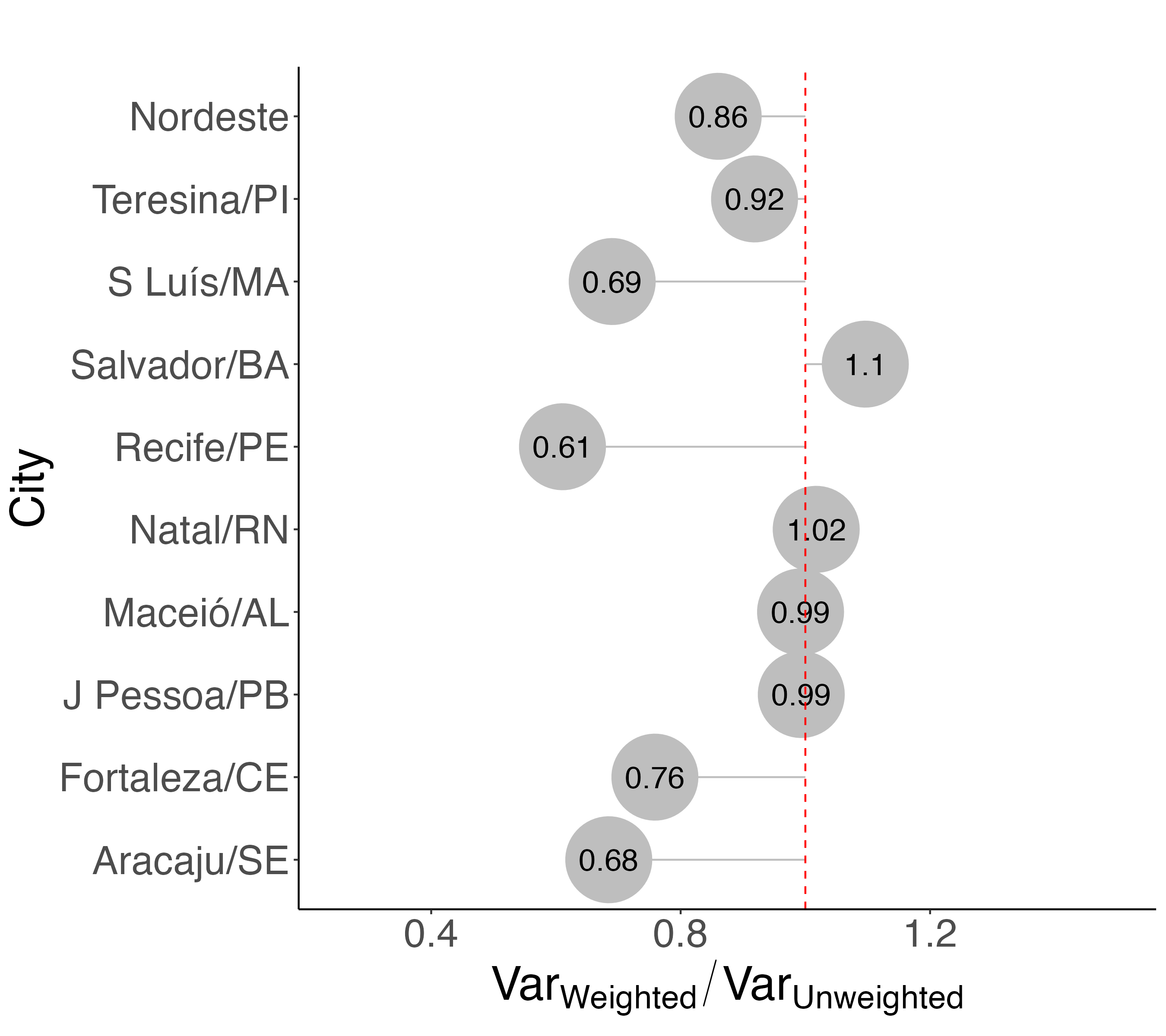}
        \caption{Sexual Violence last 12 months}
    \end{subfigure}
    ~
    \caption{Percentage Difference between Weighted and Unweighted Designs, 2017}\label{fig:VR_2017}
\end{figure}

Figures \ref{fig:VR_2016} and \ref{fig:VR_2017} plot the ratio between the weighted and unweighted designs variances, respectively, for $j = 2016 \textrm{ and } 2017$ where $VarRatio$ is defined by: $VarRatio = \nicefrac{Var(PREV_{j}^{w})}{Var(PREV_{j}^{unw})}$.


Both variances were calculated using the Taylor linearization method, with the design incorporating the effects of stratification and clustering. In cases of singleton stratum, we used the \textit{options(survey.lonely.psu=``adjust'')} in the R survey package \citep{Lumley2020}, which gives a conservative variance estimator that uses residuals from the population mean rather than the stratum mean.

It is worth reminding the reader that the PCSVDF-Mulher is a multipurpose (in the sense that it has collected a very large number of variables, see, \cite{Haziza2017}); longitudinal population-representative survey with a considerable number of ``sensitive questions'' (in the sense that these questions are either intrusive, say, touch on ``taboo'' topics, inappropriate in everyday conversation or out of bounds to be asked; or these questions involve the threat of disclosure, see, \cite{Tourangeau2007}) applied to a developing country. Also, as outlined before, the PCSVDF-Mulher sampling plan used a complex, stratified, multistage probability cluster sampling design with unequal selection probabilities to select Wave I (2016) participants. Not to mention the presence of a refreshment sample.

Last, to face these challenges and correct biases, a strategy of weighting that combines, besides the calculation of base weights, weights for nonresponse correction based on propensity score \textit{via} logistic regression improved by trimming and post-stratification (for these reasons, it is fair to say that PCSVDF-Mulher weighting design is complex, see, \cite{Sarndal2005-ca}, \cite{Haziza2017}, and \cite{Valliant2018}, among others) has been successfully deployed. Hence, the question on how to assess the effect this complex weighting design might have on the variance of the IPV prevalence (a ratio) is not a trivial task.

Up to the 1990's, the general idea on the likely impact of weighting nonresponse on estimates (basically, totals or means) had been cast as the famous ``bias-variance'' trade-off where the reduction of bias brought by weighting would come with a due price of increasing estimates variances, see, \cite{Kish1992}. However, \cite{LittleVartivarian2005} have shifted that oversimplified view, showing a more nuanced and complex setup where weighting can, indeed, decrease the variance and sometimes decrease both variance and bias. The key result is synthetized in Table 1 of \cite[p. 7]{LittleVartivarian2005}.

As long as the auxiliary variables used to estimate the propensity score of nonresponse are associated with the prediction of the outcome variable whose variance must be estimated, weighting lowers variance. If these auxiliary variables are highly associated with both the propensity score and the outcome variable, weighting lowers both bias and variance. Trimming and poststratification can also reduce variance, a point advocated by both \cite{LittleVartivarian2005}, and \cite{Chen2017}.

Considering all that, our next $VarRatio$ Figures should be taken just as an initial exploratory data analysis whose objective is to capture some descriptive patterns on the ``bias-variance'' trade-off in a set up that combine a complex multipurpose longitudinal survey  with a complex sample design and weighting.

\newpage

\noindent Figure \ref{fig:VR_2016} depicts a general pattern of variance improvement due to the weighting for nonresponse, with few exceptions for S Luís/MA and Teresina/PI; and Sub-figure (a), emotional violence ``Lifetime'', where all values are close to 1, although all are strictly less than 1. Sometimes the efficiency gain is pronounced, for instances, Sub-figures (e) and (f). This is good news, as sexual violence is the least prevalent type of IPV, posing serious practical difficulties for its measurement and analysis. This is likely due to both its low frequency in the society and its reporting being much lower due to the obvious sensitiveness of the question.

A second pattern is related to the values of variance ratio when we compare the second column (``last 12 months'' window) with the first column (``Lifetime'' window). The variance effect is lower (its $VarRatio$ is greater) at the first column \textit{vis a vis} the second one, for both decreases or increases in variance. As before in case of sexual violence, we conjectured that this pattern is related to the fact that the ``last 12 months'' window is shorter than the ``Lifetime'' window, implying prevalence values that are logically lower at former window when compared to values calculated at the later window.

Figure \ref{fig:VR_2017} deals with the 2017 figures. It does repeat both patterns observed in 2016, a general variance improvement as well as larger effects in terms of variance reduction at the ``last 12 months'' window compared to the ``Lifetime'' window. For this year; with the exception of Salvador/BA in Sub-figures (a), (e) and (f), J Pessoa/PB in Sub-figure (a) and (b), Natal/RN in Sub-figure (f),  and Aracaju/SE in Sub-figure (b) (as a matter of fact, all these four values are $\leq 1.12$); all other 53 values are $\leq 1.0$. At this juncture, we can only conjecture that the incorporation of the refreshment sample into the weighting design could have a decisive role on these efficiency gains.

\section{Final Considerations}\label{SECTION_Final}

In this paper, our main contribution has to do with devising and implementing a strategy for calculating individual cross-sectional weights and applying them to the 2016 and 2017 waves of the PCSVDF-Mulher, a large and interdisciplinary effort to build and empirical longitudinal dataset in Brazil to study of domestic violence, its causes and consequences to direct and indirect victims. By following \citep{o2002combining}, \citep{Watson_2014}, and \citep{Watson2021}, together with due adaptations to some issues idiosyncratic to the PCSVDF-Mulher design, we have come up with a set of weights that combines a refreshment sample collected in 2017 with the ongoing longitudinal sample started in 2016.

Our findings for both the weighted and unweighted designs depict values of prevalence very much in consonance to those statistics already collected in Brazil and other countries in the past and in recent years which applied a similar methodology and the same questionnaire, although without supplemental samples. For instance, \citep{MorenoJansen2005} for a set of countries, \citep{Schraiber2007} for Brazil only, \citep{Bott2019} for Latin American countries, and \citep{Sardinha2022} for more recent values calculated for a larger group of countries. This remark is worth emphasizing right way as it serves as a first important accomplishment in terms of the overall consistency of the PCSVDF-Mulher project and our calculated weights.

The weighted design points to the necessity of a general downward correction of values for prevalence calculated by the unweighted design (all over intimate partner violence types, geographies, time windows, and years), sometimes by a considerable amount. This has strong implications for the scientific analysis and public policy debate over the issue of domestic violence, and more specifically to intimate partner violence as it adds a yet underappreciated statistical concern related to the use of a correct weighting design necessary to longitudinal studies. As a matter of fact, the indiscriminate use of unweighted designs might artificially and inadvertently inflate the values of calculated IPV prevalence, which brings distortions and considerable political, social, budgetary, and scientific implications. 

As to the second and third found patterns of our weighted design, respectively, the time window effect and the ordering effect, we see another layer of consistency in our analysis. Both set up are either logically or behaviourally constructed in such a way that reflects on sub-samples values of prevalence of comparable difference.

The ``12 months’’ sub-sample prevalence is logically smaller than that of the ``Lifetime’’ sub-sample; and the literature (both at clinic set ups and population surveys) has already established a huge amount of evidence that the behavioral scripts associated with IPV translate into an ascending ordering in terms of values of prevalence (over different times, cultures and methodologies). This ordering posits the prevalence of sexual IPV first, the prevalence of physical violence second, and the prevalence of emotional violence as the third and highest one. Hence, our weights constructed by integrating a refreshment sample with an ongoing longitudinal sample seem to react to the larger variances present in smaller sub-samples.

As to the emerging literature related to the effects of weighting in the variance of estimates prompted by \cite{LittleVartivarian2005}, our analysis provides a set of new empirical results that reveals a positive (variance reduction) role that propensity score by means of a logistic regression improved by trimming and post-stratification might have in complex surveys. Of course, more theoretical refinements and empirical replications are much needed to convince. 

As wisely asserted by Rothman 2018, \textit{``the key to effective prevention [of IPV] is knowing why dating abuse happens and choosing determinants from an empirically supported theoretical model to design the strategy to address it. Though that might sound straightforward, social scientists have written about more than 20 theories to explain why partner abuse happens, so selecting a theory can be a complicated endeavor''}. Selecting a theory or looking for causal explanations of IPV seems to be a much needed priority on the agenda now, as reflected by papers like \citep{Jewkes2002}, \cite{Averett2016}, \citep{Mulla2018}, \citep{Rose2018},  and \citep{Rothman2018}. That was exactly our vision back in 2016 and it continues to be so. The breadth and interdisciplinary structure of the PCSVDF-Mulher questionnaire has been supporting our vision of approaching IPV from a modern, rigorous, and interdisciplinary perspective.

In essence, we are interested in shedding light on the causes and consequences of IPV. However, for that endeavor, collecting primary good longitudinal data and using the best statistical knowledge related to survey design are absolutely key steps as they: (1) expand the geographic and temporal scope of our analyses; (2) evaluate the selectivity of sample attrition; (3) restore statistical representativeness, (4) model socioeconomic behavior more accurately; (5) understand the process of transmission of domestic violence between generations and its implications to child development; and (6) understand transition dynamics.

Our attempt to calculate individual cross-sectional weights for the PCSVDF-Mulher project by means of innovative methodologies seems to us a modest, though important, first step to provide the necessary statistical pedigree to our efforts. Although we still need to learn from much more mature, widely supported, and well-funded longitudinal projects that have been contributing to the understanding of the causes of IPV; such as the United Kingdom Household Longitudinal Survey (UKHLS), \citep{Blom2023}, and the Panel Study of Income Dynamics (PSID), \citep{Papp2023}; we believe PCSVDF-Mulher set an innovative and viable attempt to provide an empirical benchmark of quality to back efforts to consider seriously the study of the causes and consequences of IPV in Brazil and similar developing countries.



\bibliographystyle{chicago}
\bibliography{SampleDesignPCSVDF}

\begin{thebibliography}{}

\bibitem[\protect\citeauthoryear{Andress, Golsch, and Schmidt}{Andress
  et~al.}{2015}]{Andress2015-kr}
Andress, H.-J., K.~Golsch, and A.~W. Schmidt (2015, July).
\newblock {\em Applied panel data analysis for economic and social surveys\/}
  (2013 ed.).
\newblock Berlin, Germany: Springer.

\bibitem[\protect\citeauthoryear{Averett and Wang}{Averett and
  Wang}{2016}]{Averett2016}
Averett, S.~L. and Y.~Wang (2016, January).
\newblock Identifying the causal effect of alcohol abuse on the perpetration of
  intimate partner violence by men using a natural experiment: Alcohol abuse
  and ipv.
\newblock {\em Southern Economic Journal\/}~{\em 82\/}(3), 697–724.

\bibitem[\protect\citeauthoryear{Blom and Gash}{Blom and Gash}{2023}]{Blom2023}
Blom, N. and V.~Gash (2023, November).
\newblock Measures of violence within the united kingdom household longitudinal
  survey and the crime survey for england and wales: An empirical assessment.
\newblock {\em Social Sciences\/}~{\em 12\/}(12), 649.

\bibitem[\protect\citeauthoryear{Bott, Guedes, Ruiz-Celis, and Mendoza}{Bott
  et~al.}{2019}]{Bott2019}
Bott, S., A.~Guedes, A.~P. Ruiz-Celis, and J.~A. Mendoza (2019, March).
\newblock Intimate partner violence in the americas: a systematic review and
  reanalysis of national prevalence estimates.
\newblock {\em Revista Panamericana de Salud Pública\/}~{\em 43}, 1.

\bibitem[\protect\citeauthoryear{Breiding, Basile, Smith, Black, and
  Mahendra}{Breiding et~al.}{2015}]{BreidingEtAll2015}
Breiding, M.~J., K.~C. Basile, S.~G. Smith, M.~C. Black, and R.~Mahendra
  (2015).
\newblock Intimate partner violence surveillance uniform definitions and
  recommended data elements.
\newblock Technical report, Centers for Disease Control and Prevention.

\bibitem[\protect\citeauthoryear{Burton}{Burton}{2012}]{Burton2012}
Burton, J. (2012).
\newblock Understanding society innovation panel wave 4: Results from
  methodological experiments – institute for social and economic research.
\newblock Technical report, MInstitute for Social and Economic Research (ISER).

\bibitem[\protect\citeauthoryear{Carvalho, de~Oliveira, and da~Silva}{Carvalho
  et~al.}{2018}]{Carvalho2018}
Carvalho, J.~R., V.~H. de~Oliveira, and A.~B.~R. da~Silva (2018).
\newblock The pcsvdf study: New data, prevalence and correlates of domestic
  violence in brazil.
\newblock Technical report.

\bibitem[\protect\citeauthoryear{Chapman and Gillespie}{Chapman and
  Gillespie}{2019}]{Chapman2019}
Chapman, H. and S.~M. Gillespie (2019, January).
\newblock The revised conflict tactics scales (cts2): A review of the
  properties, reliability, and validity of the cts2 as a measure of partner
  abuse in community and clinical samples.
\newblock {\em Aggression and Violent Behavior\/}~{\em 44}, 27–35.

\bibitem[\protect\citeauthoryear{Chen, Elliott, Haziza, Yang, Ghosh, Little,
  Sedransk, and Thompson}{Chen et~al.}{2017}]{Chen2017}
Chen, Q., M.~R. Elliott, D.~Haziza, Y.~Yang, M.~Ghosh, R.~J.~A. Little,
  J.~Sedransk, and M.~Thompson (2017, May).
\newblock Approaches to improving survey-weighted estimates.
\newblock {\em Statistical Science\/}~{\em 32\/}(2).

\bibitem[\protect\citeauthoryear{Costa and Barros}{Costa and
  Barros}{2016}]{Costa2016}
Costa, D. and H.~Barros (2016).
\newblock Instruments to assess intimate partner violence: A scoping review of
  the literature.
\newblock {\em Violence and Victims\/}~{\em 31\/}(4), 591–621.

\bibitem[\protect\citeauthoryear{Council}{Council}{2002}]{NAP2002}
Council, N.~R. (2002).
\newblock {\em Leveraging Longitudinal Data in Developing Countries: Report of
  a Workshop}.
\newblock National Academies Press.

\bibitem[\protect\citeauthoryear{Dal~Grande, Chittleborough, Campostrini,
  Tucker, and Taylor}{Dal~Grande et~al.}{2015}]{Grande2015}
Dal~Grande, E., C.~R. Chittleborough, S.~Campostrini, G.~Tucker, and A.~W.
  Taylor (2015, 08).
\newblock {Health Estimates Using Survey Raked-Weighting Techniques in an
  Australian Population Health Surveillance System}.
\newblock {\em American Journal of Epidemiology\/}~{\em 182\/}(6), 544--556.

\bibitem[\protect\citeauthoryear{Davis and Taylor}{Davis and
  Taylor}{1997}]{Davis1997}
Davis, R.~C. and B.~G. Taylor (1997).
\newblock A proactive response to family violence: The results of a randomized
  experiment.
\newblock {\em Criminology\/}~{\em 35\/}(2), 307--333.

\bibitem[\protect\citeauthoryear{de~Souza-Júnior, de~Freitas,
  de~Abreu~Antonaci, and Szwarcwald}{de~Souza-Júnior et~al.}{2015}]{Souza2015}
de~Souza-Júnior, P. R.~B., M.~P.~S. de~Freitas, G.~de~Abreu~Antonaci, and
  C.~L. Szwarcwald (2015, 6).
\newblock Desenho da amostra da pesquisa nacional de saúde 2013.
\newblock {\em Epidemiologia e Serviços de Saúde\/}~{\em 24}, 207--216.

\bibitem[\protect\citeauthoryear{DHS}{DHS}{2020}]{DHS2020}
DHS (2020).
\newblock The dhs program – demographic and health surveys. the dhs program
  website. funded by usaid.

\bibitem[\protect\citeauthoryear{Frees}{Frees}{2004}]{Frees2004-qp}
Frees, E.~W. (2004, August).
\newblock {\em Longitudinal and Panel Data}.
\newblock Cambridge, England: Cambridge University Press.

\bibitem[\protect\citeauthoryear{Garcia-Moreno, Jansen, Ellsberg, Heise, and
  Watts}{Garcia-Moreno et~al.}{2005}]{MorenoJansen2005}
Garcia-Moreno, C., H.~A. F.~M. Jansen, M.~Ellsberg, L.~Heise, and C.~Watts
  (2005).
\newblock Who multi-country study on women's health and domestic violence
  against women: initial results on prevalence, health outcomes and women's
  responses.
\newblock Technical report, World Health Organization.

\bibitem[\protect\citeauthoryear{Gondolf}{Gondolf}{2001}]{Gondolf2001}
Gondolf, E.~W. (2001).
\newblock Limitations of experimental evaluation of batterer programs.
\newblock {\em Trauma, Violence, \& Abuse\/}~{\em 2\/}(1), 79--88.

\bibitem[\protect\citeauthoryear{Graham}{Graham}{2009}]{graham_missing_2009}
Graham, J.~W. (2009, January).
\newblock Missing {Data} {Analysis}: {Making} {It} {Work} in the {Real}
  {World}.
\newblock {\em Annual Review of Psychology\/}~{\em 60\/}(1), 549--576.

\bibitem[\protect\citeauthoryear{Graham}{Graham}{2012}]{graham_missing_2012}
Graham, J.~W. (2012).
\newblock {\em Missing {Data}}.
\newblock New York, NY: Springer New York.

\bibitem[\protect\citeauthoryear{Haziza and Beaumont}{Haziza and
  Beaumont}{2017}]{Haziza2017}
Haziza, D. and J.-F. Beaumont (2017).
\newblock Construction of weights in surveys: A review.
\newblock {\em Statistical Science\/}~{\em 32\/}(2), 206--226.

\bibitem[\protect\citeauthoryear{Hirano, Imbens, Ridder, and Rubin}{Hirano
  et~al.}{2001}]{hirano_combining_2001}
Hirano, K., G.~W. Imbens, G.~Ridder, and D.~B. Rubin (2001, November).
\newblock Combining {Panel} {Data} {Sets} with {Attrition} and {Refreshment}
  {Samples}.
\newblock {\em Econometrica\/}~{\em 69\/}(6), 1645--1659.

\bibitem[\protect\citeauthoryear{Hsiao}{Hsiao}{2007}]{Hsiao2007}
Hsiao, C. (2007, March).
\newblock Panel data analysis—advantages and challenges.
\newblock {\em TEST\/}~{\em 16\/}(1), 1–22.

\bibitem[\protect\citeauthoryear{IBGE}{IBGE}{2016}]{IBGE2016}
IBGE (2016).
\newblock Pesquisa nacional por amostra de domicílios (pnad).

\bibitem[\protect\citeauthoryear{IBGE}{IBGE}{2022}]{IBGE2022}
IBGE (2022).
\newblock Pesquisa nacional por amostra de domicílios (pnad).

\bibitem[\protect\citeauthoryear{Jewkes}{Jewkes}{2002}]{Jewkes2002}
Jewkes, R. (2002, April).
\newblock Intimate partner violence: causes and prevention.
\newblock {\em The Lancet\/}~{\em 359\/}(9315), 1423–1429.

\bibitem[\protect\citeauthoryear{Johnson}{Johnson}{2005}]{Johnson2005}
Johnson, D. (2005, September).
\newblock Two‐wave panel analysis: Comparing statistical methods for studying
  the effects of transitions.
\newblock {\em Journal of Marriage and Family\/}~{\em 67\/}(4), 1061–1075.

\bibitem[\protect\citeauthoryear{Kang}{Kang}{2013}]{kang_prevention_2013}
Kang, H. (2013).
\newblock The prevention and handling of the missing data.
\newblock {\em Korean Journal of Anesthesiology\/}~{\em 64\/}(5), 402.

\bibitem[\protect\citeauthoryear{Kish}{Kish}{1992}]{Kish1992}
Kish, L. (1992, May).
\newblock Weighting for unequal, $p_{i}$.
\newblock {\em Journal of Official Statistics\/}~{\em 2\/}(8), 183--200.

\bibitem[\protect\citeauthoryear{Lavallée and Beaumont}{Lavallée and
  Beaumont}{2015}]{LavalleeBeaumont2015}
Lavallée, P. and J.-F. Beaumont (2015).
\newblock Why we should put some weight on weights.

\bibitem[\protect\citeauthoryear{Little and Vartivarian}{Little and
  Vartivarian}{2005}]{LittleVartivarian2005}
Little, R.~J. and S.~Vartivarian (2005).
\newblock Does weighting for nonresponse increase the variance of survey means?
\newblock {\em Survey Methodology\/}~{\em 31\/}(2), 161--168.

\bibitem[\protect\citeauthoryear{Lumley}{Lumley}{2020}]{Lumley2020}
Lumley, T. (2020).
\newblock survey: analysis of complex survey samples.
\newblock R package version 4.0.

\bibitem[\protect\citeauthoryear{Lynn}{Lynn}{2021}]{LynnLongitudinal2021}
Lynn, P. (Ed.) (2021, mar).
\newblock {\em Advances in {Longitudinal} {Survey} {Methodology}\/} (1 ed.).
\newblock Wiley {Series} in {Probability} and {Statistics}. Wiley.

\bibitem[\protect\citeauthoryear{McFarlane, Greenberg, Weltge, and
  Watson}{McFarlane et~al.}{1995}]{McFarlane1995}
McFarlane, J., L.~Greenberg, A.~Weltge, and M.~Watson (1995, October).
\newblock Identification of abuse in emergency departments: Effectiveness of a
  two-question screening tool.
\newblock {\em Journal of Emergency Nursing\/}~{\em 21\/}(5), 391–394.

\bibitem[\protect\citeauthoryear{McHugo, Kammerer, Jackson, Markoff, Gatz,
  Larson, Mazelis, and Hennigan}{McHugo et~al.}{2005}]{MCHUGO2005}
McHugo, G., N.~Kammerer, E.~Jackson, L.~Markoff, M.~Gatz, M.~Larson,
  R.~Mazelis, and K.~Hennigan (2005).
\newblock Women, co-occurring disorders, and violence study: Evaluation design
  and study population.
\newblock {\em Journal of Substance Abuse Treatment\/}~{\em 28\/}(2), 91--107.

\bibitem[\protect\citeauthoryear{Mertin and Mohr}{Mertin and
  Mohr}{2001}]{mertin2001follow}
Mertin, P. and P.~B. Mohr (2001).
\newblock A follow-up study of posttraumatic stress disorder, anxiety, and
  depression in australian victims of domestic violence.
\newblock {\em Violence and Victims\/}~{\em 16\/}(6), 645.

\bibitem[\protect\citeauthoryear{Mulla, Witte, Richardson, Hart, Kassing,
  Coffey, Hackman, and Sherwood}{Mulla et~al.}{2018}]{Mulla2018}
Mulla, M.~M., T.~H. Witte, K.~Richardson, W.~Hart, F.~L. Kassing, C.~A. Coffey,
  C.~L. Hackman, and I.~M. Sherwood (2018, September).
\newblock The causal influence of perceived social norms on intimate partner
  violence perpetration: Converging cross-sectional, longitudinal, and
  experimental support for a social disinhibition model.
\newblock {\em Personality and Social Psychology Bulletin\/}~{\em 45\/}(4),
  652–668.

\bibitem[\protect\citeauthoryear{O’Muircheartaigh and
  Pedlow}{O’Muircheartaigh and Pedlow}{2002}]{o2002combining}
O’Muircheartaigh, C. and S.~Pedlow (2002).
\newblock Combining samples vs. cumulating cases: a comparison of two weighting
  strategies in nlsy97.
\newblock In {\em American Statistical Association Proceedings of the Joint
  Statistical Meetings}, pp.\  2557--2562. American Statistical Association.

\bibitem[\protect\citeauthoryear{Papp, Mueller-Smith, Kearns, and
  Peterson}{Papp et~al.}{2023}]{Papp2023}
Papp, J., M.~Mueller-Smith, M.~C. Kearns, and C.~Peterson (2023, September).
\newblock Inventory of u.s. public data sources to measure the socioeconomic
  impact of experiencing interpersonal violence.
\newblock {\em AJPM Focus\/}~{\em 2\/}(3), 100114.

\bibitem[\protect\citeauthoryear{Pinheiro}{Pinheiro}{2010}]{PAD_MG_2010}
Pinheiro, F.~J. (2010).
\newblock Pesquisa por amostra de domicílios de minas gerais (pad-mg): Plano
  amostral, métodos de ponderação e metodologia.
\newblock Technical report, Fundação João Pinheiro, Governo de Minas Gerais.

\bibitem[\protect\citeauthoryear{{R Core Team}}{{R Core
  Team}}{2023}]{RCoreTeam}
{R Core Team} (2023).
\newblock {\em R: A Language and Environment for Statistical Computing}.
\newblock Vienna, Austria: R Foundation for Statistical Computing.

\bibitem[\protect\citeauthoryear{Renzetti, Edleson, and Bergen}{Renzetti
  et~al.}{2011}]{Renzetti2011}
Renzetti, C., J.~Edleson, and R.~Bergen (2011).
\newblock {\em Sourcebook on Violence against Women}.
\newblock SAGE Publications, Inc.

\bibitem[\protect\citeauthoryear{Rose and Corti}{Rose and
  Corti}{2000}]{Rose2000-po}
Rose, D. and L.~Corti (Eds.) (2000, nov).
\newblock {\em Researching social and economic change}.
\newblock Social research today. London, England: Routledge.

\bibitem[\protect\citeauthoryear{Rose}{Rose}{2018}]{Rose2018}
Rose, R.~A. (2018, November).
\newblock Frameworks for credible causal inference in observational studies of
  family violence.
\newblock {\em Journal of Family Violence\/}~{\em 34\/}(8), 697–710.

\bibitem[\protect\citeauthoryear{Rothman}{Rothman}{2018}]{Rothman2018}
Rothman, E.~F. (2018).
\newblock {\em Theories on the Causation of Partner Abuse Perpetration}, pp.\
  25–51.
\newblock Elsevier.

\bibitem[\protect\citeauthoryear{Ruiz-Perez, Plazaola-Castano, and
  Vives-Cases}{Ruiz-Perez et~al.}{2007}]{Ruiz-Perez2007}
Ruiz-Perez, I., J.~Plazaola-Castano, and C.~Vives-Cases (2007, dec).
\newblock Methodological issues in the study of violence against women.
\newblock {\em Journal of Epidemiology {\&} Community Health\/}~{\em
  61\/}(Supplement 2), ii26--ii31.

\bibitem[\protect\citeauthoryear{Sardinha, Maheu-Giroux, St\"{o}ckl, Meyer, and
  García-Moreno}{Sardinha et~al.}{2022}]{Sardinha2022}
Sardinha, L., M.~Maheu-Giroux, H.~St\"{o}ckl, S.~R. Meyer, and
  C.~García-Moreno (2022, February).
\newblock Global, regional, and national prevalence estimates of physical or
  sexual, or both, intimate partner violence against women in 2018.
\newblock {\em The Lancet\/}~{\em 399\/}(10327), 803–813.

\bibitem[\protect\citeauthoryear{Sarndal and Lundstrom}{Sarndal and
  Lundstrom}{2005}]{Sarndal2005-ca}
Sarndal, C.-E. and S.~Lundstrom (2005, June).
\newblock {\em Estimation in Surveys with Nonresponse}.
\newblock Wiley Series in Survey Methodology. Hoboken, NJ: Wiley-Blackwell.

\bibitem[\protect\citeauthoryear{Schraiber, D’Oliveira, Fran\c{c}a-Junior,
  Diniz, Portella, Ludermir, Valen\c{c}a, and Couto}{Schraiber
  et~al.}{2007}]{Schraiber2007}
Schraiber, L.~B., A.~F. P.~L. D’Oliveira, I.~Fran\c{c}a-Junior, S.~Diniz,
  A.~P. Portella, A.~B. Ludermir, O.~Valen\c{c}a, and M.~T. Couto (2007,
  October).
\newblock Preval\^encia da viol\^encia contra a mulher por parceiro íntimo em
  regiões do brasil.
\newblock {\em Revista de Saúde Pública\/}~{\em 41\/}(5), 797–807.

\bibitem[\protect\citeauthoryear{Singer and Willett}{Singer and
  Willett}{2003}]{Singer2003-ks}
Singer, J.~D. and J.~B. Willett (2003, April).
\newblock {\em Applied longitudinal data analysis}.
\newblock New York, NY: Oxford University Press.

\bibitem[\protect\citeauthoryear{Solon, Haider, and Wooldridge}{Solon
  et~al.}{2015}]{Solon2015}
Solon, G., S.~J. Haider, and J.~M. Wooldridge (2015).
\newblock What are we weighting for?
\newblock {\em Journal of Human Resources\/}~{\em 50\/}(2), 301–316.

\bibitem[\protect\citeauthoryear{Straus, Hamby, Boney-McCoy, and
  Sugarman}{Straus et~al.}{1996}]{Straus1996}
Straus, M.~A., S.~L. Hamby, S.~Boney-McCoy, and D.~B. Sugarman (1996).
\newblock The revised conflict tactics scales (cts2) development and
  preliminary psychometric data.
\newblock {\em Journal of family issues\/}~{\em 17\/}(3), 283--316.

\bibitem[\protect\citeauthoryear{Tarriño‐Concejero, Gil‐García,
  Barrientos‐Trigo, and García‐Carpintero‐Muñoz}{Tarriño‐Concejero
  et~al.}{2022}]{TarrioConcejero2022}
Tarriño‐Concejero, L., E.~Gil‐García, S.~Barrientos‐Trigo, and M.~d.
  l.~À. García‐Carpintero‐Muñoz (2022, July).
\newblock Instruments used to measure dating violence: A systematic review of
  psychometric properties.
\newblock {\em Journal of Advanced Nursing\/}~{\em 79\/}(4), 1267–1289.

\bibitem[\protect\citeauthoryear{Taylor, Tong, and Maxwell}{Taylor
  et~al.}{2020}]{taylor_evaluating_2020}
Taylor, L.~K., X.~Tong, and S.~E. Maxwell (2020, March).
\newblock Evaluating {Supplemental} {Samples} in {Longitudinal} {Research}:
  {Replacement} and {Refreshment} {Approaches}.
\newblock {\em Multivariate Behavioral Research\/}~{\em 55\/}(2), 277--299.

\bibitem[\protect\citeauthoryear{Thompsn, Basile, Hertz, and Sitterle}{Thompsn
  et~al.}{2006}]{Thompsn2006MeasuringIP}
Thompsn, M.~P., K.~C. Basile, M.~F. Hertz, and D.~Sitterle (2006).
\newblock Measuring intimate partner violence victimization and perpetration; a
  compendium of assessment tools.

\bibitem[\protect\citeauthoryear{Tourangeau and Yan}{Tourangeau and
  Yan}{2007}]{Tourangeau2007}
Tourangeau, R. and T.~Yan (2007).
\newblock Sensitive questions in surveys.
\newblock {\em Psychological Bulletin\/}~{\em 133\/}(5), 859–883.

\bibitem[\protect\citeauthoryear{United-Nations}{United-Nations}{2014}]{UN2014}
United-Nations (2014).
\newblock {\em Guidelines for Producing Statistics on Violence against Women}.
\newblock United Nations - Department of Economic and Social Affairs.

\bibitem[\protect\citeauthoryear{Vaisey and Miles}{Vaisey and
  Miles}{2016}]{Vaisey2016}
Vaisey, S. and A.~Miles (2016, July).
\newblock What you can—and can’t—do with three-wave panel data.
\newblock {\em Sociological Methods \&; Research\/}~{\em 46\/}(1), 44–67.

\bibitem[\protect\citeauthoryear{Valliant and Dever}{Valliant and
  Dever}{2018}]{Valliant2018-mc}
Valliant, R. and J.~A. Dever (2018, January).
\newblock {\em Survey weights}.
\newblock Philadelphia, PA: Stata Press.

\bibitem[\protect\citeauthoryear{Valliant, Dever, and Kreuter}{Valliant
  et~al.}{2018}]{Valliant2018}
Valliant, R., J.~A. Dever, and F.~Kreuter (2018, October).
\newblock {\em Practical Tools for Designing and Weighting Survey Samples}.
\newblock Springer-Verlag GmbH.

\bibitem[\protect\citeauthoryear{Vehovar}{Vehovar}{2003}]{Vehovar2003}
Vehovar, V. (2003).
\newblock Field substitutions redefined.
\newblock {\em The Survey Statistician\/}~{\em 48}, 35--37.

\bibitem[\protect\citeauthoryear{Watson}{Watson}{2014}]{Watson_2014}
Watson, N. (2014, Dec.).
\newblock Evaluation of weighting methods to integrate a top-up sample with an
  ongoing longitudinal sample.
\newblock {\em Survey Research Methods\/}~{\em 8\/}(3), 195–208.

\bibitem[\protect\citeauthoryear{Watson and Lynn}{Watson and
  Lynn}{2021}]{Watson2021}
Watson, N. and P.~Lynn (2021).
\newblock {\em Refreshment Sampling for Longitudinal Surveys}, Chapter~1, pp.\
  1--25.
\newblock John Wiley \& Sons, Ltd.

\bibitem[\protect\citeauthoryear{WHO}{WHO}{2001}]{WHO2001}
WHO (2001).
\newblock Putting women first: ethical and safety recommendations for research
  on domestic violence against women.

\bibitem[\protect\citeauthoryear{Yount, Cheong, Khan, Bergenfeld, Kaslow, and
  Clark}{Yount et~al.}{2022}]{Yount2022}
Yount, K.~M., Y.~F. Cheong, Z.~Khan, I.~Bergenfeld, N.~Kaslow, and C.~J. Clark
  (2022, March).
\newblock Global measurement of intimate partner violence to monitor
  sustainable development goal 5.
\newblock {\em BMC Public Health\/}~{\em 22\/}(1).

\end{thebibliography}

\newpage

\begin{landscape}

\section*{Appendix}

\begin{table}[!h] \centering 
  \caption{Results of the logit models used for estimating non-response in 2016 (Wave I)}
  \label{eq:logit2016} 
\begin{tabular}{@{\extracolsep{5pt}}lccccccccc} 
\\[-1.8ex]\hline 
\hline \\[-1.8ex] 
 & \multicolumn{9}{c}{\textit{Dependent variable:}} \\ 
\cline{2-10} 
\\[-1.8ex] & \multicolumn{9}{c}{Answer Violence} \\ 
 & AL & BA & CE & MA & PB & PE & PI & RN & SE \\ 
\\[-1.8ex] & (1) & (2) & (3) & (4) & (5) & (6) & (7) & (8) & (9)\\ 
\hline \\[-1.8ex] 
 Cohabitation & $-$0.113 & 0.111 & $-$1.001$^{***}$ & $-$0.727$^{***}$ & $-$0.962$^{***}$ & $-$0.320$^{*}$ & $-$0.395$^{*}$ & $-$0.402$^{**}$ & 0.221 \\ 
  & (0.194) & (0.186) & (0.220) & (0.183) & (0.294) & (0.168) & (0.214) & (0.177) & (0.187) \\ 
  & & & & & & & & & \\ 
 Know Victim & 0.026 & $-$0.311$^{*}$ & 0.338 & 0.498$^{**}$ & 0.246 & 0.318$^{*}$ & 0.937$^{***}$ & 0.454$^{**}$ & 0.158 \\ 
  & (0.207) & (0.181) & (0.225) & (0.211) & (0.283) & (0.186) & (0.269) & (0.186) & (0.197) \\ 
  & & & & & & & & & \\ 
 Children & 0.147 & 0.096 & $-$0.727$^{***}$ & 0.181 & $-$0.128 & 0.223 & $-$0.237 & 0.062 & 0.287 \\ 
  & (0.198) & (0.187) & (0.216) & (0.183) & (0.270) & (0.168) & (0.226) & (0.178) & (0.189) \\ 
  & & & & & & & & & \\ 
 Intercept & 1.388$^{***}$ & 1.612$^{***}$ & 2.703$^{***}$ & 1.447$^{***}$ & 2.990$^{***}$ & 0.881$^{***}$ & 1.537$^{***}$ & 0.712$^{***}$ & 0.924$^{***}$ \\ 
  & (0.185) & (0.168) & (0.233) & (0.167) & (0.280) & (0.157) & (0.212) & (0.145) & (0.172) \\ 
  & & & & & & & & & \\ 
\hline \\[-1.8ex] 
Observations & 739 & 922 & 897 & 794 & 836 & 783 & 632 & 630 & 745 \\ 
Log Likelihood & $-$335.927 & $-$377.446 & $-$354.657 & $-$383.626 & $-$238.991 & $-$410.701 & $-$282.201 & $-$405.811 & $-$338.702 \\ 
Akaike Inf. Crit. & 679.855 & 762.893 & 717.314 & 775.251 & 485.983 & 829.401 & 572.403 & 819.621 & 685.405 \\ 
\hline 
\hline \\[-1.8ex]
\multicolumn{10}{l}{\rule{0pt}{1em}\textit{Source: } Elaborated by the authors.}\\
\multicolumn{10}{l}{\rule{0pt}{1em}\textit{Note: } $^{*}$p$<$0.1; $^{**}$p$<$0.05; $^{***}$p$<$0.01}
\end{tabular} 
\end{table} 

\newpage

\begin{table}[!h] \centering 
  \caption{Results of the logit models used for estimating non-response in 2017 (Wave II)}
  \label{eq:logit2017} 
\begin{tabular}{@{\extracolsep{5pt}}lccccccccc} 
\\[-1.8ex]\hline 
\hline \\[-1.8ex] 
 & \multicolumn{9}{c}{\textit{Dependent variable:}} \\ 
\cline{2-10} 
\\[-1.8ex] & \multicolumn{9}{c}{Answer Violence} \\ 
 & AL & BA & CE & MA & PB & PE & PI & RN & SE \\ 
\\[-1.8ex] & (1) & (2) & (3) & (4) & (5) & (6) & (7) & (8) & (9)\\ 
\hline \\[-1.8ex] 
 Cohabitation & $-$0.435$^{**}$ & $-$0.253 & 0.650$^{***}$ & $-$0.283 & 0.262 & 0.635$^{***}$ & $-$0.761$^{***}$ & $-$0.028 & 0.702$^{***}$ \\ 
  & (0.194) & (0.163) & (0.213) & (0.190) & (0.212) & (0.140) & (0.223) & (0.148) & (0.215) \\ 
  & & & & & & & & & \\ 
 Know Victim & 0.197 & 0.568$^{***}$ & 1.266$^{***}$ & 0.164 & 1.033$^{***}$ & 0.697$^{***}$ & 0.270 & 0.923$^{***}$ & 0.510$^{*}$ \\ 
  & (0.226) & (0.176) & (0.303) & (0.230) & (0.299) & (0.168) & (0.265) & (0.176) & (0.270) \\ 
  & & & & & & & & & \\ 
 Children & 0.383$^{*}$ & 0.026 & $-$0.126 & 0.145 & $-$0.119 & $-$0.615$^{***}$ & 0.199 & $-$0.428$^{***}$ & $-$0.749$^{***}$ \\ 
  & (0.213) & (0.180) & (0.241) & (0.217) & (0.225) & (0.141) & (0.225) & (0.151) & (0.211) \\ 
  & & & & & & & & & \\ 
 Intercept & 1.582$^{***}$ & 1.426$^{***}$ & 1.699$^{***}$ & 1.700$^{***}$ & 1.585$^{***}$ & 0.308$^{***}$ & 2.021$^{***}$ & 0.489$^{***}$ & 1.592$^{***}$ \\ 
  & (0.168) & (0.131) & (0.151) & (0.152) & (0.161) & (0.113) & (0.201) & (0.121) & (0.153) \\ 
  & & & & & & & & & \\ 
\hline \\[-1.8ex] 
Observations & 825 & 1,079 & 1,107 & 840 & 856 & 1,036 & 763 & 869 & 802 \\ 
Log Likelihood & $-$394.113 & $-$513.336 & $-$346.059 & $-$392.789 & $-$296.882 & $-$608.080 & $-$334.818 & $-$600.036 & $-$327.007 \\ 
Akaike Inf. Crit. & 796.226 & 1,034.672 & 700.118 & 793.578 & 601.764 & 1,224.160 & 677.635 & 1,208.073 & 662.015 \\ 
\hline 
\hline \\[-1.8ex] 
\multicolumn{10}{l}{\rule{0pt}{1em}\textit{Source: } Elaborated by the authors.}\\
\multicolumn{10}{l}{\rule{0pt}{1em}\textit{Note: } $^{*}$p$<$0.1; $^{**}$p$<$0.05; $^{***}$p$<$0.01}
\end{tabular} 
\end{table} 

\end{landscape}
\end{document}